\documentclass[fleqn,usenatbib]{mnras}

\usepackage[T1]{fontenc}
\usepackage{ae,aecompl}

\usepackage{graphicx}	
\usepackage{amsmath}	
\usepackage{amssymb}	
\usepackage{xcolor}
\usepackage{longtable}
\usepackage{booktabs}
\usepackage{multirow}
\usepackage[utf8]{inputenc}
\usepackage{ulem}
\newcommand{\lya}{Ly$\alpha$}

\newcommand{\oxigen}{\ion{O}}
\newcommand{\oi}{\ion{O}{i}} 
 
\newcommand{\oii}{[\ion{O}{ii}]$\lambda$3727}

\newcommand{\Niii}{\ion{Ni}{ii}} 
\newcommand{\feii}{\ion{Fe}{ii}}
 
\newcommand{\caii}{\ion{Ca}{ii}} 
\newcommand{\fe}{\ion{Fe}}
\newcommand{\civ}{\ion{C}{iv}} 
\newcommand{\cii}{\ion{C}{ii}}

\newcommand{\siiv}{\ion{Si}{iv}}
\newcommand{\siii}{\ion{Si}{ii}}
\newcommand{\si}{\ion{Si}}
\newcommand{\nai}{\ion{Na}{i}}

\newcommand{\zn}{\ion{Zn}}
\newcommand{\alii}{\ion{Al}{ii}}
\newcommand{\mgii}{\ion{Mg}{ii}} 
\newcommand{\hi}{\ion{H}{i}}

\title[GRB\,160410A]{GRB\,160410A: the first Chemical Study of the Interstellar Medium of a Short GRB \thanks{Based on observations made with telescopes at the European Southern Observatory at La Silla/Paranal, Chile under programme 097.A-0036(A).}}
\author[Ag\"u\'i Fern\'andez]{
J. F. Ag\"u\'i Fern\'andez$^{1}$\thanks{E-mail: feli@iaa.es}, C. C. Th\"one$^{1}$, D. A. Kann$^{1}$, A. de Ugarte Postigo$^{1, 2}$, J. Selsing$^{3, 4}$, P. Schady$^{5}$,\newauthor R. M. Yates$^{6}$, J. Greiner$^{7}$, S. R. Oates$^{8}$, D. B. Malesani$^{9, 10}$, D. Xu$^{11}$, A. Klotz$^{12}$, S. Campana$^{13}$, A. Rossi$^{14}$, \newauthor D. A. Perley$^{15}$, M. Bla\v zek$^{1}$, P. D'Avanzo$^{13}$, A. Giunta$^{16}$, D. Hartmann$^{17}$, K. E. Heintz$^{4, 18}$, P. Jakobsson$^{18}$, \newauthor C. C. Kirkpatrick IV$^{19, 20}$, C. Kouveliotou$^{21, 22}$, A. Melandri$^{13}$, G. Pugliese$^{23}$, R. Salvaterra$^{24}$, \newauthor R. L. C. Starling$^{25}$, N. R. Tanvir$^{25}$, S. D. Vergani$^{26, 27}$ and K. Wiersema$^{28}$\\
$^{1}$Instituto de Astrof\'isica de Andaluc\'ia, Glorieta de la Astronom\'ia s/n, 18008 Granada, Spain.\\
$^{2}$Artemis, Universit\'e C\^ote d'Azur, Observatoire de la C\^ote d'Azu CNRS, F-06304, Nice, France.\\
$^{3}$DARK, Niels Bohr Institute, University of Copenhagen, Jagtvej 128, 2200 Copenhagen, Denmark.\\
$^{4}$Cosmic Dawn Center (DAWN), Niels Bohr Institute, University of Copenhagen, Jagtvej 128, 2100 Copenhagen Ø, Denmark.\\
$^{5}$Department of Physics, University of Bath, Claverton Down, Bath, BA2 7AY, UK.\\
$^{6}$Department of Physics, University of Surrey, Stag Hill, Guildford, GU2 7XH, UK.\\
$^{7}$Max-Planck Institut f\"ur Extraterrestrische Physik, Giessenbachstr 1, 85748 Garching, Germany.\\
$^{8}$School of Physics and Astronomy \& Institute for Gravitational Wave Astronomy, University of Birmingham, B15 2TT, UK.\\
$^{9}$DTU Space, National Space Institute, Technical University of Denmark, DK-2800 Kongens Lyngby, Denmark.\\
$^{10}$Department of Astrophysics/IMAPP, Radboud University Nijmegen, P.O. Box 9010, 6500 GL Nijmegen, The Netherlands.\\
$^{11}$CAS Key Laboratory of Space Astronomy and Technology, National Astronomical Observatories, Chinese Academy of Sciences, Beijing 100101, China.\\
$^{12}$IRAP, Universit\'e de Toulouse, CNRS, CNES, UPS, (Toulouse), France.\\
$^{13}$Brera Astronomical Observatory, via Bianchi 46, I-23807, Merate (LC), Italy.\\
$^{14}$INAF - Osservatorio di Astrofisica e Scienza dello Spazio, via Piero Gobetti 93/3, 40129 Bologna, Italy.\\
$^{15}$Astrophysics Research Institute, Liverpool John Moores University, IC2, Liverpool Science Park, 146 Brownlow Hill, Liverpool L3 5RF, UK.\\
$^{16}$Space Science Data Center (SSDC) - Agenzia Spaziale Italiana (ASI), Via del Politecnico, I-00133 Roma, Italy.\\
$^{17}$Clemson University, Department of Physics and Astronomy, Clemson, SC 29634-0978, USA.\\
$^{18}$Centre for Astrophysics and Cosmology, Science Institute, University of Iceland, Dunhagi 5, 107, Reykjav\'ik, Iceland.\\
$^{19}$Department of Physics, University of Helsinki, PO Box 64, FI-00014 Helsinki, Finland.\\
$^{20}$Helsinki Institute of Physics, Gustaf H\"allstr\"omin katu 2, University of Helsinki, PO Box 64, FI-00014, Helsinki, Finland.\\
$^{21}$Department of Physics, The George Washington University, 725 21st Street NW, Washington, DC 20052, USA.\\
$^{22}$Astronomy, Physics, and Statistics Institute of Sciences (APSIS), The George Washington University, Washington, DC 20052, USA.\\
$^{23}$Astronomical Institute Anton Pannekoek, University of Amsterdam, PO Box 94249, NL-1090 GE Amsterdam, The Netherlands.\\
$^{24}$INAF-Istituto di Astrofisica Spaziale e Fisica cosmica, via Alfonso Corti 12, 20133 Milano, Italy.\\
$^{25}$School of Physics and Astronomy, University of Leicester, University Road, Leicester, LE1 7RH, UK.\\
$^{26}$GEPI, Observatoire de Paris, PSL University, CNRS, 5 Place Jules Janssen, 92190 Meudon, France.\\
$^{27}$Institut d’Astrophysique de Paris, UMR 7095, CNRS-SU, 98 bis boulevard Arago, 75014 Paris, France.\\
$^{28}$Physics Department, Lancaster University, Lancaster, LA1 4YB, UK
}

\date{Accepted XXX. Received YYY; in original form ZZZ}
\pubyear{2023}
\usepackage{newtxtext,newtxmath}

\begin{document}
\label{firstpage}
\pagerange{\pageref{firstpage}--\pageref{lastpage}}
\maketitle
\clearpage
\begin{abstract}
Short gamma-ray bursts (SGRBs) are produced by the coalescence of compact binary systems which are remnants of massive stars. GRB\,160410A is classified as a short-duration GRB with extended emission and is currently the farthest SGRB with a redshift determined from an afterglow spectrum and also one of the brightest SGRBs to date. The fast reaction to the \textit{Neil Gehrels Swift Observatory} alert allowed us to obtain a spectrum of the afterglow using the X-shooter spectrograph at the Very Large Telescope (VLT). The spectrum shows several absorption features at a redshift of $z=1.7177$, in addition, we detect two intervening systems at $z=1.581$ and $z=1.444$. The spectrum shows  \lya\ in absorption with a column density of $\log (N{\rm (HI)/cm}^{2}) = 21.2 \pm 0.2$ which, together with \feii, \cii, \siii, \alii and \oi, allow us to perform the first study of chemical abundances in a SGRB host galaxy. We determine a metallicity of [X/H] = $-2.3\pm0.2$ for \feii\ and $-2.5\pm0.2$ for \siii\ and  no dust depletion. We also find no evidence for extinction in the afterglow Spectral Energy Distribution (SED) modeling. The environment has a low degree of ionisation and the \civ{} and \siiv{} lines are completely absent. We do not detect an underlying host galaxy down to deep limits. Additionally, we compare GRB\,160410A to GRB\,201221D, another high-$z$ short GRB that shows absorption lines at $z=1.045$ and an underlying massive host galaxy.

\end{abstract}




\section{Introduction}
For a brief moment, gamma-ray bursts (GRBs) are capable of outshining any other source in the Universe. Their $\gamma$-ray flashes can last from significantly less than a second to hundreds or even thousands of seconds \citep{1993ApJ...413L.101K}. According to their duration and spectral characteristics, GRBs can be divided into two classes, long/soft GRBs (LGRBs) and short/hard GRBs (SGRBs). LGRBs are associated with the collapse of very massive stars and their prompt $\gamma$-ray emission in most cases lasts for more than 2\,s. They have been shown to be linked to broad-lined Type Ic core-collapse supernovae \protect\citep[e.g.][]{1998Natur.395..670G, 2003Natur.423..847H, 2006ARA&A..44..507W, 2012grb..book..169H, 2017AdAst2017E...5C}.

SGRBs, in contrast, are associated with the merger of a binary system of compact objects, usually two neutron stars  \citep[NSs,][]{2014ARA&A..52...43B,2017ApJ...848L..13A}. They show a harder $\gamma$-ray spectrum than LGRBs and have a $T_{90}$\footnote{$T_{90}$ is defined as the time span during which from 5\% to 95\% of the total counts emitted by a GRB are detected.} of less than 2\,s  \protect\citep{1993ApJ...413L.101K}, although some events show extended emission (EE), albeit with a softer spectrum, and recently, an event, GRB 211211A, which was almost indistinguishable from a long GRB, has been associated with a compact object merger \citep{2022Natur.612..223R, 2022NatAs.tmp..264G, 2022Natur.612..232Y, 2022Natur.612..228T}. SGRB afterglows are typically less luminous than those of LGRBs \citep{Kann2011ApJ} which makes them much more difficult to detect at higher redshifts. A recent study by \protect\cite{2021ApJ...911L..28D} suggests that at high redshifts, there is a bias towards SGRBs with extended emission as they are typically brighter than regular short GRBs.

In a SGRB, during coalescence of the binary system, a relativistic jet forms producing the prompt emission and later the afterglow emission by interaction with the circumburst environment, the same way the afterglow is produced for LGRBs. SGRBs also show a so-called \textit{kilonova} emission (KN), powered by the radioactive decay of heavy elements produced via the $r$-process in a neutron-rich environment  \citep{2010MNRAS.406.2650M}. The  KN emission is normally much fainter than the afterglow and hence is only detected a few days after the GRB when the afterglow has faded  \protect\citep[e.g.][]{2019LRR....23....1M}.

Until recently, GRBs have only been detected by high-energy instruments on-board satellites such as the \textit{Neil Gehrels Swift Observatory} (\textit{Swift} hereafter)/Burst Alert Telescope \protect\citep[BAT,][]{2004ApJ...611.1005G, Barthelmy2005SSRv}, \textit{Fermi}/Gamma-Ray Burst Monitor \protect\citep[GBM,][]{2009ApJ...702..791M}, or Konus-\textit{Wind} \protect\citep{1995SSRv...71..265A}. In 2017 a gravitational wave (GW) was detected by LIGO/Virgo together with a corresponding electromagnetic counterpart at various wavelengths and the SGRB\,170817A detected by \textit{Fermi}/GBM and the SPI-ACS (SPectrometer on INTEGRAL - Anti-Coincidence Shield) on \textit{INTEGRAL} just $\sim$\,1.7\,s after the GW detection, firmly linking NS mergers to SGRBs  \protect\citep{2017ApJ...848L..14G, 2017ApJ...848L..15S, 2017ApJ...848L..13A, 2017ApJ...848L..12A, 2017ApJ...848L..27T}. However, to date this has been the only event where an electromagnetic counterpart has been detected in association with a GW signal from a NS-NS merger \protect\citep[e.g.][]{2020MNRAS.497.5518A}. Optical surveys of the sky such as the Zwicky Transient Facility (ZTF) are expected to significantly increase the KN detection rate, even in the absence of a $\gamma$-ray or GW signal  \protect\citep{2019PASP..131a8002B, 2019PASP..131g8001G,Andreoni2021arXiv}.

The presence of KN emission has been claimed for a small number of SGRBs, however, the first conclusive detection was obtained for GRB\,130603B \protect\citep{Tanvir2013Nature,Berger2013ApJ}. This SGRB was also the first showing absorption lines in the afterglow spectrum \protect\citep{deUgartePostigo2014AA}. Since then, only a few SGRBs  had a  redshift spectroscopically determined via the afterglow. This makes the study of the redshift distribution and properties of the interstellar medium (ISM) challenging for the class of SGRBs and often only an indirect redshift via association with a likely host galaxy is available \citep[e.g.][]{2010ApJ...722.1946B}.

Short bursts are associated with host galaxies featuring a wide distribution of stellar population ages and galaxy types. SGRBs do not seem to have a preferred location in their host galaxies. Some have even been detected at large distances from their putative hosts  \protect\citep[e.g.][]{2010ApJ...725.1202L, 2010ApJ...722.1946B, 2013ApJ...776...18F}. The offset distribution can be explained by the time needed for the compact objects to form from their massive star binary progenitors and the subsequent delay time for the system to merge due to GW energy loss \protect\citep{2019MNRAS.487.4847B, 2006ApJ...648.1110B, Paterson2020ApJ}.

Several studies suggest compact object mergers (NS-NS, or Neutron Star - Black Hole, NS-BH) as a major source for $r$-process element enhancement \protect\citep{2019Natur.574..497W, 2016AJ....151...82R} in dwarf galaxies \protect\citep{2015MNRAS.454.1073B} as well as in ultra-faint dwarf (UFD) galaxies \protect\citep{2016ApJ...829L..13B}. This is the case for Reticulum\,II, a UFD  containing very metal-poor stars with a higher abundance of $r$-process elements than expected from chemical evolution driven by typical core-collapse supernovae (CC-SNe). It has been suggested that a single NS-NS merger could generate the $r$-process element abundances observed in these galaxies, however, rare CC-SNe cannot be ruled out \protect\citep{2016Natur.531..610J, 2016ApJ...832..149B}.

In contrast, long GRBs are commonly found in bright, metal-poor regions within their host galaxies \protect\citep[e.g.][]{2006Natur.441..463F, 2006A&A...460L..13J, 2008ApJ...676.1151T, 2017MNRAS.467.1795L} and typically with large neutral hydrogen column densities. Most LGRB hosts have $\log (N{\rm (HI)/cm}^{2})$ > 20.3 \citep[e.g.][]{2019MNRAS.483.5380T}, which is the definition for a Damped \lya\ (DLA) system. LGRBs have proven to be ideal beacons in the study of neutral and ionised gas evolution in absorption in the ISM, the Circumgalactic Medium (CGM) and the Intergalactic Medium (IGM) \protect\citep[e.g.][]{2005ARA&A..43..861W, 2013MNRAS.431.3159S, 2019A&A...623A..92S, 2019A&A...623A..43B, 2019ApJ...884...66G}.

In this paper we present a study of the optical counterpart of GRB\,160410A and its afterglow spectrum. GRB\,160410A is the first SGRB for which the spectral observations span a large enough spectral range, together with the SGRB redshift, to cover the \lya\ absorption line and has sufficient quality to make a chemical study of the gas in its host galaxy. We also include GRB\,201221D in our analysis, a SGRB with absorption lines in the spectrum, which had a lower redshift of $z=1.045$ and a less broad spectral coverage (see Sect. \ref{sec201221dSpec}). The paper is structured as follows: In Sect.\,\ref{sec:observations} we present the observations of the afterglow and host galaxy of both GRBs, Sect.\,\ref{sec:results} presents the results on the analysis of the spectrum of the burst afterglow and its light curve as well as observations of the field to detect the host galaxy. For GRB\,201221D we also present the analysis and properties of its associated host galaxy. In Sect.\,\ref{sec:discussion} we put the results in context of similar studies for long GRBs and in Sect.\,\ref{sec:conclusions} we present our final conclusions.

Throughout this study, we adopt a cosmological model with $H_0=67.3\,\textnormal{km\,s}^{-1}\textnormal{Mpc}^{-1}$, $\Omega_M=0.315$, $\Omega_\Lambda=0.685$ \protect\citep{2014A&A...571A..16P}.

\section{Observations}
\label{sec:observations} 

\subsection{High-energy detection of GRB~160410A}

\textit{Swift}/BAT triggered on a source at $\textnormal{RA}=10^\textnormal{h}\;02^\textnormal{m}\;43^\textnormal{s}$, $\textnormal{Dec.}=+03^{\circ}\;26\arcmin\;37\arcsec$ with an uncertainty of {3\arcmin} on the $10^{\textnormal{th}}$ of April, 2016 at 05:09:48 UT \protect\citep{2016GCN.19271....1G}. The event localisation was refined by the UVOT instrument on-board \textit{Swift} that took a finding chart with the \textit{white} filter with a total exposure time of 150~s only 91~s after the BAT trigger, locating the burst to
$\textnormal{RA}=10^\textnormal{h}\;02^\textnormal{m}\;44.37^\textnormal{s}$, $\textnormal{Dec.}=+03^{\circ}\;28\arcmin\;42\farcs7$
with an uncertainty of 0\farcs49 \protect\citep{2016GCN.19275....1M}. In the refined analysis of \cite{2016GCN.19276....1S} the burst shows a duration of $\ T_{90} = 8.2\pm1.6$ s in the $15-350$ keV band and a spectral lag of $8\pm14$ ms between the $50-100$ keV and $15-2$5 keV bands and $-3\pm7$ ms between the $100-350$ keV and $25-50$ keV bands, which is consistent with zero, as expected for SGRBs \citep{NorrisBonnell2006ApJ}. The BAT light-curve analysis of \cite{2016GCN.19276....1S} shows signs of faint extended emission, $\sim$0.038 counts det$^{-1}$ s$^{-1}$ at T+10 s. 

The {\it Swift} X-Ray Telescope \citep[XRT,][]{Burrows2005SSRv} observed the GRB at five epochs starting from 83\,s after the GRB. Data were initially acquired in Windowed Timing (WT) mode due to the brightness of the source, and later in Photon Counting (PC) mode. The XRT light curve shows an initial steep decay with a power law decay index of $2.0\pm0.1$, and possibly two small flares superimposed. The light curve continued with a flattening, starting from $\sim$\,750\,s followed by another decay. However, the XRT data collected in PC mode are too sparse to provide a deeper analysis of the lightcurve.

Observations from Konus-\textit{Wind}, sensitive to higher energies in the 20 keV-10 MeV range, show a peak energy of $E_{\textnormal{peak, observed}} = 1416_{-356}^{+528}$ keV, a $\ T_{90} = 2\ $s, and an isotropic energy release of $E_{\textnormal{iso, rest}} = 4.0\times10^{52}$ erg \citep{2016GCN.19288....1F}. However, \cite{2016GCN.19288....1F} used a different cosmology than the one adopted in this work so we re-calculated it (see Sect. \ref{sec:short_long} and Appendix \ref{sec:short_long_appendix}).
Given the energy released as observed by Konus-\textit{Wind}, the initial pulse complex  \citep{2016GCN.19288....1F}, the extended softer emission \citep{2021ApJ...911L..28D}, and the negligible spectral lag \citep{2016GCN.19276....1S}, this burst has been classified as a short GRB with extended emission (see Sect. \ref{sec:short_long} and Appendix \ref{sec:short_long_appendix} for further discussion on this issue).

\subsection{X-shooter observations of GRB~160410A}\label{sec:Xsh}

\begin{figure*}
\begin{center}
	\includegraphics[width=\textwidth]{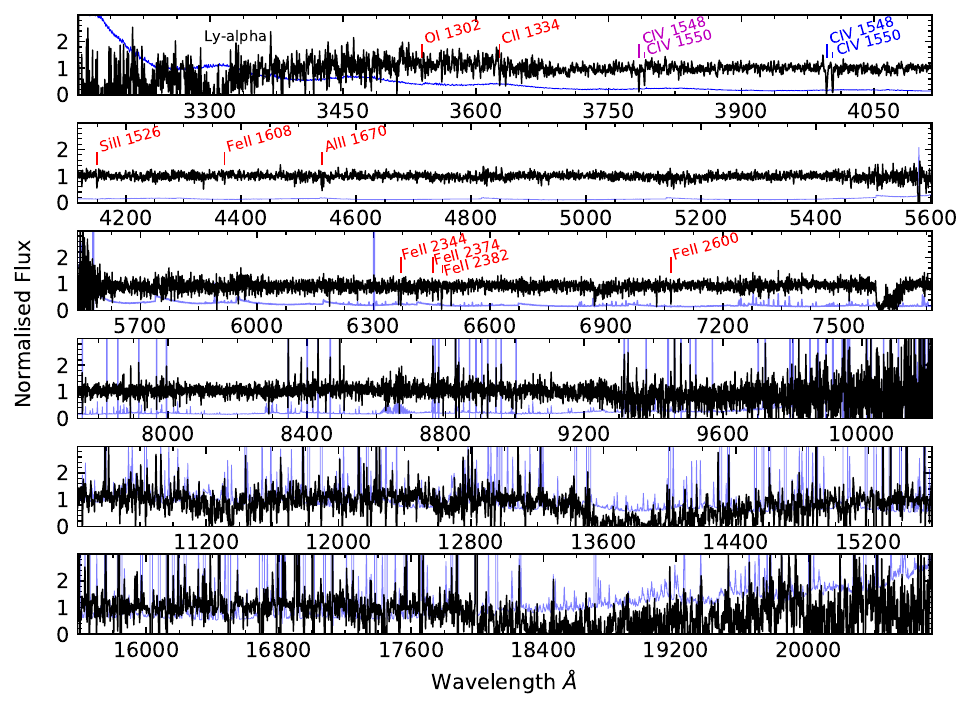} \caption{X-shooter spectrum (black) of the optical afterglow of GRB 160410A smoothed using a Gaussian kernel with $1\sigma$ for the first four panels and $2\sigma$ for the last two panels showing the NIR spectrum. The error spectrum is plotted in blue for the unbinned spectrum. Vertical lines denote the absorption lines: Red corresponds to absorption lines at the GRB redshift, blue is the intervening system at $z=1.581$, and magenta the intervening system detected at $z=1.444$. The error spectrum in the blue end is higher than the actual spectrum due to the absence of binning for the error spectrum.}
	\label{fig:AfterglowSpectrum}
\end{center}
\end{figure*}

The X-shooter spectrograph \citep{2011A&A...536A.105V} was automatically triggered after the \textit{Swift} alert using the Rapid-Response Mode (RRM). Observations of the afterglow of GRB 160410A started at 05:18:08.00 UT, 8.4 minutes after the \textit{Swift} trigger. The acquisition image showed the afterglow at $r^\prime=20.249\pm0.037$ mag (AB photometric system). The initial results of the analysis of the spectrum were reported by \cite{2016GCN.19274....1S}\footnote{We note that this spectrum was also presented as part of the X-shooter sample by \cite{2019A&A...623A..92S}.}. The observations consisted of a total of 3 exposures of 600\,s each in a dithering pattern ABB taken just before the twilight and before the telescope limit at 20$^{\circ}$ elevation was reached. The seeing was $\sim0\farcs9$ and transparency conditions were clear. Due to the high airmass ($\sim2.4$) at which the observation was performed, the spectral trace changes its position on the slit as a function of wavelength, which we modelled in the spectral extraction. The spectra were reduced using the ESO/X-shooter pipeline v.2.6.8 \citep{2010SPIE.7737E..28M} and Reflex \citep{2013A&A...559A..96F}. The final spectrum has a spectral resolution in the UVB arm of 54 km\ s$^{-1}$ and of 28 km\ s$^{-1}$ in the VIS arm. No features were detected in the NIR arm. An initial sky-subtraction was performed on the unrectified image. The spectral response function was generated using observations of a spectrophotometric standard-star \citep{2010SPIE.7735E..1IV, 1994PASP..106..566H} with an optimal extraction, as was also done for the science spectrum. We show the complete normalised spectrum in Fig. \ref{fig:AfterglowSpectrum}. The signal-to-noise (SNR) per resolution element varies between $\sim$5--15 in the UVB arm and between $\sim$7--10 for the VIS arm for the continuum in the regions where we detect absorption lines. In the NIR we find an average SNR of $\sim$1.2.

\subsection{Photometric Observations of GRB~160410A}
For our analysis of the GRB we also obtained imaging of the optical and near-IR afterglow of GRB 160410A. In addition to the observations listed below, we used the acquisition image of the X-shooter observations (see Sect.\,\ref{sec:Xsh}) as well as literature data from Skynet PROMPT \citep{2016GCN.19277....1T} and a late detection by the 2.4 m GMG telescope \citep{2016GCN.19280....1W}. For our analysis, we do not use the afterglow limits reported by \cite{2016GCN.19285....1M,2016GCN.19309....1J,2016GCN.19311....1C,Rastinejad2021ApJ}. 
Finally, we obtained late observations of the field in the optical with the 10.4 m Gran Telescopio de Canarias (GTC) and in infrared with the \textit{Spitzer} space observatory to search for an underlying host galaxy (see Sects.\,\ref{sec:GTC}, \ref{sec:Spitzer} and Fig.\,\ref{fig:FindingChart}).

\subsubsection{TAROT observations}
The 0.25 m T\'elescope \`a Action Rapide pour les Objets Transitoires (TAROT) La Silla telescope observed the location of GRB 160410A very rapidly, beginning 28 s after trigger (16.8 s after notice). Observations (originally published in \citealt{2016GCN.19287....1K}) were obtained in trailing mode \citep{Klotz2006AA}. The afterglow is detected as a very faint trail. The photometry method was based on the division of the afterglow flux by the flux of a reference star. To verify the validity of the method we chose another known star as a check. The magnitudes of the two stars ($r^\prime=13.517$ mag, $r^\prime=16.720$ mag, respectively) were taken from the Sloan Digital Sky Survey (SDSS) Data Release 12 \citep{2015ApJS..219...12A}. To evaluate the flux density uncertainties we computed the standard deviation of the background. The check star allowed us to validate the photometric method since the SDSS magnitude lies fully within the computed limits for each measurement. The TAROT best mean magnitude is 0.08 mag fainter than the SDSS value, however, the TAROT image is unfiltered, explaining this small colour effect.

The afterglow evolution shows a decay with a possible superposed flaring behaviour (see Sect. \ref{sec:ligh_curve}). We note that the columns on which the afterglow trail was located from $55-60$ s are less sensitive than the surrounding ones and we can only claim an upper limit here. Furthermore, the afterglow magnitude fell below the detection limit by the end of the trailed observation.

\subsubsection{UVOT observations}
The {\it Swift}/UltraViolet Optical Telescope (UVOT, \citealt{Roming2005SSRv}) began observing the field of GRB 160410A 91\,s after the \textit{Swift}/BAT trigger. Observations were taken in both event and image modes. The afterglow is detected in all UVOT filters except for $uvw2$ and $uvm2$, as these lie blueward of \lya\ at the redshift of GRB 160410A. Before extracting count rates from the event lists, the astrometry was refined following the methodology of \cite{oates09}. The source counts were extracted initially using a source region of 5\arcsec{} radius. When the count rate dropped to below 0.5 counts/s, we used a source region of 3\arcsec{} radius. In order to be consistent with the UVOT calibration, these count rates were then corrected to 5\arcsec{} using the curve of growth contained in the calibration files. Background counts were extracted using three circular regions of radius 10\arcsec{} located in source-free regions. The count rates were obtained from the event and image lists using the {\it Swift} tools \texttt{uvotevtlc} and \texttt{uvotsource}, respectively.
They were converted to magnitudes using the UVOT photometric zero points \protect\citep{poole,bre11}. To improve the signal-to-noise ratio, the count rates in each filter were binned using $\Delta t/t=0.2$, leading to longer but deeper exposures at later times. The early event-mode \textit{white} and \textit{u} finding charts were bright enough to be split into multiple exposures.

\subsubsection{GROND observations}
We obtained multi-band photometric observations with the Gamma-Ray burst Optical and Near-infrared Detector (GROND) \citep{2008PASP..120..405G,2019PASP..131a5002G} mounted on the 2.2~m MPG telescope at ESO La Silla observatory (originally published in \citealt{2016GCN.19272....1Y}) in the $g^\prime r^\prime i^\prime z^\prime JHK$ bands. GROND observations began about half an hour after the GRB, at very high airmass (2.7). Only a single 4M4TD\footnote{Denoting the integration time in the NIR is 4 min in total, and there are four optical images at four different dithering positions, therefore a 4M(inute)4T(elescope)D(ithers).} observation Block (OB) could be obtained before the telescope hit a pointing limit. The effective integration time was somewhat reduced by the low quality of one of the dithering positions. However, the afterglow is still detected in $g^\prime r^\prime i^\prime z^\prime $.
The following night, observations started at lower airmass but under adverse conditions. A total of 28 8M4TD OBs were obtained, but only OBs $1-21$ were usable. The afterglow had faded considerably, and was only detected in $g^\prime r^\prime i^\prime$.

Afterglow magnitudes in the optical were calibrated against standard stars in the same field from the SDSS catalogue \citep{2015ApJS..219...12A}. Near-infrared magnitudes (all upper limits) were measured against comparison stars in the field taken from the 2MASS catalogue \citep{2006AJ....131.1163S}. Reduction and analysis were performed within a custom pipeline calling upon IRAF tasks \citep{1993ASPC...52..173T}, following the methods of \cite{2008ApJ...685..376K,2008AIPC.1000..227Y}.

\subsubsection{NOT observations}

Observations were taken in the $r^\prime$ band with the Alhambra Faint Object Spectrograph and Camera (AlFOSC) at the 2.5~m Nordic Optical Telescope (NOT) at the Roque de la Muchachos Observatory on La Palma, Canary Islands, Spain (originally published by \citealt{2016GCN.19295....1M,2016GCN.19300....1M}, see Fig.\,\ref{fig:FindingChart}). The afterglow was directly calibrated against four SDSS stars in the field. In the second epoch, forced aperture photometry on the afterglow position yields a tentative $3\sigma$ detection, which, however, is in agreement with the decay extrapolated from earlier times. We therefore include it as a detection.

\subsection{Host observations of GRB 160410A}

We performed deep photometric observations with a ground-based facility in search of an underlying galaxy hosting GRB\,160410A. We also observed with the IRAC instrument on-board the \textit{Spitzer} satellite.

\subsubsection{GTC observations}\label{sec:GTC}

\begin{figure}
	\includegraphics[width=\columnwidth]{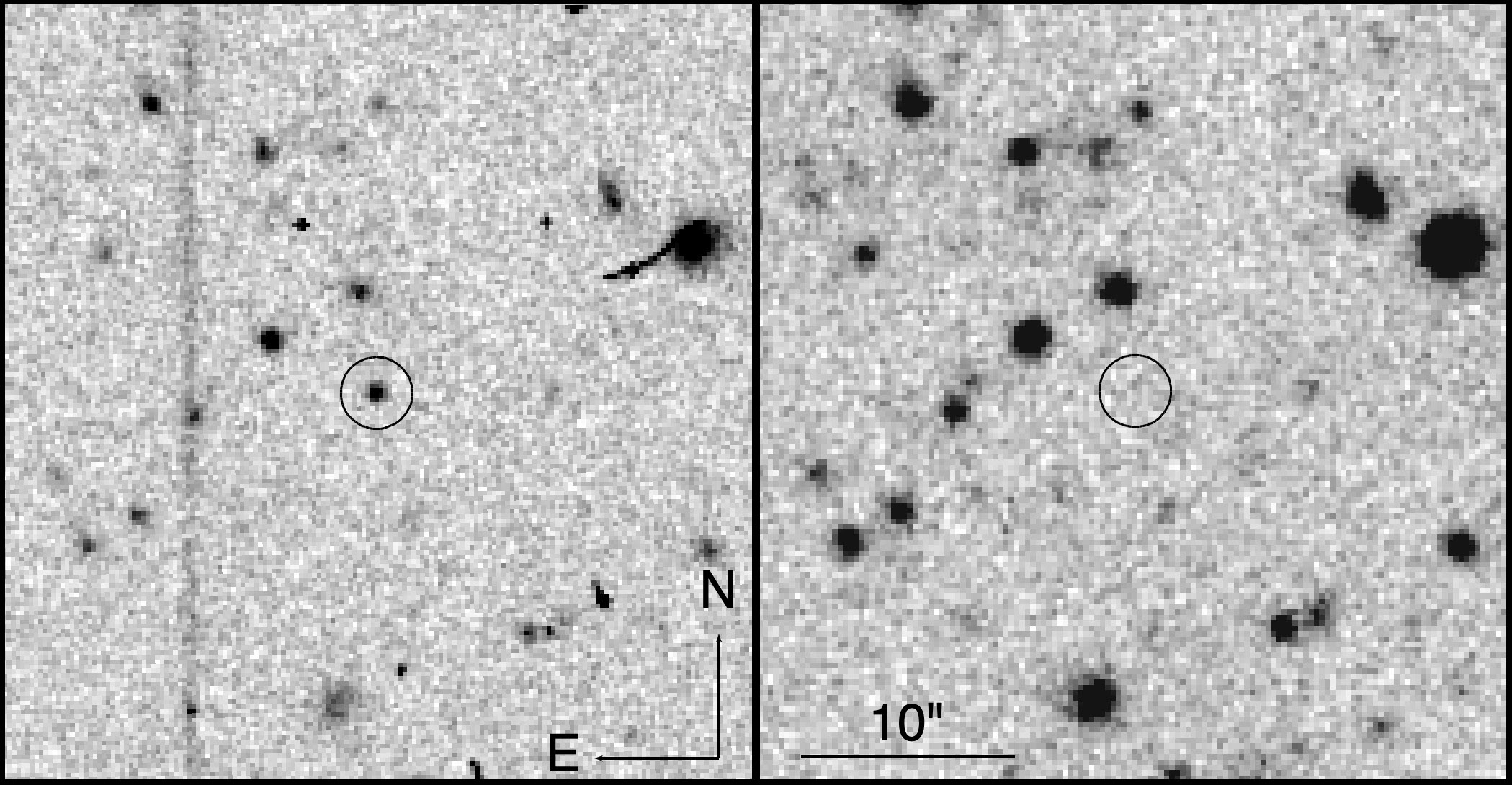} \caption{Optical observations of the field of GRB\,160410A. \textit{Left:} $r^\prime$-band image obtained by NOT 0.7 days after the burst where the afterglow is clearly detected. \textit{Right:} Deep late observation at 44.7 days in the same band by the 10.4~m GTC, where no source is detected down to a limit of $r^\prime>27.17$ mag (AB, corrected for Galactic extinction).}
	\label{fig:FindingChart}
\end{figure}

We searched for a possible underlying host galaxy at the GRB position at late times in the $r^\prime$ band using OSIRIS at GTC. Data were taken on the night of the 24th of May 2016, comprising a total of ten images with an exposure time of 180\,s each\footnote{Observations obtained under GTC programme GTC22-16A (PI: C.C. Th\"one).}.
The observations were obtained at an airmass of $\sim1.4$ and calibrations were performed using four SDSS field stars. We do not detect any source at the position of the afterglow down to a $3\sigma$ limit of 27.17 mag (AB, corrected for Galactic extinction) (see Fig.\,\ref{fig:FindingChart}).

\subsubsection{\textit{Spitzer} observations}\label{sec:Spitzer}

We obtained observations of the field of GRB\,160410A with the Infrared Array Camera (IRAC) on-board the \textit{Spitzer} Space Telescope on 2017 August 22 as part of the extended \textit{Swift}/\textit{Spitzer} GRB Host Galaxy Legacy Survey (SHOALS; \citealt{2016ApJ...817....7P}). One hour of integration ($36\times100$ s dithered images) was obtained in IRAC channel 1 ($3.6\,\mu$m). We downloaded the \textit{Spitzer} Post-Basic Calibrated Data (PBCD) co-added images from the \textit{Spitzer} Heritage Archive and used the methods of \cite{2016ApJ...817....8P} to model and subtract nearby contaminating sources within a 20\arcsec{} box around the location of the GRB. There is no source visible consistent with the location of the optical afterglow in this image.  We derive a $3\sigma$ limit on the magnitude within a 2\arcsec{} aperture centred on the GRB afterglow location of $m_{ch1}>24.74$ mag (AB).

\subsection{High-Energy detection of GRB~201221D}
\label{sec201221d_HE}

\textit{Swift}/BAT detected the short GRB\,201221D at 23:06:34 UT on 21 December 2020 \citep{2020GCN.29112....1P}. The burst is located at $\textnormal{RA}=11^\textnormal{h}\;24^\textnormal{m}\;14.09^\textnormal{s}$, $\textnormal{Dec.}=+42^{\circ}\;08\arcmin\;40.0\arcsec$ (J2000) with an uncertainty of {$\sim1\farcs1$}.
It had a duration of $T_{90}=0.16\pm0.04$ s in the 15-350 keV band \citep{2020GCN.29139....1K} and a peak energy in observer frame of $98\pm8$ keV as seen in \textit{Fermi}/GBM observations \citep{2020GCN.29140....1H}. GRB\,201221D was observed by the Konus-\textit{Wind} observatory showing a peak energy of  $E_{\textnormal{peak, observed}} = 148_{-37}^{+86}$ keV \citep{2020GCN.29130....1F}, which implies that it was a clear short GRB.

\subsection{GTC Spectroscopic observations of GRB~201221D}
\label{sec201221dSpec}  
We observed the GRB afterglow using GTC/OSIRIS starting 2.76\,hours after the trigger. The observations consisted of an acquisition image in $r^\prime$ band followed by long slit spectroscopy\footnote{Observations obtained under GTC programme GTCMULTIPLE2G-20B (PI: de Ugarte Postigo).}. Four exposures of 1200\,s were obtained covering the 3700\,\AA\ to 7800\,\AA\ spectral range, at an airmass ranging from 1.45 to 1.83.

The data were reduced using a self-developed pipeline based on IRAF routines. Data reduction included bias and response correction, and wavelength calibrations using HgAr and Ne lamps, which were also used to do a 2D distortion correction. Cosmic rays were removed using the \texttt{lacos\_spec} routine \citep[][]{2001PASP..113.1420V}. The flux calibration was performed using as reference the spectrophotometric standard star G191B2B \citep[][]{1990AJ.....99.1621O}. The 1D spectrum was obtained through optimal extraction \citep{1986PASP...98..609H}.

This is the first afterglow spectrum of a short GRB showing evidence for absorption lines obtained since GRB\,160410A and only the third spectrum of a short GRB afterglow to have them, which is why we include it in this paper for comparison. The spectrum shows emission and absorption features while the continuum is dominated by the host galaxy (see Sect. \ref{sec:grb201221D}). Unfortunately the low signal-to-noise ratio of the spectrum of GRB\,201221D limits the amount of information that can be extracted from these data.

\subsection{Host observations of GRB~201221D}

The Large Binocular Telescope (LBT) observed the underlying host galaxy of GRB\,201221D at two epochs. A first observation was obtained on the $24^{\textnormal{th}}$ of December, 2020 when we observed in the near-infrarred \textit{J} and $K_s$ filters with the LBT Utility Camera in the Infrared \citep[LUCI,][]{2003SPIE.4841..962S} imager and spectrograph under good seeing with $1\farcs0$ on average, (first reported in \citealt{Rossi2021GCN29311}). The second observation was obtained the $10^{\textnormal{th}}$ of January, 2021 in the $g^\prime r^\prime i^\prime z^\prime$ bands with the Large Binocular Camera \citep[LBC,][]{2008A&A...482..349G} under moderate seeing conditions ($1\farcs6$ on average).

LBT data were reduced using the data reduction pipeline developed at INAF-Osservatorio Astronomico di Roma \citep{2014A&A...570A..11F} that includes bias subtraction and flat-fielding, bad pixel and cosmic ray masking, astrometric calibration and coaddition. For LUCI, it includes also dark subtraction and sky subtraction. The astrometry was calibrated against field stars in the GAIA DR2 catalogue \citep{2018A&A...616A...1G} and has an astrometric precision of 0\farcs15.

All data were analysed by performing aperture photometry using \texttt{DAOPHOT} and \texttt{APPHOT} under \texttt{PyRAF/IRAF}. We have carefully selected the size of the apertures to avoid faint sources close to the host, in particular a faint source 3\farcs5 NW from the host. The photometric calibration was performed in the optical against the SDSS DR12 catalogue \citep{2015ApJS..219...12A} and in the NIR against 2MASS stars.

\section{Analysis and results}
\label{sec:results}

Our comprehensive analysis of the GRB\,160410A afterglow comprises data from high energies to optical wavelengths. Unfortunately, no near-infrared data of sufficient quality are available. For GRB\,201221D, the data coverage is significantly sparser than for GRB\,160410A.

\subsection{X-Ray analysis of GRB~160410A}

We extracted the spectrum in WT mode, avoiding the small flares (i.e., in the $89-150$\,s time-frame) collecting $\sim$\,700 source counts. This is to avoid possible contamination from the flare emission on the spectral parameters. We fitted the X-ray data with a power-law model, with the Galactic absorption fixed to $N_{\rm H,\, Gal}=1.8\times 10^{20}$ cm$^{-2}$ and a free intrinsic absorption at $z=1.72$ of the host galaxy. We adopted C-statistics and data were binned to 1 count per energy bin in the $0.3-10$ keV energy range. The best-fit power-law photon index is $\Gamma=1.6\pm0.1$ ($1\,\sigma$ confidence level). The intrinsic absorption column density is $N_H(z)=3.0^{+2.3}_{-1.9}\times 10^{21}$ cm$^{-2}$. The spectrum evolves to a softer value as the flux decreases ($\Gamma\sim$\,2), but the lower number of counts prevented us to better constrain the photon index.

\subsection{Spectral analysis}

In the spectrum of GRB\,160410A we detect several absorption features typically found in long GRB sight-lines \citep{christensen2011}. The wavelength coverage of X-shooter allows the detection of \lya\ absorption from a host galaxy DLA system at the blue end of the spectrum which, together with other metallic lines, is used to derive the metallicity of the host galaxy gas. We do not detect any fine-structure lines commonly associated with gas in the close GRB environment \citep{2007A&A...468...83V, 2011A&A...532C...3V, 2009A&A...503..437D, 2009ApJ...694..332D}. For \feii*\ $\lambda$ 2612 we derive a rest frame limit of $< 0.15$\AA\ a value 30\% lower than what has been measured for this line in a GRB composite spectrum  \citep{christensen2011}. However, the absorption lines observed are very weak so the non-detection of fine-structure lines is consistent with their relative strength in the aforementioned composite spectrum.

The lines observed are only in the UVB and VIS arms, neither emission nor absorption features are observed in the NIR arm. \cite{2016GCN.19274....1S} reported a first determination of the redshift of GRB\,160410A, based on the detection of \feii\ and \alii\ absorption lines, resulting in a value of $z = 1.717$. In our analysis, we find additional features, not mentioned in the original GCN, of  \oi, \cii\ and \siii\ at the same redshift. Contrary to the claim in \protect\cite{2016GCN.19278....1C}, we detect neither the \civ\ $\lambda$ 1548, 1550 doublet in our spectrum, nor \mgii\ $\lambda$ 2976, 2803 doublet. The non-detection of high-ionisation lines, commonly seen in other GRB sight-lines \citep{christensen2011}, point to a low ionisation environment. The \mgii\ lines fall inside the telluric A-band and can therefore not be recovered in our spectrum. Emission lines would fall as well in the NIR and are not detected, which is not surprising given the high redshift, the non-detection of a host galaxy (see Sect.\,\ref{sec:nohost}) and the fact that the emission lines fall in regions with strong telluric features or high noise levels. 

In Tab.\,\ref{tab:ew_list} we list the detected features and their corresponding equivalent widths (EW). Line identifications and measurements were obtained using the tools in the \href{http://grbspec.iaa.es/}{GRBspec} database \citep{2014SPIE.9152E..0BD, 2020SPIE11452E..18B}. The different absorption lines are centred at slightly different redshifts due to different velocity components (see Sect.\,\ref{sec:absorption}) and we obtain a non-weighted mean value of $z=1.7177 \pm 0.0001$, which matches the redshift of the strongest \feii\ component.

\subsubsection{Absorption line fitting}\label{sec:absorption}

\begin{table}
\centering
\caption{Observed absorption-line list for the spectrum of GRB\ 160410A. Top: lines at the redshift of the GRB. We include the 3$\sigma$ limits derived for \siiv\ and \civ\ (see Sect. \ref{sec:LSP}). Middle and bottom parts: \civ\ absorbers detected at $z = 1.581$ and $z = 1.444$ respectively. The first column shows the absorption line ID and its rest-frame wavelength. The second column lists the corresponding centroid for the redshifted absorption line. The third column is the measured EW in the rest frame, the fourth column is the total column density  we derive using the \texttt{VoigtFit} fitting code \protect\citep{krogager2018voigtfit} for each transition.
}
\begin{center}
\begin{tabular}{cccc}
\hline
\hline
\noalign{\smallskip}
Feature & Observed wavelength & EW & $\log(N)$\\
& ({\AA}) &  ({\AA}) & (cm$^{-2}$)\\
\hline
\lya\ 1215.670 & 3304 & -- & $21.20 \pm 0.20 $ \\

\oi\ 1302.170 & 3539.4000 & $0.32 \pm 0.08$ & $ > 15.04 $ \\
\cii\ 1334.530 & 3626.9869 & $0.26 \pm 0.08$ & $ > 14.77 $ \\
\siii\ 1526.710 & 4149.5107 & $0.33 \pm 0.06$ & $ 14.25 \pm 0.11 $ \\
\alii\ 1670.790 & 4540.6191 & $0.29 \pm 0.07$ & $ 13.13 \pm 0.21 $ \\
\feii\ 1608.450 & 4371.3224 & $0.17 \pm 0.04$ & $ 14.32 \pm 0.09 $ \\
\feii\ 2344.210 & 6370.6764 & $0.35 \pm 0.08$ & ... \\
\feii\ 2382.770 & 6475.4289 & $0.36 \pm 0.06$ & ... \\ 
\feii\ 2374.460 & 6452.9438 & $0.19 \pm 0.06$ & $ > 14.22 $ \\ 
\feii\ 2586.650 & 7029.2700 & $0.35 \pm 0.04$ & ... \\ 
\feii\ 2600.170 & 7066.1812 & $0.30 \pm 0.05$ & ... \\ 
 \\
\siiv\ 1393.760 & -- & $ < 0.36 $ & $> 13.60$ \\
\siiv\ 1402.770 & -- & $ < 0.39 $ & $> 13.93$ \\
\civ\ 1548.200 & -- & $ < 0.24 $ & $> 13.76$ \\
\civ\ 1550.770 & -- & $ < 0.24 $ & $> 14.06$ \\

\hline
\hline
\civ\  1548.200 & 3996.2713 & $0.57 \pm 0.06$ & $> 14.82$ \\
\civ\  1550.770 & 4002.9657 & $0.52 \pm 0.06$ &  ...   \\
\hline
\hline
\civ\  1548.200 & 3784.1820 & $0.38 \pm 0.07$ & $> 14.91$ \\
\civ\  1550.770 & 3790.4285 & $0.27 \pm 0.07$ &  ...  \\
\hline
\hline

\end{tabular}
\end{center}

\label{tab:ew_list}

\end{table}

We fitted Voigt profiles to the observed lines at the GRB redshift to obtain column densities, including the broad \lya\ absorption. The spectrum was normalised and as systemic redshift we adopted the strongest velocity component of the \feii\ $\lambda$\,2600 line as $v=0$\,km\,s$^{-1}$. For the line fitting we used the \texttt{VoigtFit} fitting code \citep{krogager2018voigtfit} that allows to define and tie parameters for each component such as the \textit{b}-parameter, the column density or the redshift. We find that, for most of the detected transitions, the absorption can be fitted to two components while for \alii\ and up to three \feii\ lines we find only one component. For \feii\ $\lambda$\,1608, 2344, 2382 and \siii\ the line profile is fitted well using two components and tying them for \siii\ to the ones of \feii. However, for \oi\ and \cii, the first component is fitted to a broader \textit{b}-parameter. For \feii\ $\lambda$\,2374, 2586, 2600 and \alii, only first component is well fitted. The non-detection of the secondo component for this \feii\ transitions might imply that we are underestimating the amount of \feii\ in this lines. This could be due to the low resolution of the spectrum or the low amount of \feii\ at this second component. We therefore adopt the measured column density for \feii\ $\lambda$\,2374, 2586, 2600 lines as limits and choose the measures for \feii\ $\lambda$\,1608, 2344, 2382 as values.

\begin{figure}
	\includegraphics[width=\columnwidth]{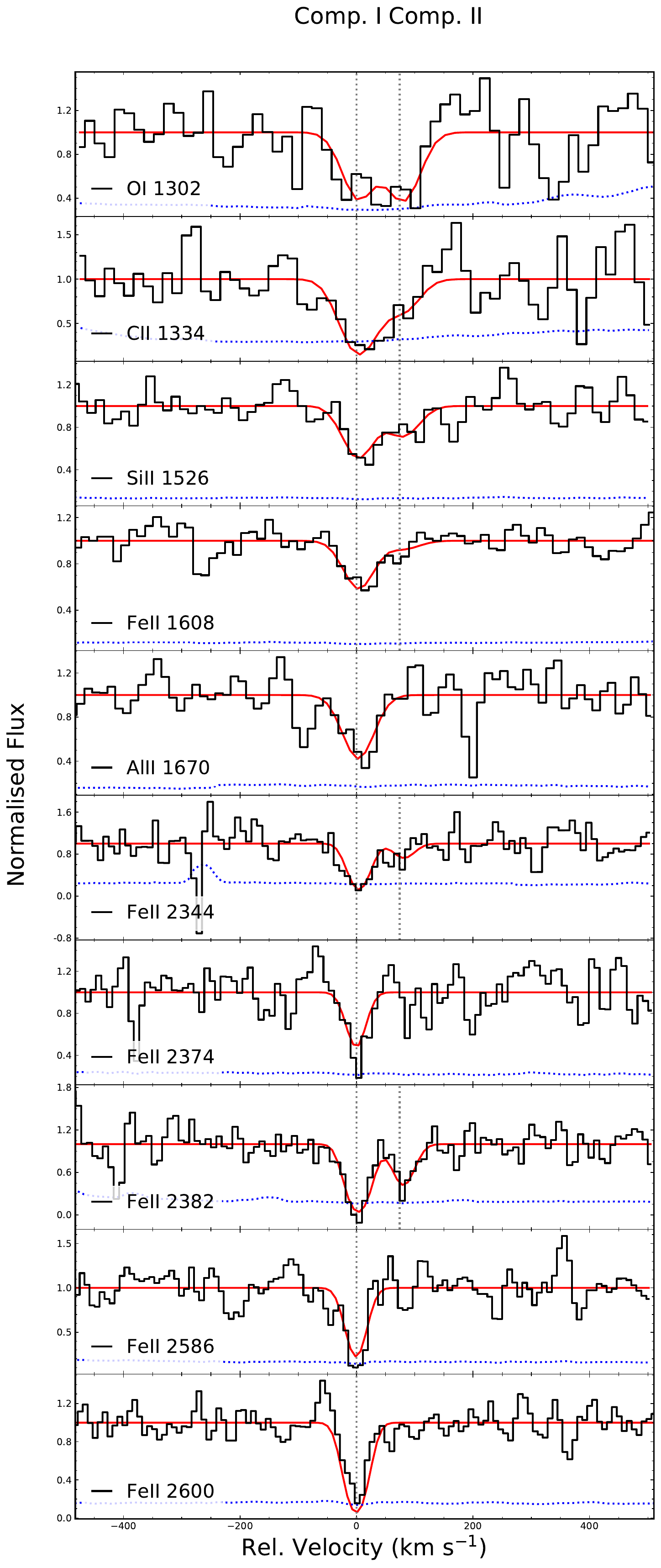}
	\caption{Absorption features detected in the GRB\,160410A afterglow emission as observed in the X-shooter afterglow spectrum. Black lines correspond to the normalised spectrum in velocity space, centered at the redshift of the GRB. We also plot the error spectrum for each line (blue dotted line) and in vertical grey we mark the components listed in Tab.\,\ref{tab:Voigt_fitting}. The red solid line shows the Voigt best-fit profile for each absorption feature.}
	\label{fig:_lines_fit}
\end{figure}

In Tab. \ref{tab:ew_list} we present the total measured column density for each line. The \oi\ and \cii\ lines are likely saturated hence we consider them as lower limits, something which could also explain the different metallicity value obtained from these two lines. The fitting results are shown in Tab.\,\ref{tab:Voigt_fitting}. We plot the lines in velocity space in the normalised spectrum in Fig.\,\ref{fig:_lines_fit}, centered at the main component. The absorption lines show a small offset towards higher velocities in the absorption lines detected in the VIS arm compared to the UVB arm (see Fig. \ref{fig:_lines_fit}). A similar shift in wavelength has been reported in the past between the VIS and NIR arms of X-shooter and also suggested for the UVB arm as a problem related to the wavelength calibration \citep[see e.g.][]{2019A&A...623A..92S, 2020A&A...634A.133G}.

\begin{figure}
	\includegraphics[width=\columnwidth]{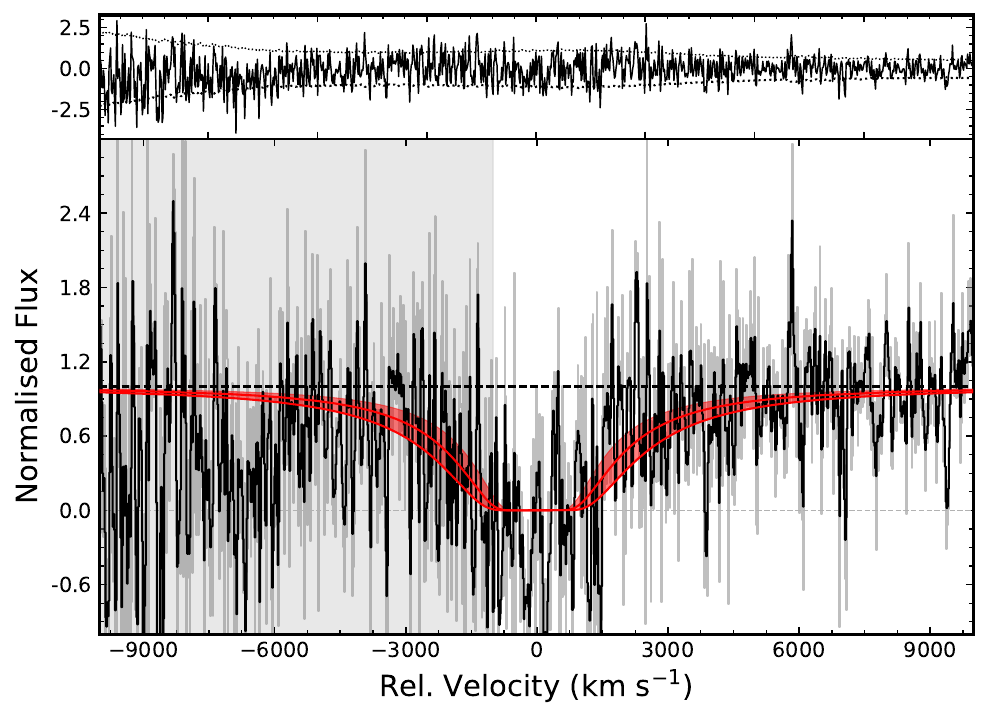}
	\caption{Voigt profile fitting of the \lya\ absorption line. The top panel shows the residuals and the $1\sigma$ error spectrum (dotted line). In the bottom panel we plot the spectrum smoothed with a Gaussian kernel of $2\sigma$ in black and the original, non-smoothed, spectrum in grey. The red solid line represents the best-fit Voigt profile. For clarity, we show in shaded red the corresponding errors.}
	\label{fig:HI_fit}
\end{figure}

In the spectrum of GRB\,160410A, the broad \lya\ absorption lies at the very blue end of the X-shooter wavelength coverage, where the continuum is rather noisy (see Fig.\,\ref{fig:HI_fit}). However, we were able to determine the column density using \texttt{VoigtFit} and masking the blue wing from $-1000$ km s$^{-1}$. We obtain a total column density of $\log (N{\rm (HI)/cm}^{2}) = 21.2\pm0.2$, which is consistent with \cite{2019A&A...623A..92S} and puts the system in the category of DLAs. The column density is also close to the median value that is found in long GRB spectra, i.e. $\log (N{\rm (HI)/cm}^{2}) = 21.59$ \citep{2019MNRAS.483.5380T}.

\begin{table*}

\caption{Column densities derived fitting a Voigt profile to the different velocity components in the absorption system at the GRB redshift. We also show the corresponding derived metallicities. $v=0$\ km\,s$^{-1}$ corresponds to the \feii\ line at 2600 \AA\ as the line with the highest S/N. Solar metallicities are photospheric, meteoric, or the average value between the two, following  \protect\cite{2009LanB...4B..712L}}.
\centering

\begin{center}
\begin{tabular}{ccccccccc}
\hline
\hline
\noalign{\smallskip}
 &  & \multicolumn{3}{c}{Component I} & \multicolumn{3}{c}{Component II} & [X/H] \\
\cmidrule(lr){3-5} \cmidrule(lr){6-8}
Ions & Transitions & \textit{v} & \textit{b} & $\log\big(\textnormal{N}\big)$ & \textit{v} & \textit{b} & $\log\big(\textnormal{N}\big)$ & \\
& ({\AA}) & (km s$^{-1}$) & (km s$^{-1}$) & (cm$^{-2}$) & (km s$^{-1}$) & (km s$^{-1}$) & (cm$^{-2}$) & \\
\midrule

\feii & 2374, 2586, 2600 & 0 & 15 & $ > 14.21 $ & -- & -- & -- & $ > -2.44$ \\
\alii & 1670       & ... & ... & $ 13.13 \pm 0.21 $ & -- & -- & -- & $ -2.50 \pm 0.30 $ \\
\feii & 1608, 2344, 2382 & 0 & 15 & $ 14.27 \pm 0.10 $ & 74 & 18 & $ 13.31 \pm 0.10 $ & $ -2.34 \pm 0.22 $ \\
\siii & 1526       & ... & ... & $ 14.10 \pm 0.15 $ & 74 & 18 & $ 13.69 \pm 0.16 $ & $ -2.46 \pm 0.23 $ \\

\oi   & 1302       & 0 & 22 & $ > 14.53 $ & 74 & 18 & $ > 14.84 $ & $>-2.85$ \\
\cii  & 1334       & ... & ... & $ > 14.65 $ & ... & ... & $ > 14.03 $ & $> -2.86$ \\

\hline
\hline
\end{tabular}
\end{center}

\label{tab:Voigt_fitting}
\end{table*}

\subsubsection{Metallicity}
\label{sec:metallicity}

GRB\,160410A is currently the only SGRB event where we are able to study the metallicity along the sight-line in its host galaxy. The different ions of C, O, Si, Al and Fe give slightly different metallicities. The values for C and O can only be considered as lower limits since the absorption lines are likely saturated. The metallicity values derived from Si, Al and Fe are very similar and consistent within errors (see Tab.\,\ref{tab:Voigt_fitting}). All the metallicity values are very low, even compared to the uncorrected metallicities for dust-depletion for LGRB environments (see Fig. \ref{fig:metallicity}).

Dust depletion can affect the observed abundances when large fractions of refractory elements are locked into dust grains. We performed a dust depletion correction on our observed metallicities following the method developed by \cite{2013A&A...560A..88D, 2016A&A...596A..97D, 2018A&A...611A..76D}. In these studies, the correction was based on the [\zn/\fe] ratio but, since we did not detect\ \zn\ we used the observed \siii\ line to derive [\zn/\fe]$_\mathrm{exp}$ using the following relation from \cite{2018A&A...611A..76D} with their corresponding fitting parameters  $A_\mathrm{1,\,Si}$ and $B_\mathrm{1,\,Si}$ as derived in \cite{2016A&A...596A..97D}.
\begin{equation*}
    [\zn/\fe]_\mathrm{exp} = \dfrac{[\si/\fe] - A_\mathrm{1,\,Si}}{B_\mathrm{1,\,Si} + 1}
    \label{eq:znfe_exp}
\end{equation*}

We found a ratio of $[\si/\fe] = -0.08 \pm 0.15$ and, with $A_\mathrm{1,\,Si} = 0.26$ and $B_\mathrm{1,\,Si} = -0.51$, we got that $[\zn/\fe]_\mathrm{exp} = -0.69 \pm 0.32$ implying no depletion. However, \citet[][see their Appendix A]{2018A&A...611A..76D} state that $A_\mathrm{1,\,Si}$ and $B_\mathrm{1,\,Si}$ might not be very well constrained.

\begin{figure}
	\includegraphics[width=\columnwidth]{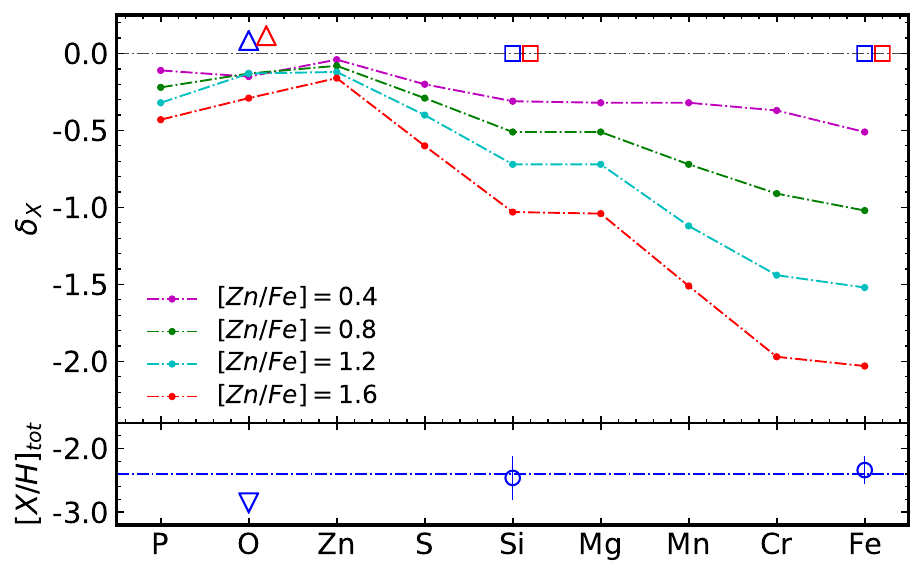}
	\caption{Dust depletion pattern and metallicity for the measured column densities of the features found in absorption in the afterglow of GRB\,160410A. \textit{Top panel:} Dust depletion sequence for GRB\,160410A. Squares (top panel) and circles (bottom panel) denote the \si\ and \fe\ lines meanwhile triangles refer to the limits we derive for the saturated \oxigen\ line. With an small offset for visualization purposes. Red squares denote the depletion $\delta_X$ values obtained for the common velocity component to all the elements observed, blue squares are the depletion considering the total column density measured. For comparison, we plot with a dash-dotted line in different colours the dust depletion sequence obtained from the fitting of QSO-DLA absorption systems in \protect\cite{2016A&A...596A..97D}. \textit{Bottom panel:} Corresponding metallicity for the observed ions considering all the velocity components. The mean metallicity is shown with a dashed-dotted line.}
	\label{fig:dust_depletion}
\end{figure}

\begin{figure}
	\includegraphics[width=\columnwidth]{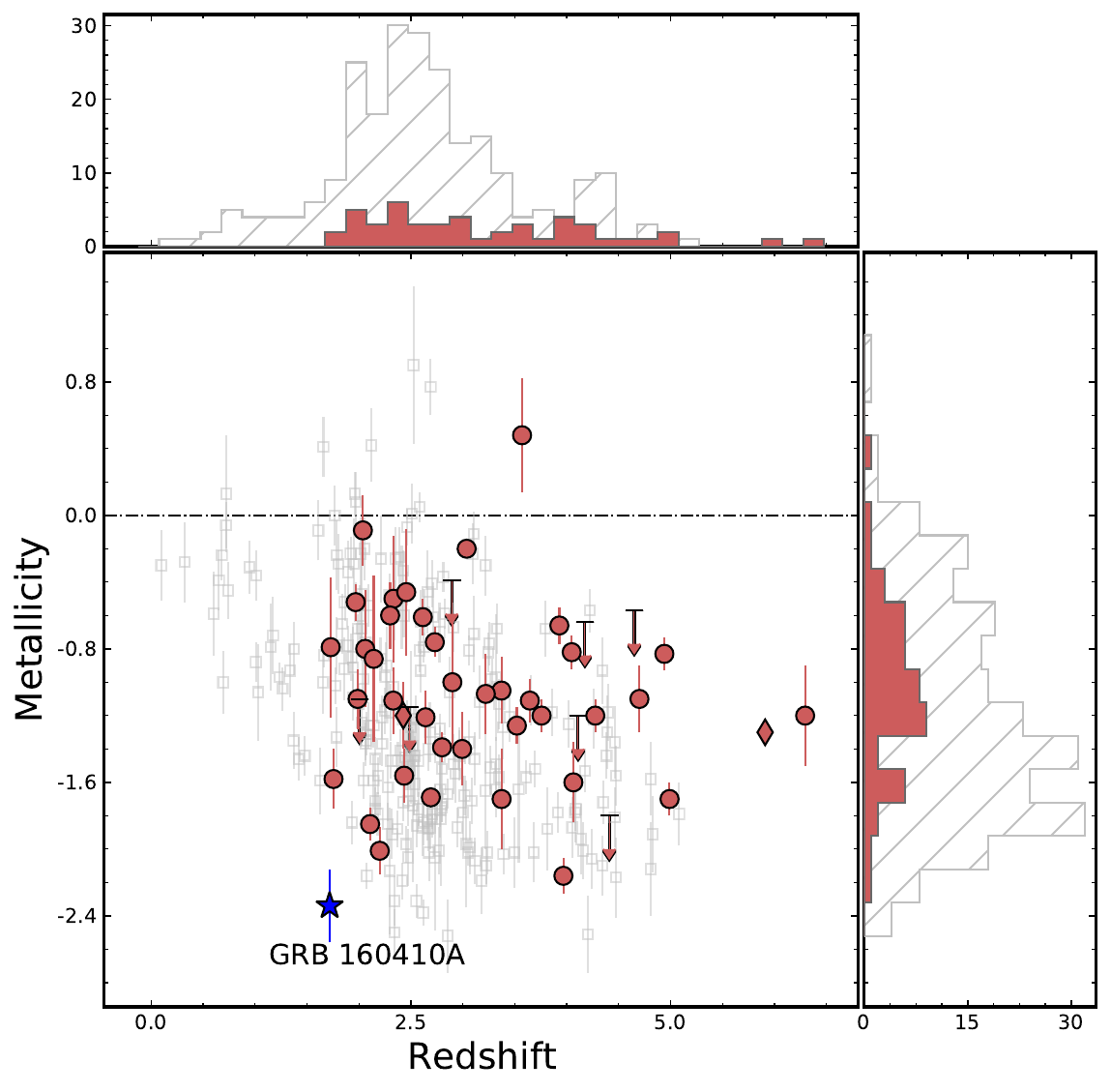}
	\caption{Metallicity of GRB host galaxies and QSO-DLAs vs. redshift. \textit{Middle panel:} Clear grey empty squares mark QSO-DLAs from the sample published in \protect\cite{2018A&A...611A..76D}. Red filled circles show the metallicities for long GRB-DLAs and a red filled diamond marks sub-DLA systems. Most of the data are from \protect\cite{2013MNRAS.428.3590T}. We note that these values are not dust-corrected. We add some more recent metallicity values from the literature (\protect\citealt{2018A&A...620A.119D, 2013A&A...557A..18K, 2015MNRAS.451..167F, 2018MNRAS.479.3456H}, and J. Greiner, priv. comm.). The values from \protect\cite{2018A&A...620A.119D} and \protect\cite{2018MNRAS.479.3456H} are dust-corrected following \protect\cite{2016A&A...596A..97D}. The value from \protect\cite{2015MNRAS.451..167F} is dust-corrected following \protect\cite{2013A&A...560A..88D}. We also show the distribution of  metallicities on the right and the distribution in redshift on the top. Hatched histograms show QSO-DLAs, red filled values show the long GRB-DLA sample. While QSO-DLA metallicities are based only on Fe and dust-corrected, the GRB-DLA metallicities are based on several elements such as Sulfur, Silicon, Zinc, Iron and Oxygen and are, generally, not dust-corrected}. As GRB\,160410A is a completely new class in itself, we do not show it in the histograms.
    \label{fig:metallicity}
\end{figure}

We performed the same analysis including only the component at $v=0$\,km\,s$^{-1}$. Using values for this first component (see Tab. \ref{tab:Voigt_fitting}) we got $[\si/\fe] = -0.18 \pm 0.18$, resulting in $[\zn/\fe]_\mathrm{exp} = -0.90 \pm 0.40$. 
Furthermore, we derived $\delta_X$, a parameter that indicates how much an observed element is depleted by dust \citep{2016A&A...596A..97D}, for both, the total absorption and only for the component defined at $v=0$\,km\,s$^{-1}$. In Fig.\,\ref{fig:dust_depletion} we plot the dust depletion pattern compared to the sequence expected for different amounts of depletion derived in \cite{2016A&A...596A..97D}. In both cases, the values that we obtain would formally result in negative depletion and hence nonphysical values. This indicates that there is no depletion in the system and, therefore, we adopt zero values for the depletion, as shown in Fig.\,\ref{fig:dust_depletion}.

Our analysis results in a very low value of $[\fe/\textnormal{H}] = -2.3 \pm 0.2$ for the metallicity along the sight-line. To put the value into the context of cosmic chemical evolution, we compared our results with those shown by \protect\cite{2018A&A...611A..76D}, which are corrected for dust depletion following the single-reference method (see Fig.\,\ref{fig:metallicity}). The metallicities in the QSO-DLA sample have been determined using the \feii\ absorption and are dust-corrected \protect\citep{2018A&A...611A..76D}. We note that \protect\cite{2018A&A...611A..76D} relax the DLA condition to a slightly lower column density ($\log (N{\rm (HI)/cm}^{2})\geq20.0$, \citealt{2016A&A...596A..97D, 2018A&A...611A..76D}) than the common definition of a DLA system ($\log (N{\rm (HI)/cm}^{2})\geq20.3$, \citealt{2005ARA&A..43..861W}), however, the number of QSOs outside the strict DLA definition is small. In addition, we include DLAs in long GRB hosts from the literature, not corrected for dust-depletion as QSO-DLA and GRB\,160410A are and, therefore, not formally comparable. We clearly see the existence of an observational bias on redshift that results from the realistic spectral coverage when obtaining GRB spectroscopy\footnote{As GRB afterglows fade within hours or days, spectroscopy is generally obtained with ground-based facilities, and the atmospheric ultraviolet cutoff implies that \lya\ is only measurable at $z\gtrsim1.5$ \citep{Updike2008ApJ}. Detection of \lya\ enabling lower redshift metallicity determinations would need an UV-capable space-based spectrograph such as HST/COS or \textit{Swift}/UVOT grism spectroscopy, with the latter needing an extremely bright afterglow such as in the case of GRB 191221B \citep{Kuin2019GCN26538}.}. We note a small tendency of long GRB-DLAs towards higher metallicities \citep[see e.g.][]{2008ApJ...683..321F, 2007ApJ...666..267P}, which is consistent with LGRBs tracing more enriched gas than QSO-DLAs as an effect of tracing gas in star-forming regions. However, since LGRB-DLA metallicities are not dust-depletion corrected, the differences between both samples might be somewhat larger. From the QSO-DLA sample, we clearly see that GRB\,160410A shows one of the lowest metallicity values for a DLA, together with a few QSO-DLAs.

\subsubsection{Line strength and host ionisation}
\label{sec:LSP}
\protect\cite{2012A&A...548A..11D} established a new parameter to compare the ISM of different GRBs by determining the relative ratio between the EWs of different lines with the average EW of (long) GRBs. We apply this method to the spectrum of GRB\,160410A using the EW measurements listed in Tab.\,\ref{tab:ew_list}. We see that all EWs measured for GRB\,160410A are lower than these of the long GRB sample by more than $1\sigma$. This, together with the large column density measured for the neutral hydrogen, are indicative of low metallicity, which is consistent with the values we derive in sec. \ref{sec:metallicity}. We obtain a value for the Line Strength Parameter of LSP = $-1.92\pm1.07$, implying that the features detected in GRB\,160410A are only $0.7\%$ as strong (or $99.3\%$ weaker) as the average strength of long GRB absorption lines.

\begin{figure}
	\includegraphics[width=\columnwidth]{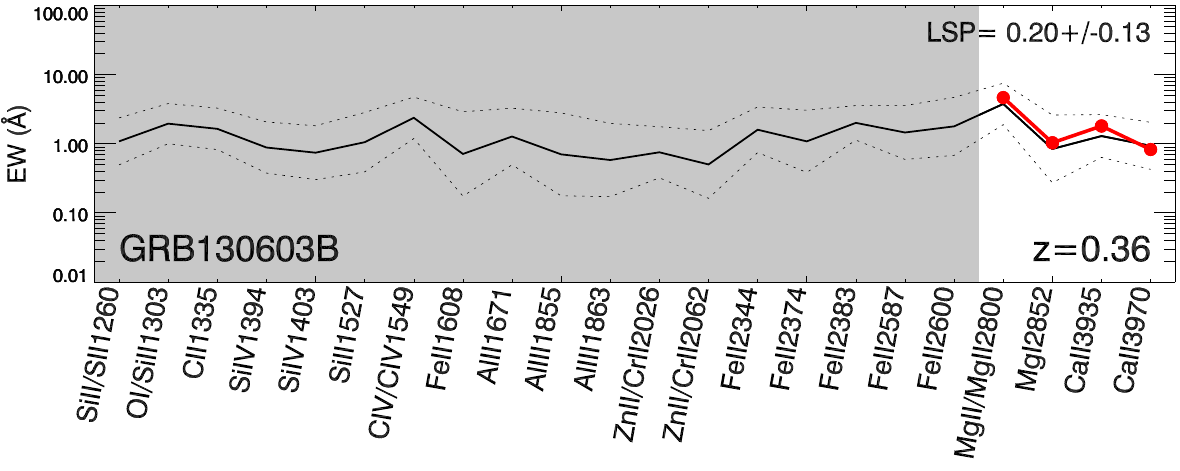}
	\includegraphics[width=\columnwidth]{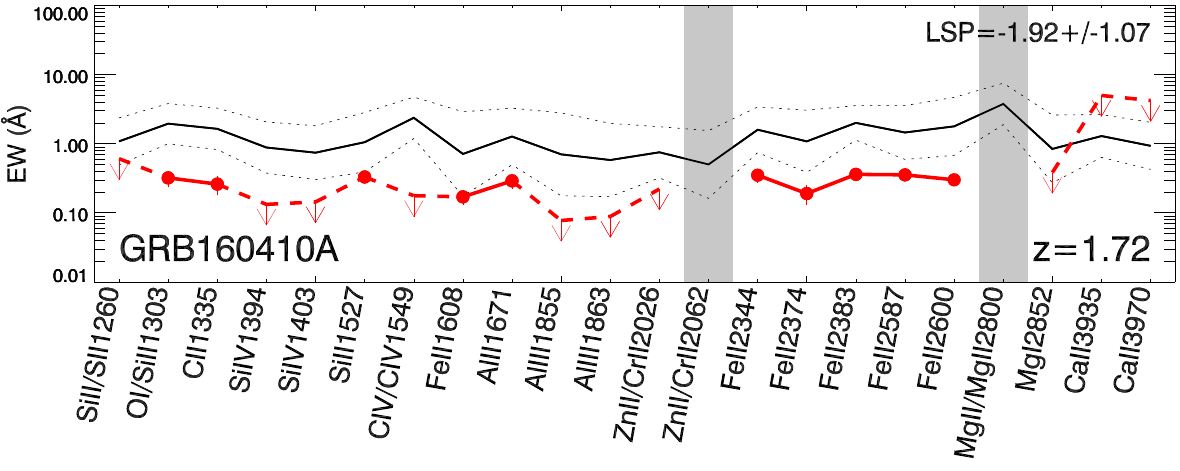}
	\includegraphics[width=\columnwidth]{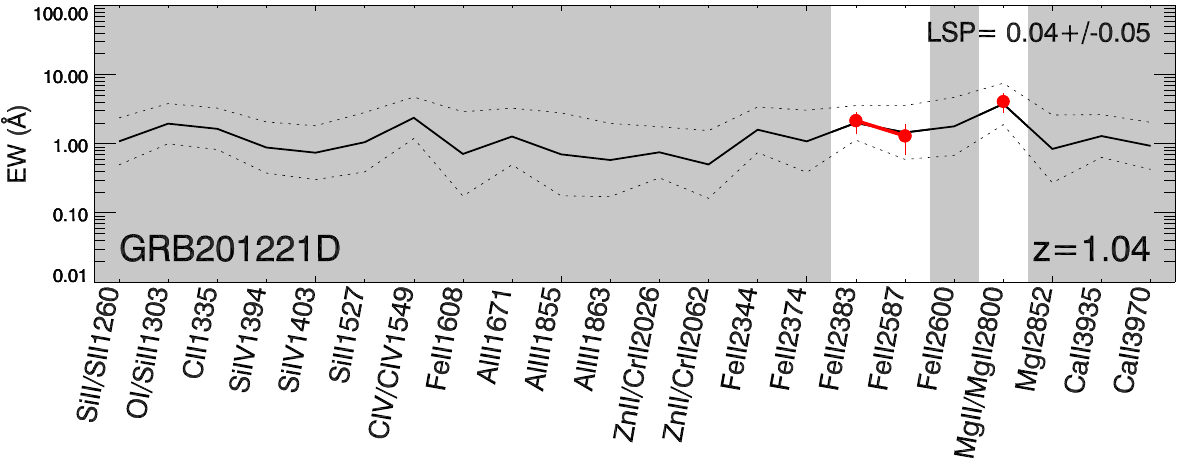}
	\caption{Equivalent width diagrams for the GRB\,130603B \citep{deUgartePostigo2014AA}, GRB\,160410A and GRB\,201221D afterglow spectra and GRB\,201221D spectrum following the process described in \protect\cite{2012A&A...548A..11D}. Red dots denote the corresponding EW for the absorption lines detected, whereas the red triangles show an inferred upper limit. The black solid line marks the average of the sample used in \protect\cite{2012A&A...548A..11D} and the upper and lower dotted lines show the standard deviation. Grey areas represent no detection of the corresponding spectral features is possible.}
	\label{fig:LSP_fitting}
\end{figure}

Fig.\,\ref{fig:LSP_fitting} shows how all the spectral features that we measured are well below the lower 1-$\sigma$ region for the EW of the sample. This difference is even stronger in the case of high-ionisation features. The \civ\ and \siiv\ limits that we derive are further away from the lower 1-$\sigma$ EWs of the sample than the \cii\ and \siii\ lines. This is indicative of a low ionisation.

\begin{figure}
	\includegraphics[width=\columnwidth]{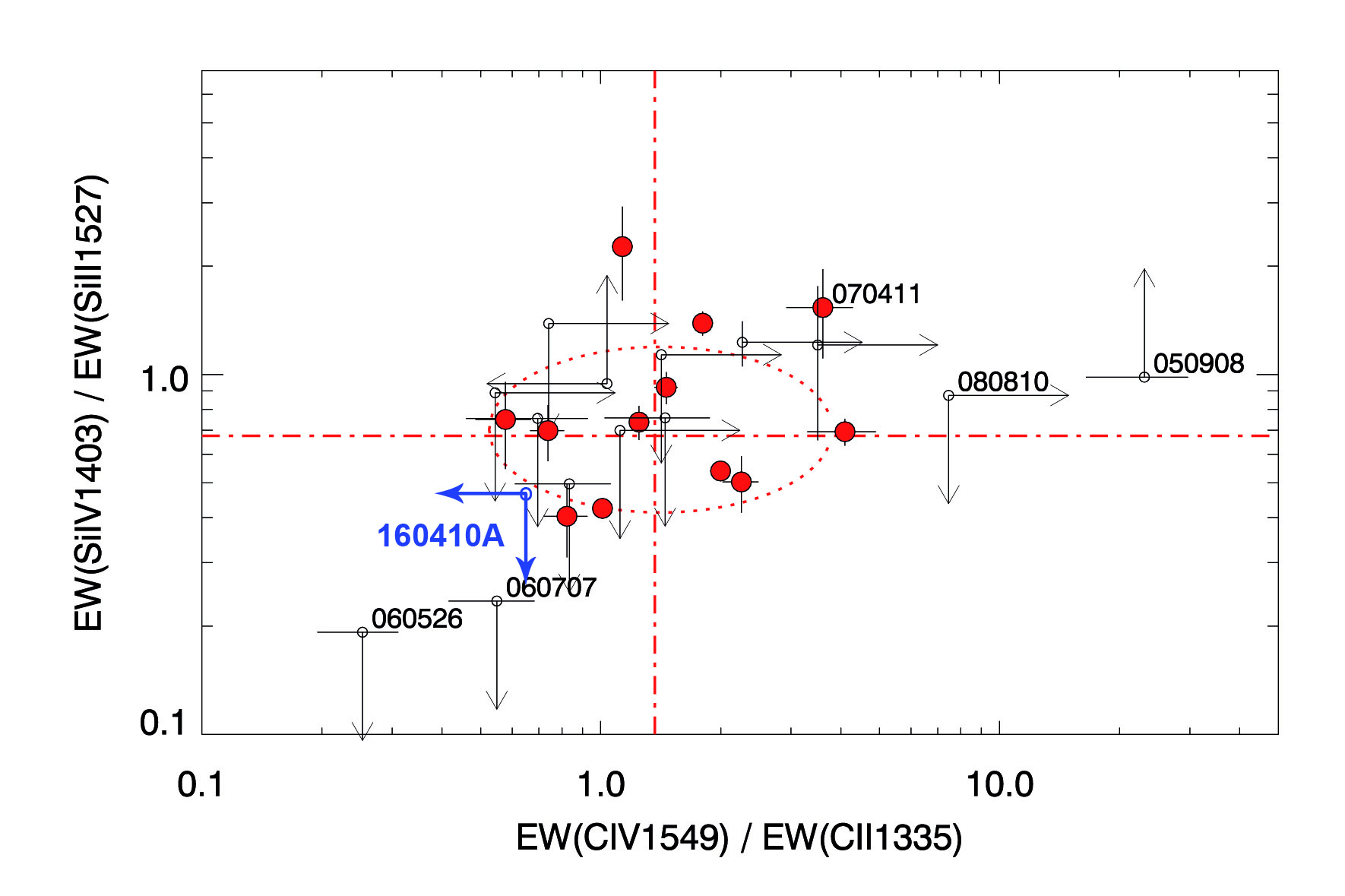}
	\caption{High-/low-ionisation C and Si line EW ratio comparison for GRB\,160410A (\textit{in blue}) compared with the sample of and using the method presented in \protect\cite{2012A&A...548A..11D}. The red filled dots mark the ratio for detections in the afterglow spectroscopy of long GRBs, empty dots with arrows show upper limits in one or both ratios. The red dash-dotted lines represent the average values of the sample, the ellipse marks the $1\sigma$ region.} \label{fig:C4_Si4_relation_fig}
\end{figure}

The low ionisation is clearly seen in Fig.\,\ref{fig:C4_Si4_relation_fig}, where we plot the line ratios \civ/\cii\ and \siiv/\siii\ of our sight-line, as compared to the sample of \cite{2012A&A...548A..11D}. In the case of GRB\,160410A, we use the 3-$\sigma$ limits to the detection of \civ\ and \siiv. GRB\,160410A is located in the lower left area of the diagram, just outside the 1-$\sigma$ region of the sample, amongst the lowest ionisation sight-lines of the sample. The moderate S/N of our spectrum prevents us from showing stronger limits to the high-ionisation lines.

\subsubsection{Intervening systems}

\begin{figure}
	\includegraphics[width=\columnwidth]{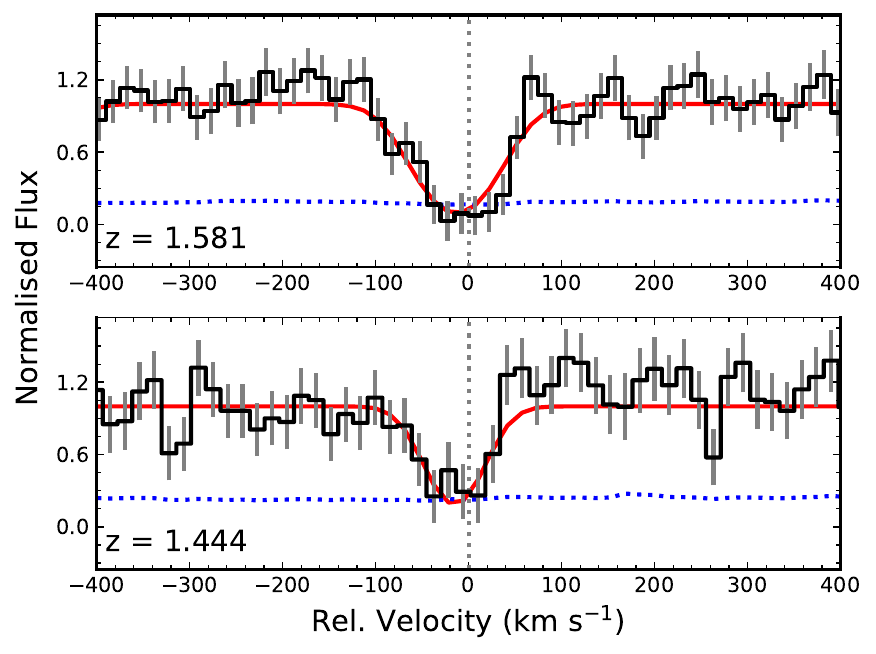}
	\caption{Absorption features detected along the line-of-sight towards GRB\,160410A for the two intervening systems. \textit{Top:} \civ\,($\lambda\lambda$\,1550\AA) line at $z=1.581$. \textit{Bottom:} Absorption line corresponding to \civ\,($\lambda\lambda$\,1550\AA) at $z=1.444$. See the caption of Fig.\,\ref{fig:_lines_fit} concerning the colour coding. The vertical dotted grey line is the fitted broad component at $v=0$\,km\,s$^{-1}$. We see that the minimum of the Voigt best-fit profile, in both cases, is clearly shifted to the left due to the lines being saturated.}
	\label{fig:intervenings}
\end{figure}

In the afterglow spectrum we detect two further systems in the line-of-sight as already mentioned by \protect\cite{2019A&A...623A..92S}, at redshifts $z=1.581$ and $z=1.444$. Both show the \civ\ $\lambda\lambda$\,1548, 1550 doublet, but no other features were detected due to the low S/N of the spectrum. We perform a Voigt fitting to both \civ\ absorbers by defining a broad component at $v=0$\,km\,s$^{-1}$ with a \textit{b}-parameter of $b=40$\,km\,s$^{-1}$ for the doublet. The fitting results are plotted in Fig.\,\ref{fig:intervenings}, resulting in a total column density of $\log (N{\rm /cm}^{2})> 14.82$ for the system at $z=1.581$ and $\log (N{\rm /cm}^{2})> 14.91$ for the one at $z=1.444$. Both values can only be considered as an upper limit since the feature is clearly saturated in both cases. We tentatively detect two possible absorption features at the corresponding wavelength for \civ\ at $z=1.663$, however, we cannot securely confirm these lines.

\subsection{A hostless burst?}\label{nohost}
\label{sec:nohost}
We do not find any source at the GRB\,160410A position in our late GTC/OSIRIS image (right panel of Fig.\,\ref{fig:FindingChart}, \citealt{2022ApJ...940...56F} also do not detect the host in other filters to shallower limits). We obtain a limiting  magnitude of $r^\prime>27.17$ mag (AB, corrected for Galactic extinction), which corresponds to an absolute magnitude limit of $M_{r^\prime}> -18.44$ mag. None of the nearby sources in the field are likely to be the possible host of GRB\,160410A. The closest object is at a projected distance of $r \sim 4\farcs9$ from the GRB position which translates to a distance\footnote{We made use of the \texttt{astropy} packages \texttt{SkyCoord} and \texttt{Cosmology}.} of $\sim42$ kiloparsecs. The host galaxy is also not detected at $3.6\ \mu$m in our deep \textit{Spitzer}/IRAC observations. Following the methods of \protect\cite{2016ApJ...817....7P,2016ApJ...817....8P}, combined with the significantly deeper optical upper limit, this yields an upper limit on the stellar mass of $M_{*}\lesssim1.14\times10^9$ M$_\odot$.

Given our upper limit on the host stellar mass we compare our results to the mass-metallicity relation (MZR) following the equations presented in \cite{2016MNRAS.456.2140M}. The low Fe-based (see Sect. \ref{sec:metallicity}) metallicity value would imply a stellar mass of $\log (M_*/M_\odot) = 6.18 \pm 0.52$. This value indicates a low stellar mass for the host galaxy of GRB\,160401A. This together with the deep limits in \textit{r}-band and $3.6\ \mu$m (see Sect. \ref{sec:Spitzer}) is telling us that the host galaxy for GRB\,160410A must be very faint

\begin{figure}
	\includegraphics[width=\columnwidth]{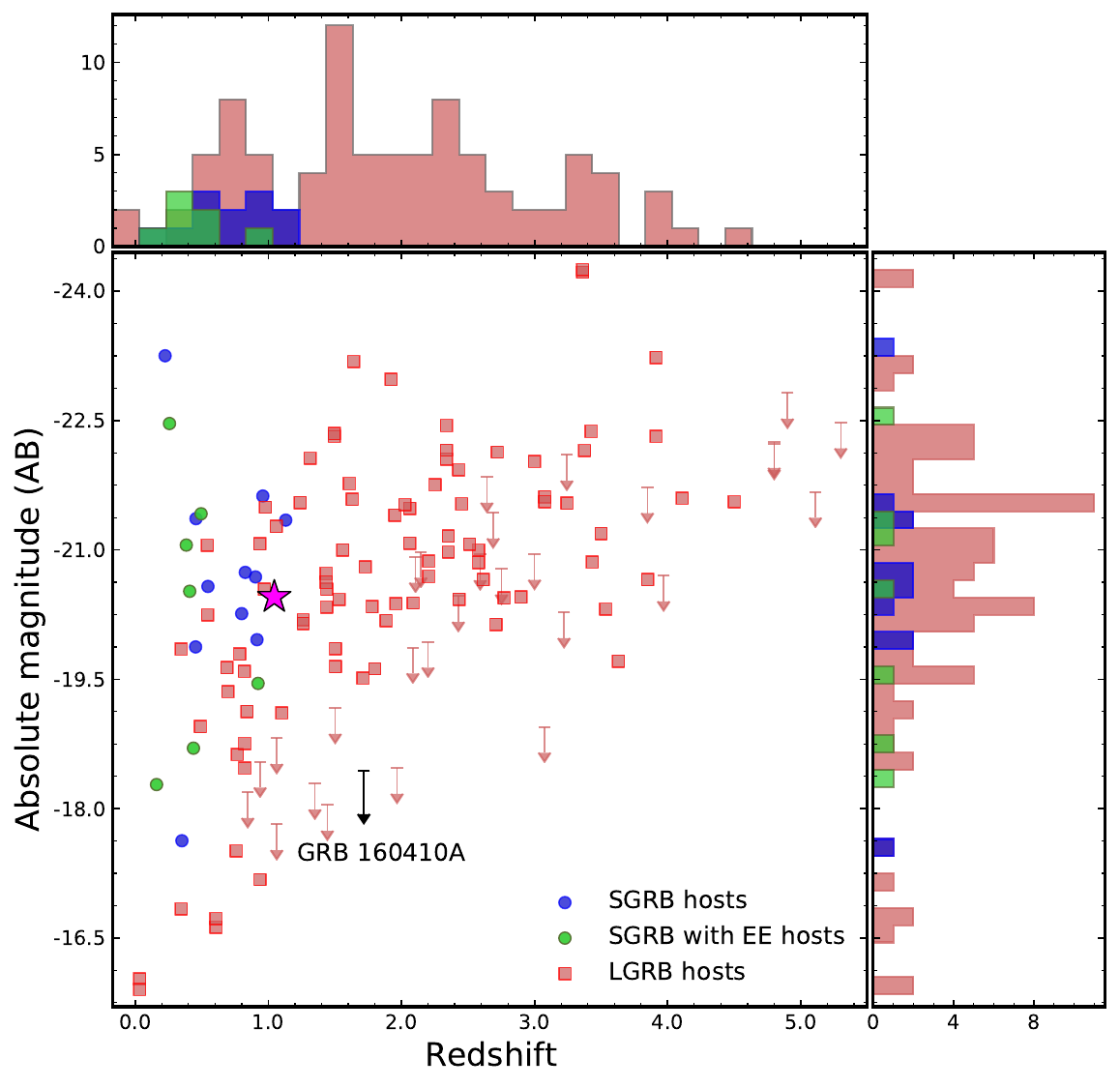}
	\caption{Long GRB host magnitudes in the $r^\prime$ and $R_C$ bands from the SHOALS sample (D. A. Perley, S. Schulze, priv. comm.). For SGRB hosts, we take $r^\prime$ and $R$ magnitudes from \protect\cite{2010ApJ...725.1202L} and \protect\cite{2022MNRAS.515.4890O} when available. The sample is the same one utilized in the ``Amati'' relation (see Sect. \ref{sec:short_long}). Note that we are using a fixed (observer-frame) band-pass filter despite a wide redshift distribution, hence the actual rest-frame band would be blue-shifted by ($1+z$). For clarity, we plot $r^\prime$ and $R_C$ with the same colour. On top and left-most we show the distribution in redshift and absolute magnitude, respectively, for each sample. In blue, we show the SGRB hosts, in clear green the SGRB with EE hosts and in clear red, the distribution for long GRB hosts. The GRB\,160410A upper limit is plotted with a black arrow and GRB\,201221D, with a magenta star. All the absolute magnitudes are corrected for Galactic extinction \protect\citep{Schlafly11}.}
	\label{fig:host_absolute_magnitude}
\end{figure}

We compare our host galaxy limit to other, securely associated, long and short GRB hosts from  the literature. For short GRBs, we use the sample from \cite{2010ApJ...725.1202L} and \cite{2022MNRAS.515.4890O}, for long GRB hosts we use the SHOALS sample (\protect\citealt{2016ApJ...817....7P},\ D. A. Perley, S. Schulze, priv. comm.). We selected only those GRB hosts with detections or upper limits in the $r^\prime$ and $R_C$ bands. In Fig.\,\ref{fig:host_absolute_magnitude} we see that, for short GRBs, the associated hosts have a broad distribution in brightness. Note, however, that only three SGRBs have a spectroscopically confirmed redshift from absorption lines in their GRB afterglow, whereas the rest get their redshifts from the associated host galaxies. At the same time, SGRBs at redshifts beyond $z\sim1$ are very sparse, hence we have to be aware of possible biases here, as had been found for LGRB hosts before the presentation of unbiased samples \citep[see e.g.][]{2016ApJ...817....7P,2016ApJ...817....8P}. However, we can conclude that any host of GRB\,160410A would be at the faint end of the distribution, even compared to long GRB hosts.

The photometric observations alone, without spectroscopic confirmation of the redshift, could have led us to think that GRB\,160410A belonged to the class of hostless SGRBs. However, the presence of a DLA as well as the metallic absorption features are indicative of the burst being hosted by a galaxy.

\subsection{GRB~160410A afterglow light curve and its spectral energy distribution.}
\label{sec:ligh_curve}

\begin{figure}
	\includegraphics[width=\columnwidth]{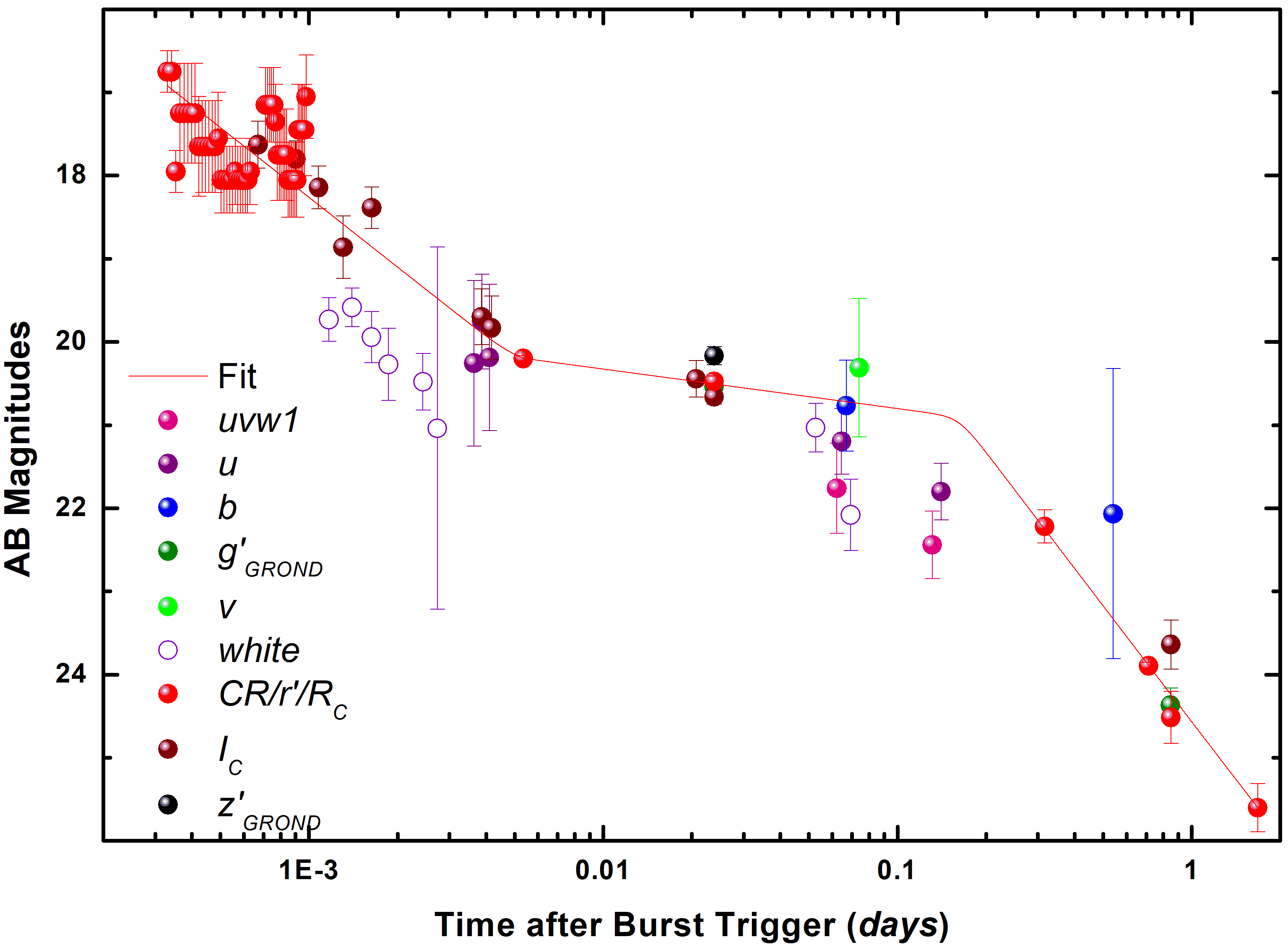}
	\caption{Light curve of the GRB\,160410A afterglow. For reasons of clarity, upper limits are omitted. Early $I_C$ data are from \protect\cite{2016GCN.19277....1T}, the $R_C$ data point at 0.31 d is from \protect\cite{2016GCN.19280....1W}, the rest from this work. Data are in the AB magnitude system and corrected for Galactic foreground extinction. The red line show the modelled light curve for the $CR/r^\prime/R_C$-band. A clear steep-shallow-steep transition is visible.}
	\label{fig:light_curve}
\end{figure}

The light curve of the optical afterglow is described by a smoothly double-broken power-law, which yields a steep/shallow/steep decay with decay indices $\alpha_\mathrm{steep}$, $\alpha_\mathrm{plateau}$, and $\alpha_\mathrm{late}$, respectively, as well as two break times $t_{b,1}$, $t_{b,2}$. We find $\alpha_\mathrm{steep}=1.11\pm0.17$,  $\alpha_\mathrm{plateau}=0.19\pm0.04$,  $t_\mathrm{b,1}=0.0052\pm0.0030$ d ($446\pm256$ s); and $\alpha_\mathrm{late}=1.86\pm0.18$, $t_\mathrm{b,2}=0.162\pm0.030$ d. The break sharpness was fixed to sharp values ($n=-10,10$ respectively), and the host galaxy was neglected, as we find no evidence for any host down to very deep limits (Sect.\,\ref{sec:nohost}). We show the optical afterglow of GRB\,160410A in Fig.\,\ref{fig:light_curve}, including the best-fit triple power-law.

\begin{figure}
	\includegraphics[width=\columnwidth]{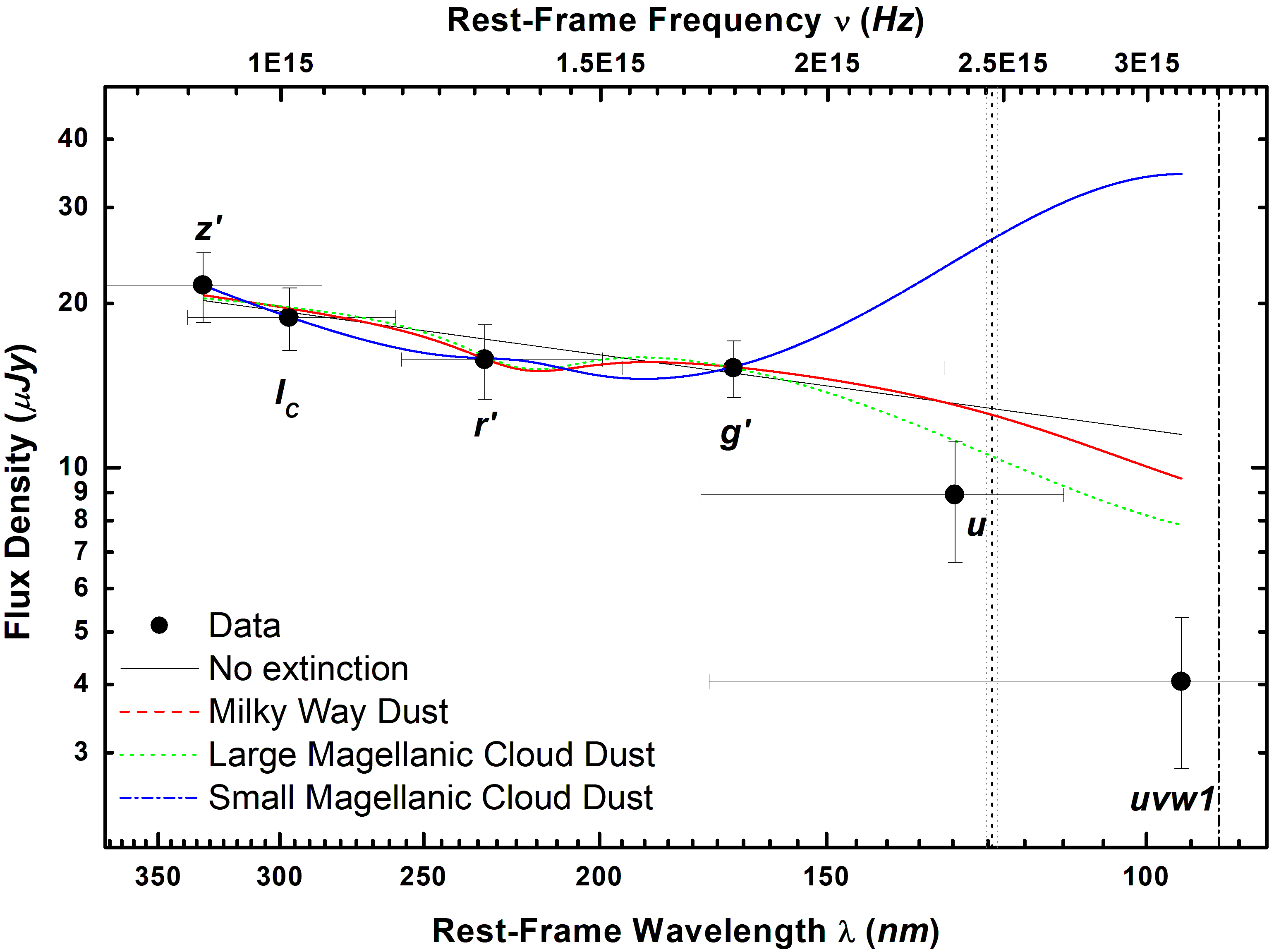}
	\caption{Spectral Energy Distribution of the afterglow of GRB\,160410A. Flux densities are determined at break time, 0.162 d. Horizontal error bars represent filter widths, these errors were not included in the fits. We show fits with no extinction (straight black line), Milky-Way dust (red dashed line), Large-Magellanic-Cloud dust (green dotted line), and Small-Magellanic-Cloud dust (blue dash-dotted line). The SMC fit is clearly unphysical, and the results of the LMC fit also show it is not realistic, see Sect. \ref{sec:ligh_curve}. The MW is possible, but a fit with no extinction is the preferred model. The $u$ and $uvw1$ data points were excluded from the fit as they are affected by \lya\ and Lyman forest/limit absorption. The \lya\ wavelength at the redshift of the GRB is marked by a vertical dotted line, with the adjacent dotted gray lines marking the FWHM of the \lya\ line. A dash-dotted vertical line marks the Lyman cutoff.}
	\label{fig:OA_SED}
\end{figure}

We use the fit to construct the Spectral Energy Distribution (SED), which stretches from $uvw1$ to $z^\prime$. We fit the SED with both a simple power-law (no dust) and dust models for Milky Way (MW), Large (LMC), and Small Magellanic Cloud (SMC) dust \citep{Pei1992ApJ}. Therefore, we exclude the $uvw1$ and $u$ bands as these lie blueward and within \lya\ respectively. The SED and the fits are shown in Fig.\,\ref{fig:OA_SED}.

For no extinction, we find $\beta=0.46\pm0.25$, and a fit with $\chi^2/{\rm d.o.f.}=0.19$. For the three dust models, we find: MW dust: $\beta=0.30\pm0.35$, $A_V=0.14\pm0.22$ mag;
LMC dust: $\beta=-0.18\pm1.21$, $A_V=0.34\pm0.63$ mag; and SMC dust: $\beta=2.55\pm2.81$, $A_V=-0.74\pm0.99$ mag. All these fits are show over-fitting ($\chi^2/{\rm d.o.f.}$ from 0.06 to 0.16). However, the SMC fit is clearly unphysical (yielding negative extinction, i.e. emissive dust), and the negative intrinsic spectral slope for LMC dust is also not expected for GRB afterglows. Only the MW fit yields a sensible result, however, dust extinction is 0 within errors, so we find no evidence for dust and hence continue to work with the no-extinction fit. The no-extinction model is also the one showing the least overfitting.

To compare this high-redshift SGRB afterglow with other SGRBs, we take SGRBs with redshifts and well-detected afterglows (as well as the deep upper limits of GRB\,050509B) from the sample of \cite{Kann2011ApJ} (there called ``Type I GRBs''), and provide additional analysis for four further SGRBs (one with an unsure classification) in Appendix \ref{SGRBS}. We show the observed (corrected for Galactic foreground extinction and, if needed, host-galaxy contribution) light curves in the Appendix, in Fig.\,\ref{fig:KPObs}.

\begin{figure}
	\includegraphics[width=\columnwidth]{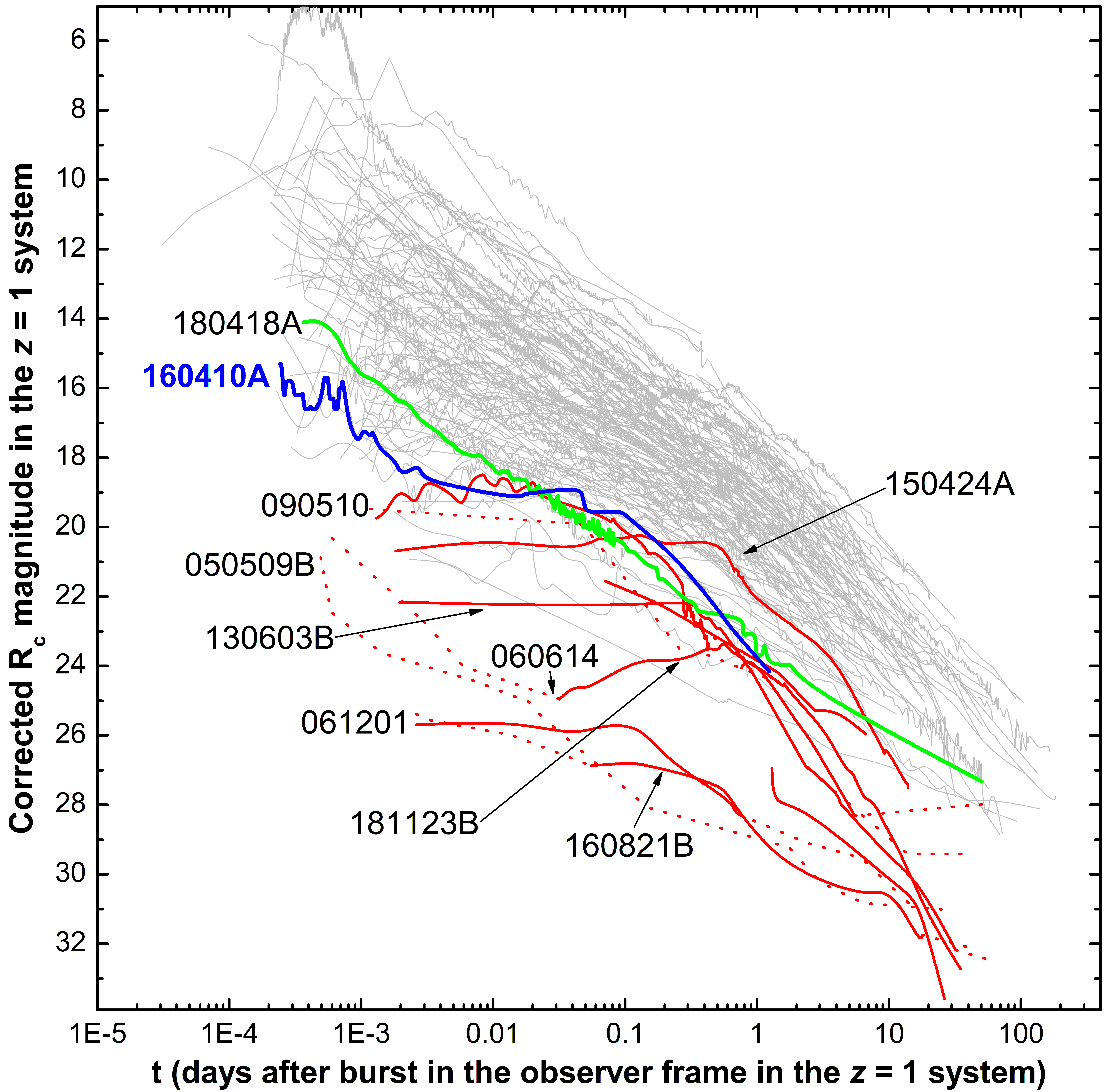}
	\caption{Afterglow of GRB\,160410A (thick blue line) in the context of a large sample of GRB afterglows. Thin grey lines are afterglows of long GRBs. Thicker red lines are a selection of afterglows of other SGRBs. GRB 180418A (green line) has an insecure classification. The afterglows are given in the $z=1$ system, see text for more details.}
	\label{fig:KPzone}
\end{figure}

GRB\,160410A lies at a higher redshift than any of these events except for GRB 181123B, which lies at a slightly higher redshift. Knowing the redshift and SED, we use the method of \cite{Kann2006ApJ} to shift all afterglows to a common redshift of $z=1$, corrected for dust extinction (see Fig.\,\ref{fig:KPzone}). The afterglow of GRB\,180418A is the brightest SGRB afterglow at very early times, but not far above that of GRB\,160410A. During the plateau phase, the afterglow of GRB\,160410A has a  similar luminosity as that of the extremely intense GRB\,090510. At later times, the very extended plateau of the GRB\,150424A afterglow makes it brighter than GRB\,160410A at the same time. Due to their (relatively) high redshifts and brightness at early times the nature of GRBs 160410A and 180418A have been heavily discussed (see also Sect. \ref{sec:discussion}). However, at 12\,h post burst, both afterglows still lie among the faintest LGRB afterglows, which is further evidence that these are likely true SGRBs.

\subsection{The GRB~201221D spectra}
\label{sec:grb201221D}
\begin{figure*}
\begin{center}
	\includegraphics[width=\textwidth]{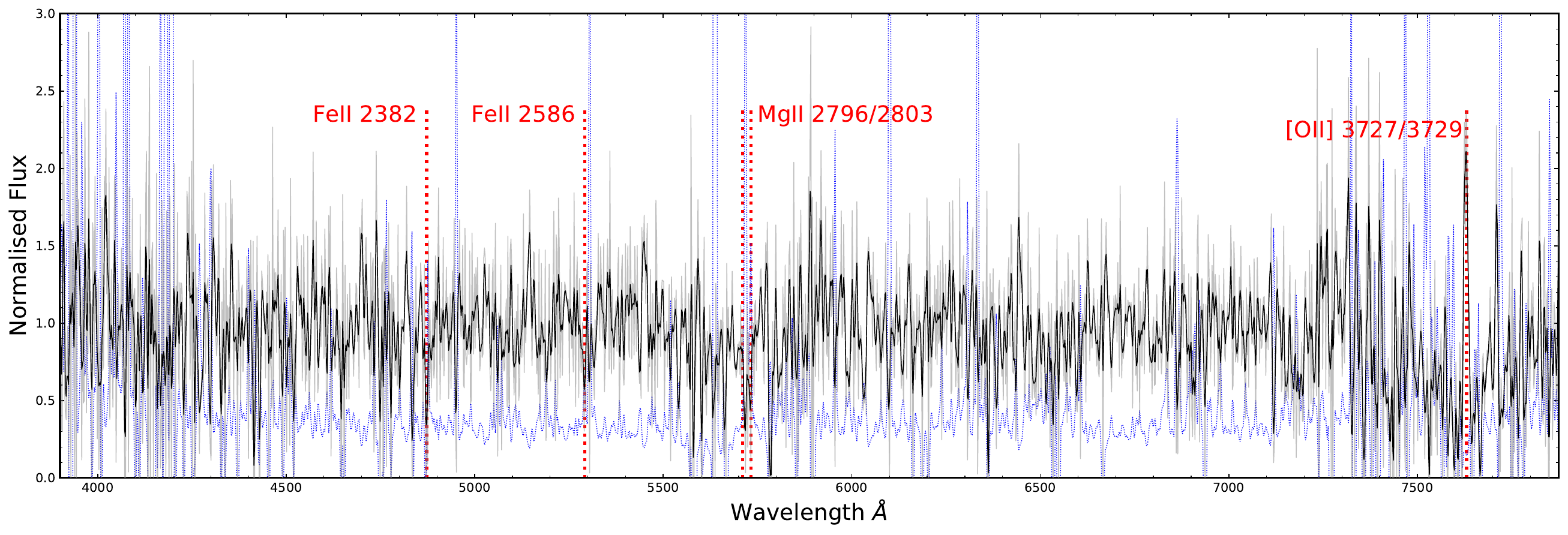}
	\caption{GTC/OSIRIS spectrum of the afterglow emission of GRB\,201221D. For plotting reasons, we smooth the spectrum by applying a Gaussian kernel of $1\sigma$ to the data. The colour coding is the same one as used in Fig.\,\ref{fig:AfterglowSpectrum}. We add the non-smoothed spectrum (grey) in the background.}
	\label{fig:AfterglowSpectrum201221D}
\end{center}
\end{figure*}

This is the only SGRB since GRB\,160410A for which we were able to detect absorption lines in the spectrum, albeit with a low S/N. The GRB afterglow is detected in the acquisition image with a magnitude of $r^\prime=23.95\pm0.20$ mag, calibrated using five Pan-STARRS field stars and not corrected for Galactic extinction \citep{Schlafly11}. In the GRB\,201221D spectrum, we find up to four absorption lines that we interpret as due to the \mgii\ doublet and to \feii\ lines. In addition, we also observe \oii\ in emission from the host. The emission falls within a telluric band and its detection tells us that the emission must therefore be rather strong \citep[see][]{2020GCN.29132....1D}. All the features are at a redshift of $z=1.0450\pm0.0008$, making GRB\,201221D another high-redshift SGRB and, in this case, with a clearly star-forming host galaxy. The spectrum is shown in Fig.\,\ref{fig:AfterglowSpectrum201221D}.

We measure EWs in the spectrum using the \href{http://grbspec.iaa.es/}{GRBspec} database and compare it to the sample of long GRBs \citep{2012A&A...548A..11D} in Fig.\,\ref{fig:LSP_fitting}. In this case, the strength of the absorption features is consistent with the average for long GRBs (LSP $= 0.04 \pm 0.05$) in contrast to GRB\,160410A where the lines were much weaker. Due to the wavelength coverage, we cannot detect the high ionisation absorption lines of \siiv\ and \civ\ and hence cannot come to a conclusion on the ionisation of the ISM. In any case, the \textit{r'}-band value measured for the GRB afterglow is consistent within errors with the one for the host galaxy (see Tab.\ref{tab:201221D_host}) and therefore, continuum contamination from the host galaxy is present in our spectrum. Thus, the EW/LSP measures might be contaminated and are not fully comparable to GRB\,160410A and GRB\,130603B.

\subsection{The host of GRB~201221D}
\label{sec:201221D_host}
In contrast to GRB\,160410A we do find a faint host candidate for GRB\,201221D at the same position as the GRB. The photometry of the host candidate is shown in Tab. \ref{tab:201221D_host} for all the bands observed with the LBT telescope.
\begin{table}
\caption{ Photometry of the GRB 201221D host galaxy. Data are given in AB magnitudes and are corrected for Galactic foreground extinction \protect\citep{Schlafly11}.}
\centering
\label{tab:201221D_host}
\begin{tabular}{ccccc}
\hline
\hline
\noalign{\smallskip}
Time after burst & Magnitude & Exposure Time & Band & Instrument \\
(days) & (AB) & (s) & &  \\
\hline
\vspace{1mm}
19.349 & $23.80	\pm	0.12$ & 120 & $g^\prime $ &	LBC	\\ \vspace{1mm}
19.349 & $23.83	\pm 0.15$ & 120 & $r^\prime $ &	LBC	\\ \vspace{1mm}
19.349 & $23.44	\pm 0.18$ & 120 & $i^\prime $ &	LBC	\\ \vspace{1mm}
19.349 & $23.11	\pm 0.25$ & 120 & $z^\prime $ &	LBC	\\ \vspace{1mm}
13.879 & $22.40	\pm 0.17$ & 60 & $J$ &	LUCIFER	\\ \vspace{1mm}
13.895 & $22.15	\pm 0.20$ & 60 & $K_S$ &	LUCIFER	\\
\hline
\hline
\end{tabular}
\end{table}
We analyse the Spectral Energy Distribution (SED) of the host galaxy with the available photometry using \texttt{CIGALE}\footnote{\url{https://cigale.lam.fr/}} \citep{2005MNRAS.360.1413B,2009A&A...507.1793N,2019A&A...622A.103B} in its most recent version. We apply a delayed star-formation history with an age for the main stellar population varying freely from 1.0 Gyr to 13 Gyr and a more recent burst whose age varies from 20 Myr to 1 Gyr. The Initial Mass Function (IMF) we use is described in \cite{2003PASP..115..763C} with a \cite{2003MNRAS.344.1000B} stellar population model, assuming a metallicity (\textit{Z}) of 0.008, 0.02 or 0.05 (where \textit{Z} $= 0.02$ is Solar metallicity \citealt{2003MNRAS.344.1000B}).

The dust attenuation is modeled with the modified attenuation law described in \cite{2000ApJ...533..682C} as implemented in \texttt{CIGALE} (see Sect. 3.4.2 in \citealt{2019A&A...622A.103B}). We consider a Milky Way (MW) \citep{1989ApJ...345..245C} extinction model with a $R_V=3.1$ and a colour excess in the nebular lines starting in 0.1 and then, varying in steps of 0.05 between 0.05 and 1.0. We also considered a Small Magellanic Cloud (SMC) and a Large Magellanic Cloud (LMC) \citep{Pei1992ApJ} extinction model but in both cases, the model was unsatisfactory compared to the one performed considering a MW model. We also allow the attenuation curve slope to vary from $-0.4$ to $0.4$, changing it in steps of $0.2$ (see Eq. 8 in \citealt{2019A&A...622A.103B}). For the re-emitted energy from dust heated by stellar photons, we use the models from \cite{2014ApJ...784...83D} and let $\alpha_{\textnormal{IR}}$, the exponent of the radiation field intensity distribution \citep{2002ApJ...576..159D}, vary between $1.0$, $2.0$ or $3.0$.

We find a best fit to the SED using a model galaxy spectrum with an intermediate mass and a moderate star-formation rate (SFR), as well as a low specific star-formation rate (sSFR). However, our modelling shows some degree of over-fitting, which might be due to the low number of data points. The results for the SED modeling are listed in Tab. \ref{table:sed_201221D_results} and a figure of the best model can be seen in Appendix \ref{sec:sed_fit}.
\begin{table}
\caption{Properties of the putative host galaxy of GRB\,201221D.}
\label{table:sed_201221D_results}
\centering
\begin{tabular}{c c}
\hline\hline
Property    & Value         \\
\hline
A$_V$ (mag) & $0.56\pm0.34$         \\
$Z$ & 0.02         \\
\noalign{\smallskip}
$\log_{10}(M) (M_{\odot})$ & 9.79$_{-0.19}^{+0.13}$   \\
\noalign{\smallskip}
$\log_{10}(SFR)\ (M_{\odot}/yr)$ & 0.81$_{-0.33}^{+0.19}$   \\
\noalign{\smallskip}
$sSFR\ (Gyr^{-1})$  & 1.06 $\pm$ 0.67   \\
Reduced $\chi^2$ & 0.45 \\
\hline
\end{tabular}
\end{table}


\section{Discussion}
\label{sec:discussion}
\subsection{On the short/long nature of GRB~160410A}
\label{sec:short_long}

Ever since the discovery of a bimodal distribution of GRB prompt emission light curves both in the temporal range \citep{Mazets1981ApSS} as well as the spectral hardness \citep{1993ApJ...413L.101K}, methods to discern between ``long/soft'' and ``short/hard'' GRBs have been extensively discussed in the literature \citep[see e.g.][for some works on this topic]{Lu2010ApJ,Lu2014MNRAS,Tsutsui2013,ShahmoradiNemiroff2015MNRAS,ZYCC2016MNRAS,2018PASP..130e4202Z,Li2016ApJS,LZY2020,Jespersen_2020}. Here we will study GRB\,160410A with several classification methods to derive clues on its nature.

\begin{figure}
	\includegraphics[width=\columnwidth]{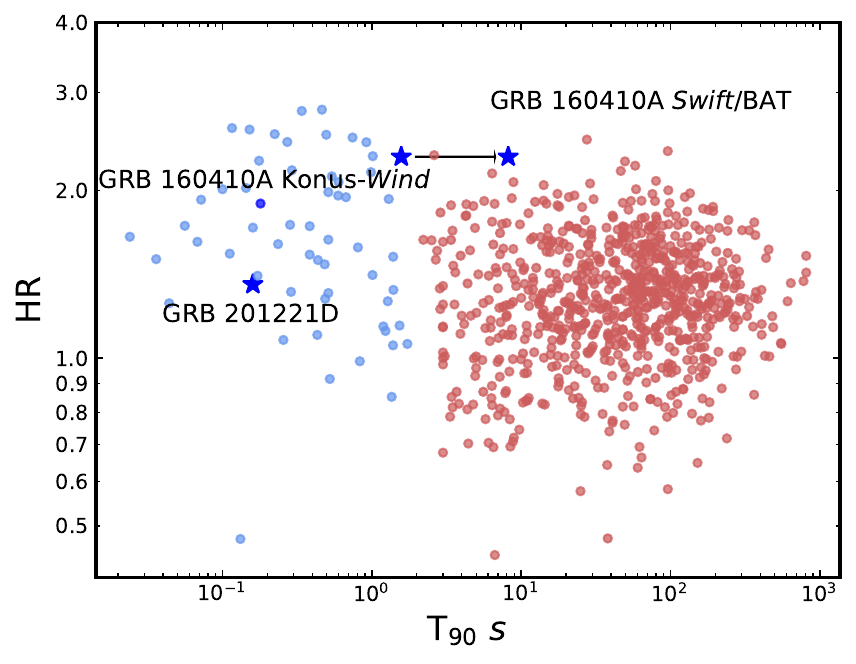}
	\caption{Hardness Ratio (HR) vs. duration $T_{90}$ using data from \protect\cite{2016ApJ...829....7L}. Based on the work from \protect\cite{1993ApJ...413L.101K}, we show SGRBs in blue and long GRBs in red. The hardness ratio of GRB\,201221D is marked with a blue stars as derived from the \textit{Swift}/BAT data products (\url{https://gcn.gsfc.nasa.gov/notices_s/682269/BA/}). For GRB\,160410A we plot the HR derived in \protect\cite{2021ApJ...911L..28D} with the $T_{90}$ from \textit{Swift}/BAT on top right and $T_{90}$ derived for the initial peak complex from Konus-\textit{Wind} data \protect\citep{2020MNRAS.492.1919M} on top left.}
	\label{fig:Hardnessratio}
\end{figure}

\begin{figure}
	\includegraphics[width=\columnwidth]{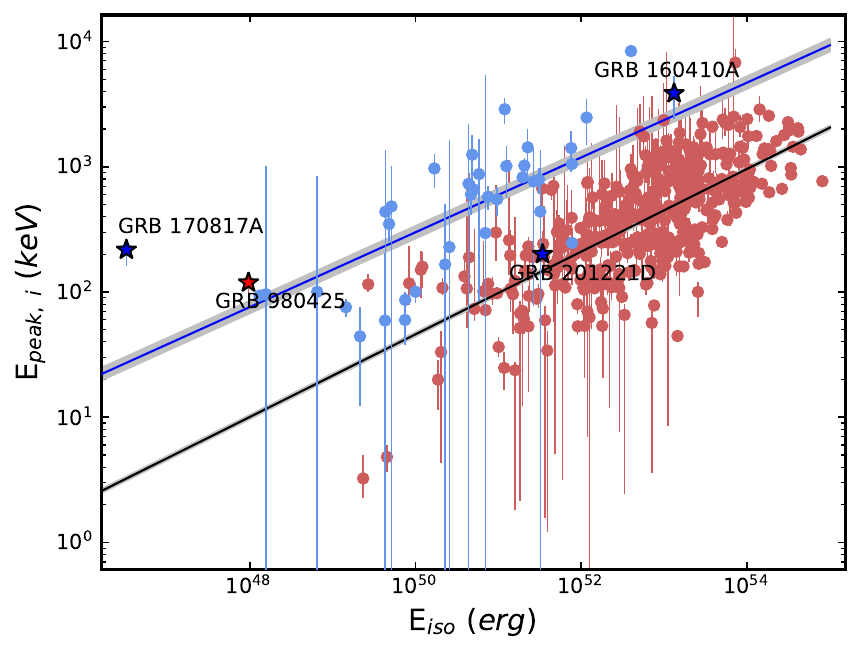}
	\caption{Modified version of the Amati relation. We highlight the positions of GRB\,160410A and GRB\,201221D with blue stars, as derived from the Konus-\textit{Wind} published data, see Sect.\,\ref{sec:short_long}. SGRBs are plotted in blue, long GRBs in red. In the corresponding colours, the best linear fits are plotted together with a shaded area around them marking the $1\sigma$ region for each fit. Here we do not distinguish between SGRBs and SGRBs with extended emission.}
	\label{fig:Amatirelation}
\end{figure}

\cite{Tsutsui2013} first found significant statistical evidence for an ``Amati'' relation  between the isotropic energy release $E_{\textnormal{iso}}$ and the rest-frame peak energy of the prompt emission spectrum $E_{\textnormal{peak, rest}}$, long known for long GRBs \citep{Amati2002AA,Amati2006MNRAS}, for SGRBs parallel to that of long GRBs but offset by a factor of $\sim100$. Using a significantly increased sample, \citet[][henceforth MP20]{2020MNRAS.492.1919M,2021MNRAS.504..926M} confirm this result, and GRB\,160410A fits with the Amati relation for SGRBs (Fig.\,\ref{fig:Amatirelation})\footnote{Note the Amati relation is also a powerful tool for the opposite case, such as the recently reported GRB\,200826A \citep{2022ApJ...932....1R,Ahumada2021NatAst,Zhang2021NatAst}, which was temporally short but fully in agreement with the long-GRB Amati relation and was shown to be accompanied by a supernova.}. Furthermore, MP20 introduce two classifiers, $EH$ and $EHD$ (Energy-Hardness and Energy-Hardness-Duration, respectively). They find that GRB\,160410A is in full agreement with the high values found for other short GRBs.

To confirm the results of MP20, we gathered GRB energetics, mostly from \cite{Tsvetkova2017ApJ,Tsvetkova2021ApJ} and compute the isotropic energy\footnote{\url{https://github.com/steveschulze/GRB_Eiso.git}}. The final sample is shown in Fig.\,\ref{fig:Amatirelation} and we perform a simple fitting to the two GRB classes following Eq.\,\ref{eq:linealfitting} as presented in MP20.

\begin{equation}
    \log\Big(\dfrac{E_{\textnormal{peak}}}{100\ \textnormal{keV}}\Big) = a \cdot \log\Big(\dfrac{E_{\textnormal{iso}}}{10^{51}\ \textnormal{erg}}\Big) + b
    \label{eq:linealfitting}
\end{equation}

For SGRBs we found a slope of $a_{\textnormal{short}}=0.30$ and $b_{\textnormal{short}}=0.78$ with a standard deviation of $\sigma_{\textnormal{short}}=0.05$ and $R^2=0.50$, for long GRBs the slope is $a_{\textnormal{long}}=0.33$, $b_{\textnormal{long}}=-0.01$ with a $\sigma_{\textnormal{long}}=0.02$ and $R^2=0.52$. We do not perform a more extended fitting as it is done in MP20 since it is not the aim of the present paper.

Another classification distinction is the \textit{hardness ratio} (HR), which is the ratio of the fluence in the 50--100 keV band over the 25--50 keV band. \cite{2021ApJ...911L..28D} obtained a $\textnormal{HR}=2.3\pm0.5$ for the main spike and claim GRB\,160410A is a short GRB with extended emission. GRB\,160410A can be classified as a long GRB (see Fig. \ref{fig:Hardnessratio}), if we only consider the $T_{90}$ as the classification criterion. However, if we consider the initial peak complex and GRB\,160410A as a short GRB with EE, as in \cite{2021ApJ...911L..28D}, we see that it matches with the short GRB scheme.

\cite{Kann2011ApJ} compare the afterglows of SGRBs to a large sample of those of long GRBs. They find $\overline{M_B}=-23.14\pm0.17$ mag (FWHM 1.61 mag) for long GRB afterglows at one day in the $z=1$ frame, and $\overline{M_B}=-17.34\pm0.50$ mag (FWHM 1.65 mag) for the short GRB afterglows at the same time. For GRB\,160410A, we derive $M_B=-19.04$ mag from our fit to the late-time light curve. This places the absolute magnitude of the afterglow at the upper edge of the FWHM of the SGRB afterglow absolute magnitude distribution, but 2.5 mag below that of long GRB afterglows, making it more likely that GRB\,160410A is a short GRB.

Further classifiers in the context of GRB\,160410A can be found in Appendix \ref{sec:short_long_appendix}. With a few exceptions, these find that GRB\,160410A is a genuine SGRB, or at least more likely to be one than a long GRB. We therefore conclude that GRB\,160410A has evidence showing it to be a member of the SGRB population.

The other SGRB studied here, GRB\,201221D, actually matches very well the location for long GRBs in the Amati relation (see Fig. \ref{fig:Amatirelation}). However, it shortness and hardness (see Sect. \ref{sec201221d_HE} and Fig. \ref{fig:Hardnessratio}) put this GRB clearly within the SGRB category.

\subsection{The GRB~160410A environment in the context of long GRBs and short GRBs}
The sample of short GRBs with detected absorption lines in its spectrum is still very small: GRB\,130603B \citep{deUgartePostigo2014AA}, GRB\,160410A, and GRB\,201221D (both in this study). GRB\,160410A is the only one that allows a chemical study of the gas in the host to be performed, as it has been commonly done for long GRBs \citep[see e.g.][]{2018A&A...620A.119D, 2013A&A...557A..18K, 2018MNRAS.479.3456H}. The spectrum shows features common to long GRBs such as \feii, \alii\ or \siii; however, it does not show any high ionisation lines of \siiv\ nor \civ, which are usually detected in long GRB environments \protect\citep[see e.g.][]{2004A&A...419..927V, christensen2011, 2012A&A...548A..11D, 2018MNRAS.479.3456H}. \cite{2016GCN.19278....1C} report the detection of \civ\ lines in their afterglow spectrum together with the \mgii\ $\lambda\lambda$2796,2803 doublet, which, however, happens to fall within the telluric A-band.

Despite the scarcity of currently available SGRB afterglow spectra, it seems that SGRB sight-lines show a large diversity, as it is the case for their host galaxies. The spectra of all three SGRBs with absorption line spectra cover the \mgii\ and the \feii\ lines as common features. GRB\,201221D and GRB\,130603B show the \mgii\ $\lambda$2796, 2803 doublet, while the \feii\ lines are only present in GRB\,201221D. In GRB\,160410A we do not observe \caii\ or \nai\ detected in GRB\,130603B. However, we detect a large number of additional absorption lines that were out of the observable range in the other two SGRBs due to their lower redshift. There is also a tentative detection of \Niii\, but the significance is very low.

GRB\,130603B happened in a spiral galaxy in what appears to be a tidally disrupted arm \citep{deUgartePostigo2014AA}. In the spectrum of GRB\,201221D we detect \oii\ emission lines, implying the presence of an underlying star-forming host galaxy.  Furthermore, both we and multiple other observing teams \citep{2020GCN.29128....1D, 2020GCN.29133....1K,Rastinejad2020GCN29142,Dimple2020GCN29148} report the detection of a faint extended source at the afterglow position. The host of GRB\,201221D (see Sect. \ref{sec:201221D_host}) is a massive galaxy consistent with the common value found in prior works on SGRB hosts \citep[see e.g.][]{2010ApJ...725.1202L}. The star-formation rate is also in agreement with the values found for SGRB hosts \citep{2009ApJ...690..231B}.

For GRB\,160410A, however, we found no host galaxy down to very deep limits (see Sect.\,\ref{sec:nohost}) and we find no hint of emission lines in the NIR. In contrast, the  detection of the broad \lya\ absorption line and the large column density tell us that the GRB happened within or behind a DLA, which are usually associated with a galaxy \citep[see e.g.][]{2005ARA&A..43..861W, 2014MNRAS.445..225C}. This implies that there must be an underlying galaxy for GRB\,160410A but it might be very faint. It is also possible that the progenitor system of GRB\,160410A has been kicked out from its host and merged outside the host or in the halo. The fact that we detect a DLA in the sight-line implies that the GRB still has to be well within the \ion{H}{i} halo of the galaxy. Given the high redshift of the system and typical kick velocities, a binary NS-NS system could move tens of kiloparsecs away from the host galaxy beyond the extension of hydrogen of the galaxy \protect\citep{2022MNRAS.514.2716M}. The distance in projection to the closest observable galaxy in the GRB\,160410A field is $\sim42$\,kpc (under the assumption this galaxy is at the redshift of GRB 160410A), making it very unlikely to be the host of the GRB while  observing such a large neutral hydrogen column density and many other absorption lines.

In prior works, the detection of the highly ionized \civ\ and \siiv\ absorption lines has been assumed to originate in the hot gas of the galactic halo of DLAs \protect\citep[see e.g.][]{1998A&A...337...51L,2000ApJ...545..591W,2003MNRAS.343..268M,2008A&A...491..189F,2018MNRAS.479.3456H}. The detection of a DLA system and the non-detection of these lines in the GRB\,160410A afterglow spectrum  would hence not favour this GRB to be in the halo or it would imply that the host has no hot gas halo.

Our analysis shows that the environment of GRB\,160410A is very different from the ones measured in the case of long GRB hosts. The probed material has a very low ionisation and very weak lines in general (see Sect.\,\ref{sec:LSP}). In contrast, GRB\,201221D (see Sect.\,\ref{sec:grb201221D}) and GRB\,130603B show LSPs of $0.04\pm0.05$ and $0.20\pm0.13$, respectively, very close to the average value for long GRB sight-lines \citep{2012A&A...548A..11D}. A high LSP value typically points toward an environment highly ionised due to star-formation and, conversely, a low value pinpoints the opposite. For long GRBs, the LSPs are usually high since these are commonly associated with star-forming environments. The SGRBs\,130603B and 201221D occurred within their hosts galaxies; however, for GRB\,160410A, no underlying galaxy was detected. We see in Fig.\,\ref{fig:LSP_fitting} that, whereas for GRBs\,130603B and 201221D the EW profiles follow the average one of long GRBs. However, we note that for GRB\,201221D, EW values might be contaminated by the host continuum (see Sect. \ref{sec:grb201221D}) and, therefore, the LSP. GRB\,160410A is completely different compared to either long GRBs or to the other two SGRBs with detected features in their afterglows. With a significantly lower value for the LSP and the non-detection of an underlying galaxy, GRB\,160410A seems to have happened in a environment that is not actively forming stars.

Furthermore, the measured [Si/Fe] ratio between an $\alpha$-element such Si, that is produced in core-collapse SNe, and the detection of Fe, mostly produced in Type Ia SNe, might be indicating that there is no $\alpha$-enhancement and therefore the absence of recent star formation in the host galaxy, which is consistent with the non detection of the galaxy, the low ionisation and the low metallicity of the traced gas (see e.g. \cite{2009MNRAS.399..574W, 2013ApJ...767..134V}).

In line with those observations, the metallicity of the material observed along the sight-line is very low (Sect.\,\ref{sec:metallicity}), comparable to the lowest values for QSO absorbers (see e.g. \protect\cite{2018A&A...611A..76D} and Fig.\,\ref{fig:metallicity}). Compared to galaxies in the local Universe, the metallicity is higher but inline with values obtained for e.g. Tucana II, an UFD galaxy with a metallicity of [Fe/H]$\sim-3.0$ \citep{2016ApJ...832L...3J}. Note, however, that this metallicity was determined from four stars, neither the ISM nor the CGM. This would further support the idea of an underlying very faint host galaxy for GRB\,160410A, however, the average metallicity of galaxies at $z\sim1.7$ is also lower than in the local Universe.

The fact that we have a low metallicity environment and possibly a very faint host galaxy contradicts general expectations for the environment of short GRBs. If the absorption lines are weak because we have an evolved galaxy with an old population that has exhausted most of the gas, we would expect a high metallicity. However, if the progenitor BNS had been kicked out from its host we should expect a low metallicity. Also \cite{2009ApJ...703.1696Z} assume in their classification scheme of short vs. long GRBs that SGRB hosts have a high metallicity \citep[see also][]{LZY2020}. The only two other GRBs with absorption line spectroscopy and hence secure host galaxy associations do seem to be different from GRB\,160410A, showing much stronger lines and luminous hosts.

The question is whether SGRBs with EE might have somewhat different progenitors and in consequence different hosts or environments. The list of SGRBs with EE is rather short \citep[see e.g.][]{2016ApJ...829....7L} and only a few of them have an associated host galaxy. \cite{2020MNRAS.492.1919M} list eleven events as SGRBs with EE (GRBs~050709, 050724, 060614, 061006, 061210, 070714B, 071227, 080123, 110402A, 150424A and 160410A). We compare the derived host properties from the GRB afterglow spectrum of GRB\,160410A as well as from the field observation with \textit{Spitzer}/IRAC with SGRBs with EE and an associated host galaxy (see Fig.\ref{fig:host_absolute_magnitude}). There is no difference in the distribution of absolute magnitudes between the hosts of normal SGRBs and those with EE, the slight shift towards lower redshifts might be an observational issue since higher redshifts make the detection of the lower luminosity EE more challenging. Taking the stellar masses obtained by SED modeling from \cite{2020ApJ...904...52N} for those seven SGRB showing EE (GRBs~050709, 050724, 061006, 061210, 070714B, 071227 and 080123, \citealt{2020MNRAS.492.1919M}), we see that their masses are typically larger than the upper limit we obtain from the \textit{Spitzer} observations (see Sect. \ref{sec:nohost}), although a few have stellar masses lower than our limit for the host of GRB\,160410A. A more recent and complete study on one SGRB with EE, GRB\,050709, shows a low-luminosity host, a subsolar metallicity and a low SFR \citep{2021A&A...650A.117N}, which would be all in line with what we see for GRB\,160410A. A more recent case is GRB\,181123B, a SGRB with EE and an associated star-forming and massive host at $z=1.754$ \citep[see ][]{Paterson2020ApJ, 2021ApJ...911L..28D}. Its association with the galaxy is still under debate \citep{Rowlinson20201MNRAS}. However, in \citet{2022Natur.612..223R}, the properties of the associated host galaxy of GRB\,211211A are in line with the limits for the host of GRB\,160410A. Despite its prompt emission shape and duration, GRB\,211211A is claimed to be a SGRB with EE and the associated host galaxy seems to be a galaxy with a low stellar mass and a low SFR.

It would be important to settle the issue of the host galaxy association by deep imaging with e.g. the Hubble Space Telescope (HST) or the James Webb Space Telescope (JWST), to determine the nature of the host and the location of the GRB within it to understand why we observe this peculiar environment in this burst. 

SGRBs have also been proposed to happen within galaxy clusters \citep[see e.g.][]{2010ApJ...722.1946B}, so that could be a possible explanation for the absence of detection of a host for GRB\,160410A. However, again, the large column density of neutral hydrogen goes against this hypothesis as DLAs are commonly associated with galaxies, as mentioned before. We also do not see a very crowded field in Fig. \ref{fig:FindingChart}.

DLA systems found in QSO absorbers have been associated with the halo of galaxies \citep{2005ARA&A..43..861W}. The non-detection of an underlying galaxy and the fact that the closest possible galaxy is at 42\,kpc from the GRB location could also mean that the GRB itself happened at a large distance from its host and that the lines we see in the afterglow spectrum are actually an intervening system. However, the non-detection of the GRB afterglow in the \textit{uvw2} and \textit{uvm2} bands and the detection in the \textit{uvw1} band (see Sect. \ref{sec:observations} and Appendix \ref{sec:photometry_table}) establish a redshift upper limit for the GRB afterglow using the  Lyman ``drop-out'' technique of $z=1.8$. This points to an association of the afterglow with the ISM detected at the absorption line redshift, making it unlikely that the DLA belongs to a foreground galaxy.

\section{Conclusions}
\label{sec:conclusions}
In this paper, we present the first study of the ISM of a SGRB host galaxy. The burst itself was one of the hardest and brightest events ever detected and at the highest redshift ever measured directly from its afterglow spectrum (and not from the association with a potential host galaxy). The X-shooter spectrum shows a broad \hi\ absorption with a large column density consistent with being a DLA. We derive a very low metallicity of$\mbox{[Fe/H]} = -2.5$, one of the lowest (dust-depletion corrected) value ever measured. GRB\,160410A shows a very low ionisation compared to what is commonly found for the environments of LGRBs. SED fitting to the light curve of the afterglow and dust depletion analysis from absorption lines finds no indication for dust extinction along the sight-line. We do not find any host galaxy down to a very deep limit; however, the presence of the DLA system in the burst afterglow spectrum indicates that there has to be an underlying host. 

The GRB\,160410A afterglow seems to be rather different compared to the other two SGRB afterglows with detected absorption lines, being significantly brighter than any other SGRB afterglow at very early times except for the controversial case of GRB 180418A (note both early detections were obtained with TAROT). GRB\,201221D is located within a more massive, star-forming ionised host galaxy, consistent with previous findings for SGRB hosts. However, our spectral coverage does not allow us to determine its metallicity. The spectrum of GRB\,160410A was obtained less than ten minutes after the GRB alert and has the largest sample of lines observed in a short GRB afterglow spectrum. This demonstrates the importance of the rapid response mode for observing these events. More data sets of this quality are needed to obtain more robust statistical conclusions on the ISM and environments of SGRBs.

\section*{Acknowledgements}
JFAF acknowledges support from the Spanish Ministerio de Ciencia, Innovaci\'on y Universidades through the grant PRE2018-086507. DAK and JFAF acknowledge support from Spanish National Research Project RTI2018-098104-J-I00 (GRBPhot). AdUP acknowledges funding from a Ram\'on y Cajal fellowship (RyC-2012-09975). MB acknowledges funding associated to a personal tecnico de apoyo fellowship (PTA2016-13192-I). DBM acknowledges research grant 19054 from VILLUM FONDEN. Part of the funding for GROND (both hardware as well as personnel) was generously granted from the Leibniz Prize to Prof. G. Hasinger (DFG grant HA 1850/28-1). A. R. acknowledges support from the INAF project Premiale Supporto Arizona \& Italia.

This work is partly based on observations made with the Gran Telescopio Canarias (GTC), installed in the Spanish Observatorio del Roque de los Muchachos of the Instituto de Astrof\' isica de Canarias, in the island of La Palma.
Partly based on observations made with the Nordic Optical Telescope, owned in collaboration by the University of Turku and Aarhus University, and operated jointly by Aarhus University, the University of Turku and the University of Oslo, representing Denmark, Finland and Norway, the University of Iceland and Stockholm University at the Observatorio del Roque de los Muchachos, La Palma, Spain, of the Instituto de Astrofisica de Canarias.
Partly based on observations made with the Large Binocular Telescope (LBT). The LBT is an international collaboration among institutions in the United States, Italy and Germany. LBT Corporation partners are: The University of Arizona on behalf of the Arizona Board of Regents; Istituto Nazionale di Astrofisica, Italy; LBT Beteiligungsgesellschaft, Germany, representing the Max-Planck Society, The Leibniz Institute for Astrophysics Potsdam, and Heidelberg University; The Ohio State University, representing OSU, University of Notre Dame, University of Minnesota and University of Virginia.

This work made use of the GRBspec database \url{https://grbspec.eu}. This work has made extensive use of IRAF and Python, particularly with \texttt{astropy} \citep[\url{http://www.astropy.org}, ][]{2013A&A...558A..33A, astropy:2018}, \texttt{matplotlib} \citep{Hunter:2007}, \texttt{photutils} \citep{larry_bradley_2020_4044744} and \texttt{numpy} \citep{harris2020array}.

\section*{Data Availability}

The spectroscopic data shown in this paper are publicly available in the GRBSpec database, at http://grbspec.eu. Raw imaging data are available in the observatory archives (ESO - Science Archive Facility, GTC Public Archive, Spitzer Heritage Archive (SHA), \textit{Swift} Archive).
Data non-publicly available on observatory archives, as well as reduced imaging data will be delivered on reasonable request to the corresponding author.




\bibliographystyle{mnras}
\bibliography{GRB160410A} 

\begin{thebibliography}{}
\makeatletter
\relax
\def\mn@urlcharsother{\let\do\@makeother \do\$\do\&\do\#\do\^\do\_\do\%\do\~}
\def\mn@doi{\begingroup\mn@urlcharsother \@ifnextchar [ {\mn@doi@}
  {\mn@doi@[]}}
\def\mn@doi@[#1]#2{\def\@tempa{#1}\ifx\@tempa\@empty \href
  {http://dx.doi.org/#2} {doi:#2}\else \href {http://dx.doi.org/#2} {#1}\fi
  \endgroup}
\def\mn@eprint#1#2{\mn@eprint@#1:#2::\@nil}
\def\mn@eprint@arXiv#1{\href {http://arxiv.org/abs/#1} {{\tt arXiv:#1}}}
\def\mn@eprint@dblp#1{\href {http://dblp.uni-trier.de/rec/bibtex/#1.xml}
  {dblp:#1}}
\def\mn@eprint@#1:#2:#3:#4\@nil{\def\@tempa {#1}\def\@tempb {#2}\def\@tempc
  {#3}\ifx \@tempc \@empty \let \@tempc \@tempb \let \@tempb \@tempa \fi \ifx
  \@tempb \@empty \def\@tempb {arXiv}\fi \@ifundefined
  {mn@eprint@\@tempb}{\@tempb:\@tempc}{\expandafter \expandafter \csname
  mn@eprint@\@tempb\endcsname \expandafter{\@tempc}}}

\bibitem[\protect\citeauthoryear{{Abbott} et~al.,}{{Abbott}
  et~al.}{2017a}]{2017ApJ...848L..12A}
{Abbott} B.~P.,  et~al., 2017a, \mn@doi [\apjl] {10.3847/2041-8213/aa91c9},
  \href {https://ui.adsabs.harvard.edu/abs/2017ApJ...848L..12A} {848, L12}

\bibitem[\protect\citeauthoryear{{Abbott} et~al.,}{{Abbott}
  et~al.}{2017b}]{2017ApJ...848L..13A}
{Abbott} B.~P.,  et~al., 2017b, \mn@doi [\apjl] {10.3847/2041-8213/aa920c},
  \href {https://ui.adsabs.harvard.edu/abs/2017ApJ...848L..13A} {848, L13}

\bibitem[\protect\citeauthoryear{{Acciari} et~al.,}{{Acciari}
  et~al.}{2021}]{MAGIC2021ApJ}
{Acciari} V.~A.,  et~al., 2021, \mn@doi [\apj] {10.3847/1538-4357/abd249},
  \href {https://ui.adsabs.harvard.edu/abs/2021ApJ...908...90A} {908, 90}

\bibitem[\protect\citeauthoryear{{Ackermann} et~al.,}{{Ackermann}
  et~al.}{2010}]{Ackermann2010ApJ}
{Ackermann} M.,  et~al., 2010, \mn@doi [\apj] {10.1088/0004-637X/716/2/1178},
  \href {https://ui.adsabs.harvard.edu/abs/2010ApJ...716.1178A} {716, 1178}

\bibitem[\protect\citeauthoryear{{Ahumada} et~al.,}{{Ahumada}
  et~al.}{2021}]{Ahumada2021NatAst}
{Ahumada} T.,  et~al., 2021, \mn@doi [Nature Astronomy]
  {10.1038/s41550-021-01428-7}, \href
  {https://ui.adsabs.harvard.edu/abs/2021NatAs...5..917A} {5, 917}

\bibitem[\protect\citeauthoryear{{Alam} et~al.,}{{Alam}
  et~al.}{2015}]{2015ApJS..219...12A}
{Alam} S.,  et~al., 2015, \mn@doi [\apjs] {10.1088/0067-0049/219/1/12}, \href
  {https://ui.adsabs.harvard.edu/abs/2015ApJS..219...12A} {219, 12}

\bibitem[\protect\citeauthoryear{{Amati}}{{Amati}}{2006}]{Amati2006MNRAS}
{Amati} L.,  2006, \mn@doi [\mnras] {10.1111/j.1365-2966.2006.10840.x}, \href
  {https://ui.adsabs.harvard.edu/abs/2006MNRAS.372..233A} {372, 233}

\bibitem[\protect\citeauthoryear{{Amati} et~al.,}{{Amati}
  et~al.}{2002}]{Amati2002AA}
{Amati} L.,  et~al., 2002, \mn@doi [\aap] {10.1051/0004-6361:20020722}, \href
  {https://ui.adsabs.harvard.edu/abs/2002A&A...390...81A} {390, 81}

\bibitem[\protect\citeauthoryear{{Anderson} et~al.,}{{Anderson}
  et~al.}{2021}]{Anderson2021MNRAS}
{Anderson} G.~E.,  et~al., 2021, \mn@doi [\mnras] {10.1093/mnras/stab727},
  \href {https://ui.adsabs.harvard.edu/abs/2021MNRAS.503.4372A} {503, 4372}

\bibitem[\protect\citeauthoryear{{Andreoni} et~al.,}{{Andreoni}
  et~al.}{2021}]{Andreoni2021arXiv}
{Andreoni} I.,  et~al., 2021, \mn@doi [\apj] {10.3847/1538-4357/ac0bc7}, \href
  {https://ui.adsabs.harvard.edu/abs/2021ApJ...918...63A} {918, 63}

\bibitem[\protect\citeauthoryear{{Antier} et~al.,}{{Antier}
  et~al.}{2020}]{2020MNRAS.497.5518A}
{Antier} S.,  et~al., 2020, \mn@doi [\mnras] {10.1093/mnras/staa1846}, \href
  {https://ui.adsabs.harvard.edu/abs/2020MNRAS.497.5518A} {497, 5518}

\bibitem[\protect\citeauthoryear{{Aptekar} et~al.,}{{Aptekar}
  et~al.}{1995}]{1995SSRv...71..265A}
{Aptekar} R.~L.,  et~al., 1995, \mn@doi [\ssr] {10.1007/BF00751332}, \href
  {https://ui.adsabs.harvard.edu/abs/1995SSRv...71..265A} {71, 265}

\bibitem[\protect\citeauthoryear{{Astropy Collaboration} et~al.,}{{Astropy
  Collaboration} et~al.}{2013}]{2013A&A...558A..33A}
{Astropy Collaboration} et~al., 2013, \mn@doi [\aap]
  {10.1051/0004-6361/201322068}, \href
  {https://ui.adsabs.harvard.edu/abs/2013A&A...558A..33A} {558, A33}

\bibitem[\protect\citeauthoryear{{Astropy Collaboration} et~al.,}{{Astropy
  Collaboration} et~al.}{2018}]{astropy:2018}
{Astropy Collaboration} et~al., 2018, \mn@doi [\aj] {10.3847/1538-3881/aabc4f},
  \href {https://ui.adsabs.harvard.edu/abs/2018AJ....156..123A} {156, 123}

\bibitem[\protect\citeauthoryear{{Barthelmy} et~al.,}{{Barthelmy}
  et~al.}{2005}]{Barthelmy2005SSRv}
{Barthelmy} S.~D.,  et~al., 2005, \mn@doi [\ssr] {10.1007/s11214-005-5096-3},
  \href {http://adsabs.harvard.edu/abs/2005SSRv..120..143B} {120, 143}

\bibitem[\protect\citeauthoryear{{Becerra} et~al.,}{{Becerra}
  et~al.}{2019}]{Becerra2019ApJ}
{Becerra} R.~L.,  et~al., 2019, \mn@doi [\apj] {10.3847/1538-4357/ab275b},
  \href {https://ui.adsabs.harvard.edu/abs/2019ApJ...881...12B} {881, 12}

\bibitem[\protect\citeauthoryear{{Belczynski}, {Perna}, {Bulik}, {Kalogera},
  {Ivanova}  \& {Lamb}}{{Belczynski} et~al.}{2006}]{2006ApJ...648.1110B}
{Belczynski} K.,  {Perna} R.,  {Bulik} T.,  {Kalogera} V.,  {Ivanova} N.,
  {Lamb} D.~Q.,  2006, \mn@doi [\apj] {10.1086/505169}, \href
  {https://ui.adsabs.harvard.edu/abs/2006ApJ...648.1110B} {648, 1110}

\bibitem[\protect\citeauthoryear{{Bellm} et~al.,}{{Bellm}
  et~al.}{2019}]{2019PASP..131a8002B}
{Bellm} E.~C.,  et~al., 2019, \mn@doi [\pasp] {10.1088/1538-3873/aaecbe}, \href
  {https://ui.adsabs.harvard.edu/abs/2019PASP..131a8002B} {131, 018002}

\bibitem[\protect\citeauthoryear{{Beniamini} \& {Piran}}{{Beniamini} \&
  {Piran}}{2019}]{2019MNRAS.487.4847B}
{Beniamini} P.,  {Piran} T.,  2019, \mn@doi [\mnras] {10.1093/mnras/stz1589},
  \href {https://ui.adsabs.harvard.edu/abs/2019MNRAS.487.4847B} {487, 4847}

\bibitem[\protect\citeauthoryear{{Beniamini}, {Nava}, {Duran}  \&
  {Piran}}{{Beniamini} et~al.}{2015}]{2015MNRAS.454.1073B}
{Beniamini} P.,  {Nava} L.,  {Duran} R.~B.,   {Piran} T.,  2015, \mn@doi
  [\mnras] {10.1093/mnras/stv2033}, \href
  {https://ui.adsabs.harvard.edu/abs/2015MNRAS.454.1073B} {454, 1073}

\bibitem[\protect\citeauthoryear{{Beniamini}, {Hotokezaka}  \&
  {Piran}}{{Beniamini} et~al.}{2016a}]{2016ApJ...829L..13B}
{Beniamini} P.,  {Hotokezaka} K.,   {Piran} T.,  2016a, \mn@doi [\apjl]
  {10.3847/2041-8205/829/1/L13}, \href
  {https://ui.adsabs.harvard.edu/abs/2016ApJ...829L..13B} {829, L13}

\bibitem[\protect\citeauthoryear{{Beniamini}, {Hotokezaka}  \&
  {Piran}}{{Beniamini} et~al.}{2016b}]{2016ApJ...832..149B}
{Beniamini} P.,  {Hotokezaka} K.,   {Piran} T.,  2016b, \mn@doi [\apj]
  {10.3847/0004-637X/832/2/149}, \href
  {https://ui.adsabs.harvard.edu/abs/2016ApJ...832..149B} {832, 149}

\bibitem[\protect\citeauthoryear{{Berger}}{{Berger}}{2009}]{2009ApJ...690..231B}
{Berger} E.,  2009, \mn@doi [\apj] {10.1088/0004-637X/690/1/231}, \href
  {https://ui.adsabs.harvard.edu/abs/2009ApJ...690..231B} {690, 231}

\bibitem[\protect\citeauthoryear{{Berger}}{{Berger}}{2010}]{2010ApJ...722.1946B}
{Berger} E.,  2010, \mn@doi [\apj] {10.1088/0004-637X/722/2/1946}, \href
  {https://ui.adsabs.harvard.edu/abs/2010ApJ...722.1946B} {722, 1946}

\bibitem[\protect\citeauthoryear{{Berger}}{{Berger}}{2014}]{2014ARA&A..52...43B}
{Berger} E.,  2014, \mn@doi [\araa] {10.1146/annurev-astro-081913-035926},
  \href {https://ui.adsabs.harvard.edu/abs/2014ARA&A..52...43B} {52, 43}

\bibitem[\protect\citeauthoryear{{Berger}, {Fong}  \& {Chornock}}{{Berger}
  et~al.}{2013}]{Berger2013ApJ}
{Berger} E.,  {Fong} W.,   {Chornock} R.,  2013, \mn@doi [\apjl]
  {10.1088/2041-8205/774/2/L23}, \href
  {https://ui.adsabs.harvard.edu/abs/2013ApJ...774L..23B} {774, L23}

\bibitem[\protect\citeauthoryear{{Bla{\v{z}}ek}, {de Ugarte Postigo}, {Kann},
  {Th{\"o}ne}, {Ag{\"u}{\'\i} Fern{\'a}ndez}  \& {Izzo}}{{Bla{\v{z}}ek}
  et~al.}{2020}]{2020SPIE11452E..18B}
{Bla{\v{z}}ek} M.,  {de Ugarte Postigo} A.,  {Kann} D.~A.,  {Th{\"o}ne} C.~C.,
  {Ag{\"u}{\'\i} Fern{\'a}ndez} J.~F.,   {Izzo} L.,  2020, in Society of
  Photo-Optical Instrumentation Engineers (SPIE) Conference Series. p. 1145218,
  \mn@doi{10.1117/12.2562420}

\bibitem[\protect\citeauthoryear{{Bolmer} et~al.,}{{Bolmer}
  et~al.}{2019}]{2019A&A...623A..43B}
{Bolmer} J.,  et~al., 2019, \mn@doi [\aap] {10.1051/0004-6361/201834422}, \href
  {https://ui.adsabs.harvard.edu/abs/2019A&A...623A..43B} {623, A43}

\bibitem[\protect\citeauthoryear{{Boquien}, {Burgarella}, {Roehlly}, {Buat},
  {Ciesla}, {Corre}, {Inoue}  \& {Salas}}{{Boquien}
  et~al.}{2019}]{2019A&A...622A.103B}
{Boquien} M.,  {Burgarella} D.,  {Roehlly} Y.,  {Buat} V.,  {Ciesla} L.,
  {Corre} D.,  {Inoue} A.~K.,   {Salas} H.,  2019, \mn@doi [\aap]
  {10.1051/0004-6361/201834156}, \href
  {https://ui.adsabs.harvard.edu/abs/2019A&A...622A.103B} {622, A103}

\bibitem[\protect\citeauthoryear{Bradley et~al.,}{Bradley
  et~al.}{2020}]{larry_bradley_2020_4044744}
Bradley L.,  et~al., 2020, astropy/photutils: 1.0.0,
  \mn@doi{10.5281/zenodo.4044744}, \url
  {https://doi.org/10.5281/zenodo.4044744}

\bibitem[\protect\citeauthoryear{{Breeveld} \& {Siegel}}{{Breeveld} \&
  {Siegel}}{2016}]{Breeveld2016GCN19839}
{Breeveld} A.~A.,  {Siegel} M.~H.,  2016, GRB Coordinates Network, \href
  {https://ui.adsabs.harvard.edu/abs/2016GCN.19839....1B} {19839}

\bibitem[\protect\citeauthoryear{{Breeveld}, {Landsman}, {Holland}, {Roming},
  {Kuin}  \& {Page}}{{Breeveld} et~al.}{2011}]{bre11}
{Breeveld} A.~A.,  {Landsman} W.,  {Holland} S.~T.,  {Roming} P.,  {Kuin}
  N.~P.~M.,   {Page} M.~J.,  2011, in {J.~E.~McEnery, J.~L.~Racusin, \&
  N.~Gehrels} ed.,  American Institute of Physics Conference Series Vol. 1358,
  American Institute of Physics Conference Series. pp 373--376 (\mn@eprint
  {arXiv} {1102.4717}), \mn@doi{10.1063/1.3621807}

\bibitem[\protect\citeauthoryear{{Bruzual} \& {Charlot}}{{Bruzual} \&
  {Charlot}}{2003}]{2003MNRAS.344.1000B}
{Bruzual} G.,  {Charlot} S.,  2003, \mn@doi [\mnras]
  {10.1046/j.1365-8711.2003.06897.x}, \href
  {https://ui.adsabs.harvard.edu/abs/2003MNRAS.344.1000B} {344, 1000}

\bibitem[\protect\citeauthoryear{{Burgarella}, {Buat}  \&
  {Iglesias-P{\'a}ramo}}{{Burgarella} et~al.}{2005}]{2005MNRAS.360.1413B}
{Burgarella} D.,  {Buat} V.,   {Iglesias-P{\'a}ramo} J.,  2005, \mn@doi
  [\mnras] {10.1111/j.1365-2966.2005.09131.x}, \href
  {https://ui.adsabs.harvard.edu/abs/2005MNRAS.360.1413B} {360, 1413}

\bibitem[\protect\citeauthoryear{{Burrows} et~al.,}{{Burrows}
  et~al.}{2005}]{Burrows2005SSRv}
{Burrows} D.~N.,  et~al., 2005, \mn@doi [\ssr] {10.1007/s11214-005-5097-2},
  \href {http://adsabs.harvard.edu/abs/2005SSRv..120..165B} {120, 165}

\bibitem[\protect\citeauthoryear{{Butler} et~al.,}{{Butler}
  et~al.}{2015}]{Butler2015GCN17762}
{Butler} N.,  et~al., 2015, GRB Coordinates Network, \href
  {https://ui.adsabs.harvard.edu/abs/2015GCN.17762....1B} {17762}

\bibitem[\protect\citeauthoryear{{Calzetti}, {Armus}, {Bohlin}, {Kinney},
  {Koornneef}  \& {Storchi-Bergmann}}{{Calzetti}
  et~al.}{2000}]{2000ApJ...533..682C}
{Calzetti} D.,  {Armus} L.,  {Bohlin} R.~C.,  {Kinney} A.~L.,  {Koornneef} J.,
   {Storchi-Bergmann} T.,  2000, \mn@doi [\apj] {10.1086/308692}, \href
  {https://ui.adsabs.harvard.edu/abs/2000ApJ...533..682C} {533, 682}

\bibitem[\protect\citeauthoryear{{Cano}, {Wang}, {Dai}  \& {Wu}}{{Cano}
  et~al.}{2017}]{2017AdAst2017E...5C}
{Cano} Z.,  {Wang} S.-Q.,  {Dai} Z.-G.,   {Wu} X.-F.,  2017, \mn@doi [Advances
  in Astronomy] {10.1155/2017/8929054}, \href
  {https://ui.adsabs.harvard.edu/abs/2017AdAst2017E...5C} {2017, 8929054}

\bibitem[\protect\citeauthoryear{{Cao}, {Kulkarni}, {Yan}, {Ravi}, {Vedantham}
  \& {Kasliwal}}{{Cao} et~al.}{2016}]{2016GCN.19278....1C}
{Cao} Y.,  {Kulkarni} S.~R.,  {Yan} L.,  {Ravi} V.,  {Vedantham} H.~K.,
  {Kasliwal} M.~M.,  2016, GRB Coordinates Network, \href
  {https://ui.adsabs.harvard.edu/abs/2016GCN.19278....1C} {19278}

\bibitem[\protect\citeauthoryear{{Cardelli}, {Clayton}  \& {Mathis}}{{Cardelli}
  et~al.}{1989}]{1989ApJ...345..245C}
{Cardelli} J.~A.,  {Clayton} G.~C.,   {Mathis} J.~S.,  1989, \mn@doi [\apj]
  {10.1086/167900}, \href
  {https://ui.adsabs.harvard.edu/abs/1989ApJ...345..245C} {345, 245}

\bibitem[\protect\citeauthoryear{{Chabrier}}{{Chabrier}}{2003}]{2003PASP..115..763C}
{Chabrier} G.,  2003, \mn@doi [\pasp] {10.1086/376392}, \href
  {https://ui.adsabs.harvard.edu/abs/2003PASP..115..763C} {115, 763}

\bibitem[\protect\citeauthoryear{{Choi}, {Kim}, {Park}, {Shin}  \& {Im}}{{Choi}
  et~al.}{2018}]{Choi2018GCN22668}
{Choi} C.,  {Kim} Y.,  {Park} W.,  {Shin} S.,   {Im} M.,  2018, GRB Coordinates
  Network, \href {https://ui.adsabs.harvard.edu/abs/2018GCN.22668....1C}
  {22668}

\bibitem[\protect\citeauthoryear{{Christensen}, {Fynbo}, {Prochaska},
  {Th{\"o}ne}, {de Ugarte Postigo}  \& {Jakobsson}}{{Christensen}
  et~al.}{2011}]{christensen2011}
{Christensen} L.,  {Fynbo} J.~P.~U.,  {Prochaska} J.~X.,  {Th{\"o}ne} C.~C.,
  {de Ugarte Postigo} A.,   {Jakobsson} P.,  2011, \mn@doi [\apj]
  {10.1088/0004-637X/727/2/73}, \href
  {https://ui.adsabs.harvard.edu/abs/2011ApJ...727...73C} {727, 73}

\bibitem[\protect\citeauthoryear{{Christensen}, {M{\o}ller}, {Fynbo}  \&
  {Zafar}}{{Christensen} et~al.}{2014}]{2014MNRAS.445..225C}
{Christensen} L.,  {M{\o}ller} P.,  {Fynbo} J.~P.~U.,   {Zafar} T.,  2014,
  \mn@doi [\mnras] {10.1093/mnras/stu1726}, \href
  {https://ui.adsabs.harvard.edu/abs/2014MNRAS.445..225C} {445, 225}

\bibitem[\protect\citeauthoryear{{Cobb}}{{Cobb}}{2016}]{2016GCN.19311....1C}
{Cobb} B.~E.,  2016, GRB Coordinates Network, \href
  {https://ui.adsabs.harvard.edu/abs/2016GCN.19311....1C} {19311}

\bibitem[\protect\citeauthoryear{{Cucchiara} et~al.,}{{Cucchiara}
  et~al.}{2013}]{Cucchiara2013ApJ}
{Cucchiara} A.,  et~al., 2013, \mn@doi [\apj] {10.1088/0004-637X/777/2/94},
  \href {https://ui.adsabs.harvard.edu/abs/2013ApJ...777...94C} {777, 94}

\bibitem[\protect\citeauthoryear{{D'Elia} et~al.,}{{D'Elia}
  et~al.}{2009a}]{2009A&A...503..437D}
{D'Elia} V.,  et~al., 2009a, \mn@doi [\aap] {10.1051/0004-6361/200911674},
  \href {https://ui.adsabs.harvard.edu/abs/2009A&A...503..437D} {503, 437}

\bibitem[\protect\citeauthoryear{{D'Elia} et~al.,}{{D'Elia}
  et~al.}{2009b}]{2009ApJ...694..332D}
{D'Elia} V.,  et~al., 2009b, \mn@doi [\apj] {10.1088/0004-637X/694/1/332},
  \href {https://ui.adsabs.harvard.edu/abs/2009ApJ...694..332D} {694, 332}

\bibitem[\protect\citeauthoryear{{Dale} \& {Helou}}{{Dale} \&
  {Helou}}{2002}]{2002ApJ...576..159D}
{Dale} D.~A.,  {Helou} G.,  2002, \mn@doi [\apj] {10.1086/341632}, \href
  {https://ui.adsabs.harvard.edu/abs/2002ApJ...576..159D} {576, 159}

\bibitem[\protect\citeauthoryear{{Dale}, {Helou}, {Magdis}, {Armus},
  {D{\'\i}az-Santos}  \& {Shi}}{{Dale} et~al.}{2014}]{2014ApJ...784...83D}
{Dale} D.~A.,  {Helou} G.,  {Magdis} G.~E.,  {Armus} L.,  {D{\'\i}az-Santos}
  T.,   {Shi} Y.,  2014, \mn@doi [\apj] {10.1088/0004-637X/784/1/83}, \href
  {https://ui.adsabs.harvard.edu/abs/2014ApJ...784...83D} {784, 83}

\bibitem[\protect\citeauthoryear{{De Cia}, {Ledoux}, {Savaglio}, {Schady}  \&
  {Vreeswijk}}{{De Cia} et~al.}{2013}]{2013A&A...560A..88D}
{De Cia} A.,  {Ledoux} C.,  {Savaglio} S.,  {Schady} P.,   {Vreeswijk} P.~M.,
  2013, \mn@doi [\aap] {10.1051/0004-6361/201321834}, \href
  {https://ui.adsabs.harvard.edu/abs/2013A&A...560A..88D} {560, A88}

\bibitem[\protect\citeauthoryear{{De Cia}, {Ledoux}, {Mattsson}, {Petitjean},
  {Srianand}, {Gavignaud}  \& {Jenkins}}{{De Cia}
  et~al.}{2016}]{2016A&A...596A..97D}
{De Cia} A.,  {Ledoux} C.,  {Mattsson} L.,  {Petitjean} P.,  {Srianand} R.,
  {Gavignaud} I.,   {Jenkins} E.~B.,  2016, \mn@doi [\aap]
  {10.1051/0004-6361/201527895}, \href
  {https://ui.adsabs.harvard.edu/abs/2016A&A...596A..97D} {596, A97}

\bibitem[\protect\citeauthoryear{{De Cia}, {Ledoux}, {Petitjean}  \&
  {Savaglio}}{{De Cia} et~al.}{2018}]{2018A&A...611A..76D}
{De Cia} A.,  {Ledoux} C.,  {Petitjean} P.,   {Savaglio} S.,  2018, \mn@doi
  [\aap] {10.1051/0004-6361/201731970}, \href
  {https://ui.adsabs.harvard.edu/abs/2018A&A...611A..76D} {611, A76}

\bibitem[\protect\citeauthoryear{{Dichiara}, {Troja}, {Cenko}, {O'Connor},
  {Gatkine}, {Durbak}, {Kutyrev}  \& {Veilleux}}{{Dichiara}
  et~al.}{2020}]{2020GCN.29128....1D}
{Dichiara} S.,  {Troja} E.,  {Cenko} S.~B.,  {O'Connor} B.,  {Gatkine} P.,
  {Durbak} J.~M.,  {Kutyrev} A.,   {Veilleux} S.,  2020, GRB Coordinates
  Network, \href {https://ui.adsabs.harvard.edu/abs/2020GCN.29128....1D}
  {29128}

\bibitem[\protect\citeauthoryear{{Dichiara} et~al.,}{{Dichiara}
  et~al.}{2021}]{2021ApJ...911L..28D}
{Dichiara} S.,  et~al., 2021, \mn@doi [\apjl] {10.3847/2041-8213/abf562}, \href
  {https://ui.adsabs.harvard.edu/abs/2021ApJ...911L..28D} {911, L28}

\bibitem[\protect\citeauthoryear{{Dimple}, {Panchal}, {Gangopadhyay}, {Ghosh},
  {Gupta}, {Kumar}, {Misra}  \& {Pandey}}{{Dimple}
  et~al.}{2020}]{Dimple2020GCN29148}
{Dimple} A.,  {Panchal} A.,  {Gangopadhyay} A.,  {Ghosh} A.,  {Gupta} R.,
  {Kumar} A.,  {Misra} K.,   {Pandey} S.~B.,  2020, GRB Coordinates Network,
  \href {https://ui.adsabs.harvard.edu/abs/2020GCN.29148....1P} {29148}

\bibitem[\protect\citeauthoryear{{Fong} \& {Berger}}{{Fong} \&
  {Berger}}{2013}]{2013ApJ...776...18F}
{Fong} W.,  {Berger} E.,  2013, \mn@doi [\apj] {10.1088/0004-637X/776/1/18},
  \href {https://ui.adsabs.harvard.edu/abs/2013ApJ...776...18F} {776, 18}

\bibitem[\protect\citeauthoryear{{Fong} et~al.,}{{Fong}
  et~al.}{2014}]{Fong2014ApJ}
{Fong} W.,  et~al., 2014, \mn@doi [\apj] {10.1088/0004-637X/780/2/118}, \href
  {https://ui.adsabs.harvard.edu/abs/2014ApJ...780..118F} {780, 118}

\bibitem[\protect\citeauthoryear{{Fong} et~al.,}{{Fong}
  et~al.}{2022}]{2022ApJ...940...56F}
{Fong} W.-f.,  et~al., 2022, \mn@doi [\apj] {10.3847/1538-4357/ac91d0}, \href
  {https://ui.adsabs.harvard.edu/abs/2022ApJ...940...56F} {940, 56}

\bibitem[\protect\citeauthoryear{{Fontana} et~al.,}{{Fontana}
  et~al.}{2014}]{2014A&A...570A..11F}
{Fontana} A.,  et~al., 2014, \mn@doi [\aap] {10.1051/0004-6361/201423543},
  \href {https://ui.adsabs.harvard.edu/abs/2014A&A...570A..11F} {570, A11}

\bibitem[\protect\citeauthoryear{{Fox}, {Ledoux}, {Vreeswijk}, {Smette}  \&
  {Jaunsen}}{{Fox} et~al.}{2008}]{2008A&A...491..189F}
{Fox} A.~J.,  {Ledoux} C.,  {Vreeswijk} P.~M.,  {Smette} A.,   {Jaunsen} A.~O.,
   2008, \mn@doi [\aap] {10.1051/0004-6361:200810286}, \href
  {https://ui.adsabs.harvard.edu/abs/2008A&A...491..189F} {491, 189}

\bibitem[\protect\citeauthoryear{{Frederiks} et~al.,}{{Frederiks}
  et~al.}{2016}]{2016GCN.19288....1F}
{Frederiks} D.,  et~al., 2016, GRB Coordinates Network, \href
  {https://ui.adsabs.harvard.edu/abs/2016GCN.19288....1F} {19288}

\bibitem[\protect\citeauthoryear{{Frederiks} et~al.,}{{Frederiks}
  et~al.}{2020}]{2020GCN.29130....1F}
{Frederiks} D.,  et~al., 2020, GRB Coordinates Network, \href
  {https://ui.adsabs.harvard.edu/abs/2020GCN.29130....1F} {29130}

\bibitem[\protect\citeauthoryear{{Freudling}, {Romaniello}, {Bramich},
  {Ballester}, {Forchi}, {Garc{\'\i}a-Dabl{\'o}}, {Moehler}  \&
  {Neeser}}{{Freudling} et~al.}{2013}]{2013A&A...559A..96F}
{Freudling} W.,  {Romaniello} M.,  {Bramich} D.~M.,  {Ballester} P.,  {Forchi}
  V.,  {Garc{\'\i}a-Dabl{\'o}} C.~E.,  {Moehler} S.,   {Neeser} M.~J.,  2013,
  \mn@doi [\aap] {10.1051/0004-6361/201322494}, \href
  {https://ui.adsabs.harvard.edu/abs/2013A&A...559A..96F} {559, A96}

\bibitem[\protect\citeauthoryear{{Friis} et~al.,}{{Friis}
  et~al.}{2015}]{2015MNRAS.451..167F}
{Friis} M.,  et~al., 2015, \mn@doi [\mnras] {10.1093/mnras/stv960}, \href
  {https://ui.adsabs.harvard.edu/abs/2015MNRAS.451..167F} {451, 167}

\bibitem[\protect\citeauthoryear{{Fruchter} et~al.,}{{Fruchter}
  et~al.}{2006}]{2006Natur.441..463F}
{Fruchter} A.~S.,  et~al., 2006, \mn@doi [\nat] {10.1038/nature04787}, \href
  {https://ui.adsabs.harvard.edu/abs/2006Natur.441..463F} {441, 463}

\bibitem[\protect\citeauthoryear{{Fynbo}, {Prochaska}, {Sommer-Larsen},
  {Dessauges-Zavadsky}  \& {M{\o}ller}}{{Fynbo}
  et~al.}{2008}]{2008ApJ...683..321F}
{Fynbo} J. P.~U.,  {Prochaska} J.~X.,  {Sommer-Larsen} J.,
  {Dessauges-Zavadsky} M.,   {M{\o}ller} P.,  2008, \mn@doi [\apj]
  {10.1086/589555}, \href
  {https://ui.adsabs.harvard.edu/abs/2008ApJ...683..321F} {683, 321}

\bibitem[\protect\citeauthoryear{{Gaia Collaboration} et~al.,}{{Gaia
  Collaboration} et~al.}{2018}]{2018A&A...616A...1G}
{Gaia Collaboration} et~al., 2018, \mn@doi [\aap]
  {10.1051/0004-6361/201833051}, \href
  {https://ui.adsabs.harvard.edu/abs/2018A&A...616A...1G} {616, A1}

\bibitem[\protect\citeauthoryear{{Galama} et~al.,}{{Galama}
  et~al.}{1998}]{1998Natur.395..670G}
{Galama} T.~J.,  et~al., 1998, \mn@doi [\nat] {10.1038/27150}, \href
  {https://ui.adsabs.harvard.edu/abs/1998Natur.395..670G} {395, 670}

\bibitem[\protect\citeauthoryear{{Gatkine}, {Veilleux}  \&
  {Cucchiara}}{{Gatkine} et~al.}{2019}]{2019ApJ...884...66G}
{Gatkine} P.,  {Veilleux} S.,   {Cucchiara} A.,  2019, \mn@doi [\apj]
  {10.3847/1538-4357/ab31ae}, \href
  {https://ui.adsabs.harvard.edu/abs/2019ApJ...884...66G} {884, 66}

\bibitem[\protect\citeauthoryear{{Gehrels} et~al.,}{{Gehrels}
  et~al.}{2004}]{2004ApJ...611.1005G}
{Gehrels} N.,  et~al., 2004, \mn@doi [\apj] {10.1086/422091}, \href
  {https://ui.adsabs.harvard.edu/abs/2004ApJ...611.1005G} {611, 1005}

\bibitem[\protect\citeauthoryear{{Gehrels} et~al.,}{{Gehrels}
  et~al.}{2006}]{Gehrels2006Nature}
{Gehrels} N.,  et~al., 2006, \mn@doi [\nat] {10.1038/nature05376}, \href
  {http://adsabs.harvard.edu/abs/2006Natur.444.1044G} {444, 1044}

\bibitem[\protect\citeauthoryear{{Giallongo} et~al.,}{{Giallongo}
  et~al.}{2008}]{2008A&A...482..349G}
{Giallongo} E.,  et~al., 2008, \mn@doi [\aap] {10.1051/0004-6361:20078402},
  \href {https://ui.adsabs.harvard.edu/abs/2008A&A...482..349G} {482, 349}

\bibitem[\protect\citeauthoryear{{Gibson}, {Malesani}, {Page}, {Palmer}  \&
  {Siegel}}{{Gibson} et~al.}{2016}]{2016GCN.19271....1G}
{Gibson} S.~L.,  {Malesani} D.,  {Page} K.~L.,  {Palmer} D.~M.,   {Siegel}
  M.~H.,  2016, GRB Coordinates Network, \href
  {https://ui.adsabs.harvard.edu/abs/2016GCN.19271....1G} {19271}

\bibitem[\protect\citeauthoryear{{Goldstein} et~al.,}{{Goldstein}
  et~al.}{2017}]{2017ApJ...848L..14G}
{Goldstein} A.,  et~al., 2017, \mn@doi [\apjl] {10.3847/2041-8213/aa8f41},
  \href {https://ui.adsabs.harvard.edu/abs/2017ApJ...848L..14G} {848, L14}

\bibitem[\protect\citeauthoryear{{Gompertz} et~al.,}{{Gompertz}
  et~al.}{2022}]{2022NatAs.tmp..264G}
{Gompertz} B.~P.,  et~al., 2022, \mn@doi [Nature Astronomy]
  {10.1038/s41550-022-01819-4}, \href
  {https://ui.adsabs.harvard.edu/abs/2022NatAs.tmp..264G} {}

\bibitem[\protect\citeauthoryear{{Gonneau} et~al.,}{{Gonneau}
  et~al.}{2020}]{2020A&A...634A.133G}
{Gonneau} A.,  et~al., 2020, \mn@doi [\aap] {10.1051/0004-6361/201936825},
  \href {https://ui.adsabs.harvard.edu/abs/2020A&A...634A.133G} {634, A133}

\bibitem[\protect\citeauthoryear{{Graham} et~al.,}{{Graham}
  et~al.}{2019}]{2019PASP..131g8001G}
{Graham} M.~J.,  et~al., 2019, \mn@doi [\pasp] {10.1088/1538-3873/ab006c},
  \href {https://ui.adsabs.harvard.edu/abs/2019PASP..131g8001G} {131, 078001}

\bibitem[\protect\citeauthoryear{{Greiner}}{{Greiner}}{2019}]{2019PASP..131a5002G}
{Greiner} J.,  2019, \mn@doi [\pasp] {10.1088/1538-3873/aaec5d}, \href
  {https://ui.adsabs.harvard.edu/abs/2019PASP..131a5002G} {131, 015002}

\bibitem[\protect\citeauthoryear{{Greiner} et~al.,}{{Greiner}
  et~al.}{2008}]{2008PASP..120..405G}
{Greiner} J.,  et~al., 2008, \mn@doi [\pasp] {10.1086/587032}, \href
  {https://ui.adsabs.harvard.edu/abs/2008PASP..120..405G} {120, 405}

\bibitem[\protect\citeauthoryear{{Guidorzi}, {Martone}, {Kobayashi}, {Mundell},
  {Gomboc}  \& {Steele}}{{Guidorzi} et~al.}{2018}]{Guidorzi2018GCN22648}
{Guidorzi} C.,  {Martone} R.,  {Kobayashi} S.,  {Mundell} C.~G.,  {Gomboc} A.,
   {Steele} I.~A.,  2018, GRB Coordinates Network, \href
  {https://ui.adsabs.harvard.edu/abs/2018GCN.22648....1G} {22648}

\bibitem[\protect\citeauthoryear{{Hamburg}, {Malacaria}, {Meegan}  \& {Fermi
  GBM Team}}{{Hamburg} et~al.}{2020}]{2020GCN.29140....1H}
{Hamburg} R.,  {Malacaria} C.,  {Meegan} C.,   {Fermi GBM Team} 2020, GRB
  Coordinates Network, \href
  {https://ui.adsabs.harvard.edu/abs/2020GCN.29140....1H} {29140}

\bibitem[\protect\citeauthoryear{{Hamuy}, {Suntzeff}, {Heathcote}, {Walker},
  {Gigoux}  \& {Phillips}}{{Hamuy} et~al.}{1994}]{1994PASP..106..566H}
{Hamuy} M.,  {Suntzeff} N.~B.,  {Heathcote} S.~R.,  {Walker} A.~R.,  {Gigoux}
  P.,   {Phillips} M.~M.,  1994, \mn@doi [\pasp] {10.1086/133417}, \href
  {https://ui.adsabs.harvard.edu/abs/1994PASP..106..566H} {106, 566}

\bibitem[\protect\citeauthoryear{Harris et~al.,}{Harris
  et~al.}{2020}]{harris2020array}
Harris C.~R.,  et~al., 2020, \mn@doi [Nature] {10.1038/s41586-020-2649-2}, 585,
  357

\bibitem[\protect\citeauthoryear{{Heintz} et~al.,}{{Heintz}
  et~al.}{2018}]{2018MNRAS.479.3456H}
{Heintz} K.~E.,  et~al., 2018, \mn@doi [\mnras] {10.1093/mnras/sty1447}, \href
  {https://ui.adsabs.harvard.edu/abs/2018MNRAS.479.3456H} {479, 3456}

\bibitem[\protect\citeauthoryear{{Hjorth} \& {Bloom}}{{Hjorth} \&
  {Bloom}}{2012}]{2012grb..book..169H}
{Hjorth} J.,  {Bloom} J.~S.,  2012, {The Gamma-Ray Burst - Supernova
  Connection}.
pp 169--190

\bibitem[\protect\citeauthoryear{{Hjorth} et~al.,}{{Hjorth}
  et~al.}{2003}]{2003Natur.423..847H}
{Hjorth} J.,  et~al., 2003, \mn@doi [\nat] {10.1038/nature01750}, \href
  {https://ui.adsabs.harvard.edu/abs/2003Natur.423..847H} {423, 847}

\bibitem[\protect\citeauthoryear{{Horiuchi}, {Hanayama}, {Honma}, {Itoh},
  {Shiraishi}, {Murata}, {Tachibana}  \& {Kawai}}{{Horiuchi}
  et~al.}{2018}]{Horiuchi2018GCN22670}
{Horiuchi} T.,  {Hanayama} H.,  {Honma} M.,  {Itoh} R.,  {Shiraishi} K.,
  {Murata} K.,  {Tachibana} Y.,   {Kawai} N.,  2018, GRB Coordinates Network,
  \href {https://ui.adsabs.harvard.edu/abs/2018GCN.22670....1H} {22670}

\bibitem[\protect\citeauthoryear{{Horne}}{{Horne}}{1986}]{1986PASP...98..609H}
{Horne} K.,  1986, \mn@doi [\pasp] {10.1086/131801}, \href
  {https://ui.adsabs.harvard.edu/abs/1986PASP...98..609H} {98, 609}

\bibitem[\protect\citeauthoryear{Hunter}{Hunter}{2007}]{Hunter:2007}
Hunter J.~D.,  2007, \mn@doi [Computing in Science \& Engineering]
  {10.1109/MCSE.2007.55}, 9, 90

\bibitem[\protect\citeauthoryear{{Jakobsson} et~al.,}{{Jakobsson}
  et~al.}{2006}]{2006A&A...460L..13J}
{Jakobsson} P.,  et~al., 2006, \mn@doi [\aap] {10.1051/0004-6361:20066405},
  \href {https://ui.adsabs.harvard.edu/abs/2006A&A...460L..13J} {460, L13}

\bibitem[\protect\citeauthoryear{{Japelj} et~al.,}{{Japelj}
  et~al.}{2015}]{Japelj2015A&A}
{Japelj} J.,  et~al., 2015, \mn@doi [\aap] {10.1051/0004-6361/201525665}, \href
  {https://ui.adsabs.harvard.edu/abs/2015A&A...579A..74J} {579, A74}

\bibitem[\protect\citeauthoryear{{Jespersen}, {Severin}, {Steinhardt},
  {Vinther}, {Fynbo}, {Selsing}  \& {Watson}}{{Jespersen}
  et~al.}{2020}]{Jespersen_2020}
{Jespersen} C.~K.,  {Severin} J.~B.,  {Steinhardt} C.~L.,  {Vinther} J.,
  {Fynbo} J. P.~U.,  {Selsing} J.,   {Watson} D.,  2020, \mn@doi [\apjl]
  {10.3847/2041-8213/ab964d}, \href
  {https://ui.adsabs.harvard.edu/abs/2020ApJ...896L..20J} {896, L20}

\bibitem[\protect\citeauthoryear{{Ji}, {Frebel}, {Chiti}  \& {Simon}}{{Ji}
  et~al.}{2016a}]{2016Natur.531..610J}
{Ji} A.~P.,  {Frebel} A.,  {Chiti} A.,   {Simon} J.~D.,  2016a, \mn@doi [\nat]
  {10.1038/nature17425}, \href
  {https://ui.adsabs.harvard.edu/abs/2016Natur.531..610J} {531, 610}

\bibitem[\protect\citeauthoryear{{Ji}, {Frebel}, {Ezzeddine}  \& {Casey}}{{Ji}
  et~al.}{2016b}]{2016ApJ...832L...3J}
{Ji} A.~P.,  {Frebel} A.,  {Ezzeddine} R.,   {Casey} A.~R.,  2016b, \mn@doi
  [\apjl] {10.3847/2041-8205/832/1/L3}, \href
  {https://ui.adsabs.harvard.edu/abs/2016ApJ...832L...3J} {832, L3}

\bibitem[\protect\citeauthoryear{{Jin} et~al.,}{{Jin}
  et~al.}{2018}]{2018ApJ...857..128J}
{Jin} Z.-P.,  et~al., 2018, \mn@doi [\apj] {10.3847/1538-4357/aab76d}, \href
  {https://ui.adsabs.harvard.edu/abs/2018ApJ...857..128J} {857, 128}

\bibitem[\protect\citeauthoryear{{Juvan} et~al.,}{{Juvan}
  et~al.}{2016}]{2016GCN.19309....1J}
{Juvan} I.,  et~al., 2016, GRB Coordinates Network, \href
  {https://ui.adsabs.harvard.edu/abs/2016GCN.19309....1J} {19309}

\bibitem[\protect\citeauthoryear{{Kann}, {Klose}  \& {Zeh}}{{Kann}
  et~al.}{2006}]{Kann2006ApJ}
{Kann} D.~A.,  {Klose} S.,   {Zeh} A.,  2006, \mn@doi [\apj] {10.1086/500652},
  \href {http://adsabs.harvard.edu/abs/2006ApJ...641..993K} {641, 993}

\bibitem[\protect\citeauthoryear{{Kann} et~al.,}{{Kann}
  et~al.}{2011}]{Kann2011ApJ}
{Kann} D.~A.,  et~al., 2011, \mn@doi [\apj] {10.1088/0004-637X/734/2/96}, \href
  {https://ui.adsabs.harvard.edu/abs/2011ApJ...734...96K} {734, 96}

\bibitem[\protect\citeauthoryear{{Kann}, {Tanga}  \& {Greiner}}{{Kann}
  et~al.}{2015}]{Kann2015GCN17757}
{Kann} D.~A.,  {Tanga} M.,   {Greiner} J.,  2015, GRB Coordinates Network,
  \href {https://ui.adsabs.harvard.edu/abs/2015GCN.17757....1K} {17757}

\bibitem[\protect\citeauthoryear{{Kasliwal}, {Korobkin}, {Lau}, {Wollaeger}  \&
  {Fryer}}{{Kasliwal} et~al.}{2017}]{Kasliwal2017ApJ}
{Kasliwal} M.~M.,  {Korobkin} O.,  {Lau} R.~M.,  {Wollaeger} R.,   {Fryer}
  C.~L.,  2017, \mn@doi [\apjl] {10.3847/2041-8213/aa799d}, \href
  {https://ui.adsabs.harvard.edu/abs/2017ApJ...843L..34K} {843, L34}

\bibitem[\protect\citeauthoryear{{Kilpatrick}, {Malesani}  \&
  {Fong}}{{Kilpatrick} et~al.}{2020}]{2020GCN.29133....1K}
{Kilpatrick} C.~D.,  {Malesani} D.~B.,   {Fong} W.,  2020, GRB Coordinates
  Network, \href {https://ui.adsabs.harvard.edu/abs/2020GCN.29133....1K}
  {29133}

\bibitem[\protect\citeauthoryear{{Klose} et~al.,}{{Klose}
  et~al.}{2019}]{2019ApJ...887..206K}
{Klose} S.,  et~al., 2019, \mn@doi [\apj] {10.3847/1538-4357/ab528a}, \href
  {https://ui.adsabs.harvard.edu/abs/2019ApJ...887..206K} {887, 206}

\bibitem[\protect\citeauthoryear{{Klotz}, {Gendre}, {Stratta}, {Atteia},
  {Bo{\"e}r}, {Malacrino}, {Damerdji}  \& {Behrend}}{{Klotz}
  et~al.}{2006}]{Klotz2006AA}
{Klotz} A.,  {Gendre} B.,  {Stratta} G.,  {Atteia} J.~L.,  {Bo{\"e}r} M.,
  {Malacrino} F.,  {Damerdji} Y.,   {Behrend} R.,  2006, \mn@doi [\aap]
  {10.1051/0004-6361:20065158}, \href
  {https://ui.adsabs.harvard.edu/abs/2006A&A...451L..39K} {451, L39}

\bibitem[\protect\citeauthoryear{{Klotz}, {Turpin}, {Atteia}, {Boer}, {Laugier}
   \& {Gendre}}{{Klotz} et~al.}{2016}]{2016GCN.19287....1K}
{Klotz} A.,  {Turpin} D.,  {Atteia} J.~L.,  {Boer} M.,  {Laugier} R.,
  {Gendre} B.,  2016, GRB Coordinates Network, \href
  {https://ui.adsabs.harvard.edu/abs/2016GCN.19287....1K} {19287}

\bibitem[\protect\citeauthoryear{{Knust} et~al.,}{{Knust}
  et~al.}{2017}]{2017A&A...607A..84K}
{Knust} F.,  et~al., 2017, \mn@doi [\aap] {10.1051/0004-6361/201730578}, \href
  {https://ui.adsabs.harvard.edu/abs/2017A&A...607A..84K} {607, A84}

\bibitem[\protect\citeauthoryear{{Kouveliotou}, {Meegan}, {Fishman}, {Bhat},
  {Briggs}, {Koshut}, {Paciesas}  \& {Pendleton}}{{Kouveliotou}
  et~al.}{1993}]{1993ApJ...413L.101K}
{Kouveliotou} C.,  {Meegan} C.~A.,  {Fishman} G.~J.,  {Bhat} N.~P.,  {Briggs}
  M.~S.,  {Koshut} T.~M.,  {Paciesas} W.~S.,   {Pendleton} G.~N.,  1993,
  \mn@doi [\apjl] {10.1086/186969}, \href
  {https://ui.adsabs.harvard.edu/abs/1993ApJ...413L.101K} {413, L101}

\bibitem[\protect\citeauthoryear{{Krimm} et~al.,}{{Krimm}
  et~al.}{2020}]{2020GCN.29139....1K}
{Krimm} H.~A.,  et~al., 2020, GRB Coordinates Network, \href
  {https://ui.adsabs.harvard.edu/abs/2020GCN.29139....1K} {29139}

\bibitem[\protect\citeauthoryear{Krogager}{Krogager}{2018}]{krogager2018voigtfit}
Krogager J.-K.,  2018, VoigtFit: A Python package for Voigt profile fitting
  (\mn@eprint {arXiv} {1803.01187})

\bibitem[\protect\citeauthoryear{{Kr{\"u}hler} et~al.,}{{Kr{\"u}hler}
  et~al.}{2008}]{2008ApJ...685..376K}
{Kr{\"u}hler} T.,  et~al., 2008, \mn@doi [\apj] {10.1086/590240}, \href
  {https://ui.adsabs.harvard.edu/abs/2008ApJ...685..376K} {685, 376}

\bibitem[\protect\citeauthoryear{{Kr{\"u}hler} et~al.,}{{Kr{\"u}hler}
  et~al.}{2013}]{2013A&A...557A..18K}
{Kr{\"u}hler} T.,  et~al., 2013, \mn@doi [\aap] {10.1051/0004-6361/201321772},
  \href {https://ui.adsabs.harvard.edu/abs/2013A&A...557A..18K} {557, A18}

\bibitem[\protect\citeauthoryear{{Kuin} \& {Swift/UVOT Team}}{{Kuin} \&
  {Swift/UVOT Team}}{2019}]{Kuin2019GCN26538}
{Kuin} N.~P.~M.,  {Swift/UVOT Team} 2019, GRB Coordinates Network, \href
  {https://ui.adsabs.harvard.edu/abs/2019GCN.26538....1K} {26538}

\bibitem[\protect\citeauthoryear{{Lamb} et~al.,}{{Lamb}
  et~al.}{2019}]{Lamb2019ApJ}
{Lamb} G.~P.,  et~al., 2019, \mn@doi [\apj] {10.3847/1538-4357/ab38bb}, \href
  {https://ui.adsabs.harvard.edu/abs/2019ApJ...883...48L} {883, 48}

\bibitem[\protect\citeauthoryear{{Ledoux}, {Petitjean}, {Bergeron}, {Wampler}
  \& {Srianand}}{{Ledoux} et~al.}{1998}]{1998A&A...337...51L}
{Ledoux} C.,  {Petitjean} P.,  {Bergeron} J.,  {Wampler} E.~J.,   {Srianand}
  R.,  1998, \aap, \href
  {https://ui.adsabs.harvard.edu/abs/1998A&A...337...51L} {337, 51}

\bibitem[\protect\citeauthoryear{{Leibler} \& {Berger}}{{Leibler} \&
  {Berger}}{2010}]{2010ApJ...725.1202L}
{Leibler} C.~N.,  {Berger} E.,  2010, \mn@doi [\apj]
  {10.1088/0004-637X/725/1/1202}, \href
  {https://ui.adsabs.harvard.edu/abs/2010ApJ...725.1202L} {725, 1202}

\bibitem[\protect\citeauthoryear{{Li}, {Zhang}  \& {L{\"u}}}{{Li}
  et~al.}{2016}]{Li2016ApJS}
{Li} Y.,  {Zhang} B.,   {L{\"u}} H.-J.,  2016, \mn@doi [\apjs]
  {10.3847/0067-0049/227/1/7}, \href
  {https://ui.adsabs.harvard.edu/abs/2016ApJS..227....7L} {227, 7}

\bibitem[\protect\citeauthoryear{{Li}, {Zhang}  \& {Yuan}}{{Li}
  et~al.}{2020}]{LZY2020}
{Li} Y.,  {Zhang} B.,   {Yuan} Q.,  2020, \mn@doi [\apj]
  {10.3847/1538-4357/ab96b8}, \href
  {https://ui.adsabs.harvard.edu/abs/2020ApJ...897..154L} {897, 154}

\bibitem[\protect\citeauthoryear{{Lien} et~al.,}{{Lien}
  et~al.}{2016}]{2016ApJ...829....7L}
{Lien} A.,  et~al., 2016, \mn@doi [\apj] {10.3847/0004-637X/829/1/7}, \href
  {https://ui.adsabs.harvard.edu/abs/2016ApJ...829....7L} {829, 7}

\bibitem[\protect\citeauthoryear{{Lodders}, {Palme}  \& {Gail}}{{Lodders}
  et~al.}{2009}]{2009LanB...4B..712L}
{Lodders} K.,  {Palme} H.,   {Gail} H.~P.,  2009, \mn@doi [Landolt
  B\&ouml;rnstein] {10.1007/978-3-540-88055-4\_34}, \href
  {https://ui.adsabs.harvard.edu/abs/2009LanB...4B..712L} {4B, 712}

\bibitem[\protect\citeauthoryear{{L{\"u}}, {Liang}, {Zhang}  \&
  {Zhang}}{{L{\"u}} et~al.}{2010}]{Lu2010ApJ}
{L{\"u}} H.-J.,  {Liang} E.-W.,  {Zhang} B.-B.,   {Zhang} B.,  2010, \mn@doi
  [\apj] {10.1088/0004-637X/725/2/1965}, \href
  {https://ui.adsabs.harvard.edu/abs/2010ApJ...725.1965L} {725, 1965}

\bibitem[\protect\citeauthoryear{{L{\"u}}, {Zhang}, {Liang}, {Zhang}  \&
  {Sakamoto}}{{L{\"u}} et~al.}{2014}]{Lu2014MNRAS}
{L{\"u}} H.-J.,  {Zhang} B.,  {Liang} E.-W.,  {Zhang} B.-B.,   {Sakamoto} T.,
  2014, \mn@doi [\mnras] {10.1093/mnras/stu982}, \href
  {https://ui.adsabs.harvard.edu/abs/2014MNRAS.442.1922L} {442, 1922}

\bibitem[\protect\citeauthoryear{{Lyman} et~al.,}{{Lyman}
  et~al.}{2017}]{2017MNRAS.467.1795L}
{Lyman} J.~D.,  et~al., 2017, \mn@doi [\mnras] {10.1093/mnras/stx220}, \href
  {https://ui.adsabs.harvard.edu/abs/2017MNRAS.467.1795L} {467, 1795}

\bibitem[\protect\citeauthoryear{{Ma}, {Hopkins}, {Faucher-Gigu{\`e}re},
  {Zolman}, {Muratov}, {Kere{\v{s}}}  \& {Quataert}}{{Ma}
  et~al.}{2016}]{2016MNRAS.456.2140M}
{Ma} X.,  {Hopkins} P.~F.,  {Faucher-Gigu{\`e}re} C.-A.,  {Zolman} N.,
  {Muratov} A.~L.,  {Kere{\v{s}}} D.,   {Quataert} E.,  2016, \mn@doi [\mnras]
  {10.1093/mnras/stv2659}, \href
  {https://ui.adsabs.harvard.edu/abs/2016MNRAS.456.2140M} {456, 2140}

\bibitem[\protect\citeauthoryear{{Malesani} \& {Kirkpatrick}}{{Malesani} \&
  {Kirkpatrick}}{2016}]{2016GCN.19295....1M}
{Malesani} D.,  {Kirkpatrick} C.,  2016, GRB Coordinates Network, \href
  {https://ui.adsabs.harvard.edu/abs/2016GCN.19295....1M} {19295}

\bibitem[\protect\citeauthoryear{{Malesani}, {Xu}, {Watson}  \&
  {Blay}}{{Malesani} et~al.}{2015}]{Malesani2015GCN17756}
{Malesani} D.,  {Xu} D.,  {Watson} D.~J.,   {Blay} P.,  2015, GRB Coordinates
  Network, \href {https://ui.adsabs.harvard.edu/abs/2015GCN.17756....1M}
  {17756}

\bibitem[\protect\citeauthoryear{{Malesani}, {Xu}  \& {Kuutma}}{{Malesani}
  et~al.}{2016}]{2016GCN.19300....1M}
{Malesani} D.,  {Xu} D.,   {Kuutma} T.,  2016, GRB Coordinates Network, \href
  {https://ui.adsabs.harvard.edu/abs/2016GCN.19300....1M} {19300}

\bibitem[\protect\citeauthoryear{{Malesani}, {Heintz}, {Stone}  \&
  {Stone}}{{Malesani} et~al.}{2018}]{Malesani2018GCN22660}
{Malesani} D.,  {Heintz} K.~E.,  {Stone} M.,   {Stone} J.,  2018, GRB
  Coordinates Network, \href
  {https://ui.adsabs.harvard.edu/abs/2018GCN.22660....1M} {22660}

\bibitem[\protect\citeauthoryear{{Maller}, {Prochaska}, {Somerville}  \&
  {Primack}}{{Maller} et~al.}{2003}]{2003MNRAS.343..268M}
{Maller} A.~H.,  {Prochaska} J.~X.,  {Somerville} R.~S.,   {Primack} J.~R.,
  2003, \mn@doi [\mnras] {10.1046/j.1365-8711.2003.06660.x}, \href
  {https://ui.adsabs.harvard.edu/abs/2003MNRAS.343..268M} {343, 268}

\bibitem[\protect\citeauthoryear{{Mandhai}, {Lamb}, {Tanvir}, {Bray}, {Nixon},
  {Eyles-Ferris}, {Levan}  \& {Gompertz}}{{Mandhai}
  et~al.}{2022}]{2022MNRAS.514.2716M}
{Mandhai} S.,  {Lamb} G.~P.,  {Tanvir} N.~R.,  {Bray} J.,  {Nixon} C.~J.,
  {Eyles-Ferris} R.~A.~J.,  {Levan} A.~J.,   {Gompertz} B.~P.,  2022, \mn@doi
  [\mnras] {10.1093/mnras/stac1473}, \href
  {https://ui.adsabs.harvard.edu/abs/2022MNRAS.514.2716M} {514, 2716}

\bibitem[\protect\citeauthoryear{{Marshall} \& {Gibson}}{{Marshall} \&
  {Gibson}}{2016}]{2016GCN.19275....1M}
{Marshall} F.~E.,  {Gibson} S.~L.,  2016, GRB Coordinates Network, \href
  {https://ui.adsabs.harvard.edu/abs/2016GCN.19275....1M} {19275}

\bibitem[\protect\citeauthoryear{{Mazets} et~al.,}{{Mazets}
  et~al.}{1981}]{Mazets1981ApSS}
{Mazets} E.~P.,  et~al., 1981, \mn@doi [\apss] {10.1007/BF00649140}, \href
  {http://adsabs.harvard.edu/abs/1981Ap&SS..80....3M} {80, 3}

\bibitem[\protect\citeauthoryear{{Meegan} et~al.,}{{Meegan}
  et~al.}{2009}]{2009ApJ...702..791M}
{Meegan} C.,  et~al., 2009, \mn@doi [\apj] {10.1088/0004-637X/702/1/791}, \href
  {https://ui.adsabs.harvard.edu/abs/2009ApJ...702..791M} {702, 791}

\bibitem[\protect\citeauthoryear{{Metzger}}{{Metzger}}{2019}]{2019LRR....23....1M}
{Metzger} B.~D.,  2019, \mn@doi [Living Reviews in Relativity]
  {10.1007/s41114-019-0024-0}, \href
  {https://ui.adsabs.harvard.edu/abs/2019LRR....23....1M} {23, 1}

\bibitem[\protect\citeauthoryear{{Metzger} et~al.,}{{Metzger}
  et~al.}{2010}]{2010MNRAS.406.2650M}
{Metzger} B.~D.,  et~al., 2010, \mn@doi [\mnras]
  {10.1111/j.1365-2966.2010.16864.x}, \href
  {https://ui.adsabs.harvard.edu/abs/2010MNRAS.406.2650M} {406, 2650}

\bibitem[\protect\citeauthoryear{{Minaev} \& {Pozanenko}}{{Minaev} \&
  {Pozanenko}}{2020}]{2020MNRAS.492.1919M}
{Minaev} P.~Y.,  {Pozanenko} A.~S.,  2020, \mn@doi [\mnras]
  {10.1093/mnras/stz3611}, \href
  {https://ui.adsabs.harvard.edu/abs/2020MNRAS.492.1919M} {492, 1919}

\bibitem[\protect\citeauthoryear{{Minaev} \& {Pozanenko}}{{Minaev} \&
  {Pozanenko}}{2021}]{2021MNRAS.504..926M}
{Minaev} P.~Y.,  {Pozanenko} A.~S.,  2021, \mn@doi [\mnras]
  {10.1093/mnras/stab1031ERRATUM: 10.1093/mnras/stz3611}, \href
  {https://ui.adsabs.harvard.edu/abs/2021MNRAS.504..926M} {504, 926}

\bibitem[\protect\citeauthoryear{{Misra}, {Paswan}, {Singh}, {Pandey}, {Kumar}
  \& {Omar}}{{Misra} et~al.}{2018}]{Misra2018GCN22663}
{Misra} K.,  {Paswan} A.,  {Singh} M.,  {Pandey} S.~B.,  {Kumar} T.~S.,
  {Omar} A.,  2018, GRB Coordinates Network, \href
  {https://ui.adsabs.harvard.edu/abs/2018GCN.22663....1M} {22663}

\bibitem[\protect\citeauthoryear{{Modigliani} et~al.,}{{Modigliani}
  et~al.}{2010}]{2010SPIE.7737E..28M}
{Modigliani} A.,  et~al., 2010, in Observatory Operations: Strategies,
  Processes, and Systems III. p. 773728, \mn@doi{10.1117/12.857211}

\bibitem[\protect\citeauthoryear{{Muraki} et~al.,}{{Muraki}
  et~al.}{2016}]{2016GCN.19285....1M}
{Muraki} Y.,  et~al., 2016, GRB Coordinates Network, \href
  {https://ui.adsabs.harvard.edu/abs/2016GCN.19285....1M} {19285}

\bibitem[\protect\citeauthoryear{{Nicuesa Guelbenzu} et~al.,}{{Nicuesa
  Guelbenzu} et~al.}{2021}]{2021A&A...650A.117N}
{Nicuesa Guelbenzu} A.~M.,  et~al., 2021, \mn@doi [\aap]
  {10.1051/0004-6361/202039689}, \href
  {https://ui.adsabs.harvard.edu/abs/2021A&A...650A.117N} {650, A117}

\bibitem[\protect\citeauthoryear{{Noll}, {Burgarella}, {Giovannoli}, {Buat},
  {Marcillac}  \& {Mu{\~n}oz-Mateos}}{{Noll}
  et~al.}{2009}]{2009A&A...507.1793N}
{Noll} S.,  {Burgarella} D.,  {Giovannoli} E.,  {Buat} V.,  {Marcillac} D.,
  {Mu{\~n}oz-Mateos} J.~C.,  2009, \mn@doi [\aap]
  {10.1051/0004-6361/200912497}, \href
  {https://ui.adsabs.harvard.edu/abs/2009A&A...507.1793N} {507, 1793}

\bibitem[\protect\citeauthoryear{{Norris} \& {Bonnell}}{{Norris} \&
  {Bonnell}}{2006}]{NorrisBonnell2006ApJ}
{Norris} J.~P.,  {Bonnell} J.~T.,  2006, \mn@doi [\apj] {10.1086/502796}, \href
  {https://ui.adsabs.harvard.edu/abs/2006ApJ...643..266N} {643, 266}

\bibitem[\protect\citeauthoryear{{Nugent} et~al.,}{{Nugent}
  et~al.}{2020}]{2020ApJ...904...52N}
{Nugent} A.~E.,  et~al., 2020, \mn@doi [\apj] {10.3847/1538-4357/abc24a}, \href
  {https://ui.adsabs.harvard.edu/abs/2020ApJ...904...52N} {904, 52}

\bibitem[\protect\citeauthoryear{{O'Connor} et~al.,}{{O'Connor}
  et~al.}{2022}]{2022MNRAS.515.4890O}
{O'Connor} B.,  et~al., 2022, \mn@doi [\mnras] {10.1093/mnras/stac1982}, \href
  {https://ui.adsabs.harvard.edu/abs/2022MNRAS.515.4890O} {515, 4890}

\bibitem[\protect\citeauthoryear{{Oates} et~al.,}{{Oates}
  et~al.}{2009}]{oates09}
{Oates} S.~R.,  et~al., 2009, \mn@doi [\mnras]
  {10.1111/j.1365-2966.2009.14544.x}, \href
  {https://ui.adsabs.harvard.edu/abs/2009MNRAS.395..490O} {395, 490}

\bibitem[\protect\citeauthoryear{{Oke}}{{Oke}}{1990}]{1990AJ.....99.1621O}
{Oke} J.~B.,  1990, \mn@doi [\aj] {10.1086/115444}, \href
  {https://ui.adsabs.harvard.edu/abs/1990AJ.....99.1621O} {99, 1621}

\bibitem[\protect\citeauthoryear{{Page} et~al.,}{{Page}
  et~al.}{2020}]{2020GCN.29112....1P}
{Page} K.~L.,  et~al., 2020, GRB Coordinates Network, \href
  {https://ui.adsabs.harvard.edu/abs/2020GCN.29112....1P} {29112}

\bibitem[\protect\citeauthoryear{{Pandey} et~al.,}{{Pandey}
  et~al.}{2019}]{Pandey2019MNRAS}
{Pandey} S.~B.,  et~al., 2019, \mn@doi [\mnras] {10.1093/mnras/stz530}, \href
  {https://ui.adsabs.harvard.edu/abs/2019MNRAS.485.5294P} {485, 5294}

\bibitem[\protect\citeauthoryear{{Paterson} et~al.,}{{Paterson}
  et~al.}{2020}]{Paterson2020ApJ}
{Paterson} K.,  et~al., 2020, \mn@doi [\apjl] {10.3847/2041-8213/aba4b0}, \href
  {https://ui.adsabs.harvard.edu/abs/2020ApJ...898L..32P} {898, L32}

\bibitem[\protect\citeauthoryear{{Pei}}{{Pei}}{1992}]{Pei1992ApJ}
{Pei} Y.~C.,  1992, \mn@doi [\apj] {10.1086/171637}, \href
  {http://adsabs.harvard.edu/abs/1992ApJ...395..130P} {395, 130}

\bibitem[\protect\citeauthoryear{{Perley} et~al.,}{{Perley}
  et~al.}{2016a}]{2016ApJ...817....7P}
{Perley} D.~A.,  et~al., 2016a, \mn@doi [\apj] {10.3847/0004-637X/817/1/7},
  \href {https://ui.adsabs.harvard.edu/abs/2016ApJ...817....7P} {817, 7}

\bibitem[\protect\citeauthoryear{{Perley} et~al.,}{{Perley}
  et~al.}{2016b}]{2016ApJ...817....8P}
{Perley} D.~A.,  et~al., 2016b, \mn@doi [\apj] {10.3847/0004-637X/817/1/8},
  \href {https://ui.adsabs.harvard.edu/abs/2016ApJ...817....8P} {817, 8}

\bibitem[\protect\citeauthoryear{{Planck Collaboration} et~al.,}{{Planck
  Collaboration} et~al.}{2014}]{2014A&A...571A..16P}
{Planck Collaboration} et~al., 2014, \mn@doi [\aap]
  {10.1051/0004-6361/201321591}, \href
  {https://ui.adsabs.harvard.edu/abs/2014A&A...571A..16P} {571, A16}

\bibitem[\protect\citeauthoryear{{Poole} et~al.,}{{Poole} et~al.}{2008}]{poole}
{Poole} T.~S.,  et~al., 2008, \mn@doi [MNRAS]
  {10.1111/j.1365-2966.2007.12563.x}, \href
  {http://adsabs.harvard.edu/abs/2008MNRAS.383..627P} {383, 627}

\bibitem[\protect\citeauthoryear{{Prochaska}, {Chen}, {Dessauges-Zavadsky}  \&
  {Bloom}}{{Prochaska} et~al.}{2007}]{2007ApJ...666..267P}
{Prochaska} J.~X.,  {Chen} H.-W.,  {Dessauges-Zavadsky} M.,   {Bloom} J.~S.,
  2007, \mn@doi [\apj] {10.1086/520042}, \href
  {https://ui.adsabs.harvard.edu/abs/2007ApJ...666..267P} {666, 267}

\bibitem[\protect\citeauthoryear{{Rastinejad}, {Paterson}, {Kilpatrick}  \&
  {Fong}}{{Rastinejad} et~al.}{2020}]{Rastinejad2020GCN29142}
{Rastinejad} J.,  {Paterson} K.,  {Kilpatrick} C.~D.,   {Fong} W.,  2020, GRB
  Coordinates Network, \href
  {https://ui.adsabs.harvard.edu/abs/2020GCN.29142....1R} {29142}

\bibitem[\protect\citeauthoryear{{Rastinejad} et~al.,}{{Rastinejad}
  et~al.}{2021}]{Rastinejad2021ApJ}
{Rastinejad} J.~C.,  et~al., 2021, \mn@doi [\apj] {10.3847/1538-4357/ac04b4},
  \href {https://ui.adsabs.harvard.edu/abs/2021ApJ...916...89R} {916, 89}

\bibitem[\protect\citeauthoryear{{Rastinejad} et~al.,}{{Rastinejad}
  et~al.}{2022}]{2022Natur.612..223R}
{Rastinejad} J.~C.,  et~al., 2022, \mn@doi [\nat] {10.1038/s41586-022-05390-w},
  \href {https://ui.adsabs.harvard.edu/abs/2022Natur.612..223R} {612, 223}

\bibitem[\protect\citeauthoryear{{Roederer} et~al.,}{{Roederer}
  et~al.}{2016}]{2016AJ....151...82R}
{Roederer} I.~U.,  et~al., 2016, \mn@doi [\aj] {10.3847/0004-6256/151/3/82},
  \href {https://ui.adsabs.harvard.edu/abs/2016AJ....151...82R} {151, 82}

\bibitem[\protect\citeauthoryear{{Roming} et~al.,}{{Roming}
  et~al.}{2005}]{Roming2005SSRv}
{Roming} P. W.~A.,  et~al., 2005, \mn@doi [\ssr] {10.1007/s11214-005-5095-4},
  \href {https://ui.adsabs.harvard.edu/abs/2005SSRv..120...95R} {120, 95}

\bibitem[\protect\citeauthoryear{{Rossi} \& {CIBO Collaboration}}{{Rossi} \&
  {CIBO Collaboration}}{2021}]{Rossi2021GCN29311}
{Rossi} A.,  {CIBO Collaboration} 2021, GRB Coordinates Network, \href
  {https://ui.adsabs.harvard.edu/abs/2021GCN.29311....1R} {29311}

\bibitem[\protect\citeauthoryear{{Rossi} et~al.,}{{Rossi}
  et~al.}{2022}]{2022ApJ...932....1R}
{Rossi} A.,  et~al., 2022, \mn@doi [\apj] {10.3847/1538-4357/ac60a2}, \href
  {https://ui.adsabs.harvard.edu/abs/2022ApJ...932....1R} {932, 1}

\bibitem[\protect\citeauthoryear{{Rouco Escorial} et~al.,}{{Rouco Escorial}
  et~al.}{2021}]{RoucoEscorial2021ApJ}
{Rouco Escorial} A.,  et~al., 2021, \mn@doi [\apj] {10.3847/1538-4357/abee85},
  \href {https://ui.adsabs.harvard.edu/abs/2021ApJ...912...95R} {912, 95}

\bibitem[\protect\citeauthoryear{{Rowlinson} et~al.,}{{Rowlinson}
  et~al.}{2021}]{Rowlinson20201MNRAS}
{Rowlinson} A.,  et~al., 2021, \mn@doi [\mnras] {10.1093/mnras/stab2060}, \href
  {https://ui.adsabs.harvard.edu/abs/2021MNRAS.506.5268R} {506, 5268}

\bibitem[\protect\citeauthoryear{{Sakamoto} et~al.,}{{Sakamoto}
  et~al.}{2016}]{2016GCN.19276....1S}
{Sakamoto} T.,  et~al., 2016, GRB Coordinates Network, \href
  {https://ui.adsabs.harvard.edu/abs/2016GCN.19276....1S} {19276}

\bibitem[\protect\citeauthoryear{{Savchenko} et~al.,}{{Savchenko}
  et~al.}{2017}]{2017ApJ...848L..15S}
{Savchenko} V.,  et~al., 2017, \mn@doi [\apjl] {10.3847/2041-8213/aa8f94},
  \href {https://ui.adsabs.harvard.edu/abs/2017ApJ...848L..15S} {848, L15}

\bibitem[\protect\citeauthoryear{{Schady}}{{Schady}}{2018}]{Schady2018GCN22662}
{Schady} P.,  2018, GRB Coordinates Network, \href
  {https://ui.adsabs.harvard.edu/abs/2018GCN.22662....1S} {22662}

\bibitem[\protect\citeauthoryear{{Schady} \& {Chen}}{{Schady} \&
  {Chen}}{2018}]{Schady2018GCN22666}
{Schady} P.,  {Chen} T.~W.,  2018, GRB Coordinates Network, \href
  {https://ui.adsabs.harvard.edu/abs/2018GCN.22666....1S} {22666}

\bibitem[\protect\citeauthoryear{{Schlafly} \& {Finkbeiner}}{{Schlafly} \&
  {Finkbeiner}}{2011}]{Schlafly11}
{Schlafly} E.~F.,  {Finkbeiner} D.~P.,  2011, \mn@doi [\apj]
  {10.1088/0004-637X/737/2/103}, \href
  {https://ui.adsabs.harvard.edu/abs/2011ApJ...737..103S} {737, 103}

\bibitem[\protect\citeauthoryear{{Seifert} et~al.,}{{Seifert}
  et~al.}{2003}]{2003SPIE.4841..962S}
{Seifert} W.,  et~al., 2003, in {Iye} M.,  {Moorwood} A. F.~M.,  eds,  Society
  of Photo-Optical Instrumentation Engineers (SPIE) Conference Series Vol.
  4841, Instrument Design and Performance for Optical/Infrared Ground-based
  Telescopes. pp 962--973, \mn@doi{10.1117/12.459494}

\bibitem[\protect\citeauthoryear{{Selsing} et~al.,}{{Selsing}
  et~al.}{2016}]{2016GCN.19274....1S}
{Selsing} J.,  et~al., 2016, GRB Coordinates Network, \href
  {https://ui.adsabs.harvard.edu/abs/2016GCN.19274....1S} {19274}

\bibitem[\protect\citeauthoryear{{Selsing} et~al.,}{{Selsing}
  et~al.}{2019}]{2019A&A...623A..92S}
{Selsing} J.,  et~al., 2019, \mn@doi [\aap] {10.1051/0004-6361/201832835},
  \href {https://ui.adsabs.harvard.edu/abs/2019A&A...623A..92S} {623, A92}

\bibitem[\protect\citeauthoryear{{Shahmoradi} \& {Nemiroff}}{{Shahmoradi} \&
  {Nemiroff}}{2015}]{ShahmoradiNemiroff2015MNRAS}
{Shahmoradi} A.,  {Nemiroff} R.~J.,  2015, \mn@doi [\mnras]
  {10.1093/mnras/stv714}, \href
  {https://ui.adsabs.harvard.edu/abs/2015MNRAS.451..126S} {451, 126}

\bibitem[\protect\citeauthoryear{{Skrutskie} et~al.,}{{Skrutskie}
  et~al.}{2006}]{2006AJ....131.1163S}
{Skrutskie} M.~F.,  et~al., 2006, \mn@doi [\aj] {10.1086/498708}, \href
  {https://ui.adsabs.harvard.edu/abs/2006AJ....131.1163S} {131, 1163}

\bibitem[\protect\citeauthoryear{{Starling}, {Willingale}, {Tanvir}, {Scott},
  {Wiersema}, {O'Brien}, {Levan}  \& {Stewart}}{{Starling}
  et~al.}{2013}]{2013MNRAS.431.3159S}
{Starling} R.~L.~C.,  {Willingale} R.,  {Tanvir} N.~R.,  {Scott} A.~E.,
  {Wiersema} K.,  {O'Brien} P.~T.,  {Levan} A.~J.,   {Stewart} G.~C.,  2013,
  \mn@doi [\mnras] {10.1093/mnras/stt400}, \href
  {https://ui.adsabs.harvard.edu/abs/2013MNRAS.431.3159S} {431, 3159}

\bibitem[\protect\citeauthoryear{{Tanvir}, {Levan}, {Fruchter}, {Hjorth},
  {Hounsell}, {Wiersema}  \& {Tunnicliffe}}{{Tanvir}
  et~al.}{2013}]{Tanvir2013Nature}
{Tanvir} N.~R.,  {Levan} A.~J.,  {Fruchter} A.~S.,  {Hjorth} J.,  {Hounsell}
  R.~A.,  {Wiersema} K.,   {Tunnicliffe} R.~L.,  2013, \mn@doi [\nat]
  {10.1038/nature12505}, \href
  {https://ui.adsabs.harvard.edu/abs/2013Natur.500..547T} {500, 547}

\bibitem[\protect\citeauthoryear{{Tanvir} et~al.,}{{Tanvir}
  et~al.}{2017}]{2017ApJ...848L..27T}
{Tanvir} N.~R.,  et~al., 2017, \mn@doi [\apjl] {10.3847/2041-8213/aa90b6},
  \href {https://ui.adsabs.harvard.edu/abs/2017ApJ...848L..27T} {848, L27}

\bibitem[\protect\citeauthoryear{{Tanvir} et~al.,}{{Tanvir}
  et~al.}{2019}]{2019MNRAS.483.5380T}
{Tanvir} N.~R.,  et~al., 2019, \mn@doi [\mnras] {10.1093/mnras/sty3460}, \href
  {https://ui.adsabs.harvard.edu/abs/2019MNRAS.483.5380T} {483, 5380}

\bibitem[\protect\citeauthoryear{{Th{\"o}ne} et~al.,}{{Th{\"o}ne}
  et~al.}{2008}]{2008ApJ...676.1151T}
{Th{\"o}ne} C.~C.,  et~al., 2008, \mn@doi [\apj] {10.1086/528943}, \href
  {https://ui.adsabs.harvard.edu/abs/2008ApJ...676.1151T} {676, 1151}

\bibitem[\protect\citeauthoryear{{Th{\"o}ne} et~al.,}{{Th{\"o}ne}
  et~al.}{2013}]{2013MNRAS.428.3590T}
{Th{\"o}ne} C.~C.,  et~al., 2013, \mn@doi [\mnras] {10.1093/mnras/sts303},
  \href {https://ui.adsabs.harvard.edu/abs/2013MNRAS.428.3590T} {428, 3590}

\bibitem[\protect\citeauthoryear{{Tody}}{{Tody}}{1993}]{1993ASPC...52..173T}
{Tody} D.,  1993, in {Hanisch} R.~J.,  {Brissenden} R.~J.~V.,   {Barnes} J.,
  eds,  Astronomical Society of the Pacific Conference Series Vol. 52,
  Astronomical Data Analysis Software and Systems II. p.~173

\bibitem[\protect\citeauthoryear{{Troja} et~al.,}{{Troja}
  et~al.}{2018}]{Troja2018GCN22664}
{Troja} E.,  et~al., 2018, GRB Coordinates Network, \href
  {https://ui.adsabs.harvard.edu/abs/2018GCN.22664....1T} {22664}

\bibitem[\protect\citeauthoryear{{Troja} et~al.,}{{Troja}
  et~al.}{2019}]{Troja2019MNRAS}
{Troja} E.,  et~al., 2019, \mn@doi [\mnras] {10.1093/mnras/stz2255}, \href
  {https://ui.adsabs.harvard.edu/abs/2019MNRAS.489.2104T} {489, 2104}

\bibitem[\protect\citeauthoryear{{Troja} et~al.,}{{Troja}
  et~al.}{2022}]{2022Natur.612..228T}
{Troja} E.,  et~al., 2022, \mn@doi [\nat] {10.1038/s41586-022-05327-3}, \href
  {https://ui.adsabs.harvard.edu/abs/2022Natur.612..228T} {612, 228}

\bibitem[\protect\citeauthoryear{{Trotter} et~al.,}{{Trotter}
  et~al.}{2016}]{2016GCN.19277....1T}
{Trotter} A.,  et~al., 2016, GRB Coordinates Network, \href
  {https://ui.adsabs.harvard.edu/abs/2016GCN.19277....1T} {19277}

\bibitem[\protect\citeauthoryear{{Tsutsui}, {Yonetoku}, {Nakamura}, {Takahashi}
   \& {Morihara}}{{Tsutsui} et~al.}{2013}]{Tsutsui2013}
{Tsutsui} R.,  {Yonetoku} D.,  {Nakamura} T.,  {Takahashi} K.,   {Morihara} Y.,
   2013, \mn@doi [\mnras] {10.1093/mnras/stt262}, \href
  {https://ui.adsabs.harvard.edu/abs/2013MNRAS.431.1398T} {431, 1398}

\bibitem[\protect\citeauthoryear{{Tsvetkova} et~al.,}{{Tsvetkova}
  et~al.}{2017}]{Tsvetkova2017ApJ}
{Tsvetkova} A.,  et~al., 2017, \mn@doi [\apj] {10.3847/1538-4357/aa96af}, \href
  {https://ui.adsabs.harvard.edu/abs/2017ApJ...850..161T} {850, 161}

\bibitem[\protect\citeauthoryear{{Tsvetkova} et~al.,}{{Tsvetkova}
  et~al.}{2021}]{Tsvetkova2021ApJ}
{Tsvetkova} A.,  et~al., 2021, \mn@doi [\apj] {10.3847/1538-4357/abd569}, \href
  {https://ui.adsabs.harvard.edu/abs/2021ApJ...908...83T} {908, 83}

\bibitem[\protect\citeauthoryear{{Updike} et~al.,}{{Updike}
  et~al.}{2008}]{Updike2008ApJ}
{Updike} A.~C.,  et~al., 2008, \mn@doi [\apj] {10.1086/590236}, \href
  {https://ui.adsabs.harvard.edu/abs/2008ApJ...685..361U} {685, 361}

\bibitem[\protect\citeauthoryear{{Vargas}, {Geha}, {Kirby}  \&
  {Simon}}{{Vargas} et~al.}{2013}]{2013ApJ...767..134V}
{Vargas} L.~C.,  {Geha} M.,  {Kirby} E.~N.,   {Simon} J.~D.,  2013, \mn@doi
  [\apj] {10.1088/0004-637X/767/2/134}, \href
  {https://ui.adsabs.harvard.edu/abs/2013ApJ...767..134V} {767, 134}

\bibitem[\protect\citeauthoryear{{Vernet} et~al.,}{{Vernet}
  et~al.}{2010}]{2010SPIE.7735E..1IV}
{Vernet} J.,  et~al., 2010, in Ground-based and Airborne Instrumentation for
  Astronomy III. p. 77351I, \mn@doi{10.1117/12.857124}

\bibitem[\protect\citeauthoryear{{Vernet} et~al.,}{{Vernet}
  et~al.}{2011}]{2011A&A...536A.105V}
{Vernet} J.,  et~al., 2011, \mn@doi [\aap] {10.1051/0004-6361/201117752}, \href
  {https://ui.adsabs.harvard.edu/abs/2011A&A...536A.105V} {536, A105}

\bibitem[\protect\citeauthoryear{{Vreeswijk} et~al.,}{{Vreeswijk}
  et~al.}{2004}]{2004A&A...419..927V}
{Vreeswijk} P.~M.,  et~al., 2004, \mn@doi [\aap] {10.1051/0004-6361:20040086},
  \href {https://ui.adsabs.harvard.edu/abs/2004A&A...419..927V} {419, 927}

\bibitem[\protect\citeauthoryear{{Vreeswijk} et~al.,}{{Vreeswijk}
  et~al.}{2007}]{2007A&A...468...83V}
{Vreeswijk} P.~M.,  et~al., 2007, \mn@doi [\aap] {10.1051/0004-6361:20066780},
  \href {https://ui.adsabs.harvard.edu/abs/2007A&A...468...83V} {468, 83}

\bibitem[\protect\citeauthoryear{{Vreeswijk} et~al.,}{{Vreeswijk}
  et~al.}{2011}]{2011A&A...532C...3V}
{Vreeswijk} P.~M.,  et~al., 2011, \mn@doi [\aap] {10.1051/0004-6361/20066780e},
  \href {https://ui.adsabs.harvard.edu/abs/2011A&A...532C...3V} {532, C3}

\bibitem[\protect\citeauthoryear{{Wang}, {Mao}  \& {Bai}}{{Wang}
  et~al.}{2016}]{2016GCN.19280....1W}
{Wang} C.~J.,  {Mao} J.,   {Bai} J.~M.,  2016, GRB Coordinates Network, \href
  {https://ui.adsabs.harvard.edu/abs/2016GCN.19280....1W} {19280}

\bibitem[\protect\citeauthoryear{{Watson} et~al.,}{{Watson}
  et~al.}{2019}]{2019Natur.574..497W}
{Watson} D.,  et~al., 2019, \mn@doi [\nat] {10.1038/s41586-019-1676-3}, \href
  {https://ui.adsabs.harvard.edu/abs/2019Natur.574..497W} {574, 497}

\bibitem[\protect\citeauthoryear{{Wiersma}, {Schaye}, {Theuns}, {Dalla Vecchia}
   \& {Tornatore}}{{Wiersma} et~al.}{2009}]{2009MNRAS.399..574W}
{Wiersma} R. P.~C.,  {Schaye} J.,  {Theuns} T.,  {Dalla Vecchia} C.,
  {Tornatore} L.,  2009, \mn@doi [\mnras] {10.1111/j.1365-2966.2009.15331.x},
  \href {https://ui.adsabs.harvard.edu/abs/2009MNRAS.399..574W} {399, 574}

\bibitem[\protect\citeauthoryear{{Wolfe} \& {Prochaska}}{{Wolfe} \&
  {Prochaska}}{2000}]{2000ApJ...545..591W}
{Wolfe} A.~M.,  {Prochaska} J.~X.,  2000, \mn@doi [\apj] {10.1086/317860},
  \href {https://ui.adsabs.harvard.edu/abs/2000ApJ...545..591W} {545, 591}

\bibitem[\protect\citeauthoryear{{Wolfe}, {Gawiser}  \& {Prochaska}}{{Wolfe}
  et~al.}{2005}]{2005ARA&A..43..861W}
{Wolfe} A.~M.,  {Gawiser} E.,   {Prochaska} J.~X.,  2005, \mn@doi [\araa]
  {10.1146/annurev.astro.42.053102.133950}, \href
  {https://ui.adsabs.harvard.edu/abs/2005ARA&A..43..861W} {43, 861}

\bibitem[\protect\citeauthoryear{{Woosley} \& {Bloom}}{{Woosley} \&
  {Bloom}}{2006}]{2006ARA&A..44..507W}
{Woosley} S.~E.,  {Bloom} J.~S.,  2006, \mn@doi [\araa]
  {10.1146/annurev.astro.43.072103.150558}, \href
  {https://ui.adsabs.harvard.edu/abs/2006ARA&A..44..507W} {44, 507}

\bibitem[\protect\citeauthoryear{{Yang} et~al.,}{{Yang}
  et~al.}{2022}]{2022Natur.612..232Y}
{Yang} J.,  et~al., 2022, \mn@doi [\nat] {10.1038/s41586-022-05403-8}, \href
  {https://ui.adsabs.harvard.edu/abs/2022Natur.612..232Y} {612, 232}

\bibitem[\protect\citeauthoryear{{Yates}, {Kruehler}  \& {Greiner}}{{Yates}
  et~al.}{2016}]{2016GCN.19272....1Y}
{Yates} R.,  {Kruehler} T.,   {Greiner} J.,  2016, GRB Coordinates Network,
  \href {https://ui.adsabs.harvard.edu/abs/2016GCN.19272....1Y} {19272}

\bibitem[\protect\citeauthoryear{{Yolda{\textcommabelow s}}, {Kr{\"u}hler},
  {Greiner}, {Yolda{\textcommabelow s}}, {Clemens}, {Szokoly}, {Primak}  \&
  {Klose}}{{Yolda{\textcommabelow s}} et~al.}{2008}]{2008AIPC.1000..227Y}
{Yolda{\textcommabelow s}} A.~K.,  {Kr{\"u}hler} T.,  {Greiner} J.,
  {Yolda{\textcommabelow s}} A.,  {Clemens} C.,  {Szokoly} G.,  {Primak} N.,
  {Klose} S.,  2008, in {Galassi} M.,  {Palmer} D.,   {Fenimore} E.,  eds,
  American Institute of Physics Conference Series Vol. 1000, Gamma-ray Bursts
  2007. pp 227--231, \mn@doi{10.1063/1.2943450}

\bibitem[\protect\citeauthoryear{{Zhang}}{{Zhang}}{2006}]{Zhang2006Nature}
{Zhang} B.,  2006, \mn@doi [\nat] {10.1038/4441010a}, \href
  {https://ui.adsabs.harvard.edu/abs/2006Natur.444.1010Z} {444, 1010}

\bibitem[\protect\citeauthoryear{{Zhang} et~al.,}{{Zhang}
  et~al.}{2009}]{2009ApJ...703.1696Z}
{Zhang} B.,  et~al., 2009, \mn@doi [\apj] {10.1088/0004-637X/703/2/1696}, \href
  {https://ui.adsabs.harvard.edu/abs/2009ApJ...703.1696Z} {703, 1696}

\bibitem[\protect\citeauthoryear{{Zhang}, {Yang}, {Choi}  \& {Chang}}{{Zhang}
  et~al.}{2016}]{ZYCC2016MNRAS}
{Zhang} Z.-B.,  {Yang} E.-B.,  {Choi} C.-S.,   {Chang} H.-Y.,  2016, \mn@doi
  [\mnras] {10.1093/mnras/stw1835}, \href
  {https://ui.adsabs.harvard.edu/abs/2016MNRAS.462.3243Z} {462, 3243}

\bibitem[\protect\citeauthoryear{{Zhang}, {Zhang}, {Zhao}, {Luo}, {Jiang},
  {Wang}, {Han}  \& {Terheide}}{{Zhang} et~al.}{2018}]{2018PASP..130e4202Z}
{Zhang} Z.~B.,  {Zhang} C.~T.,  {Zhao} Y.~X.,  {Luo} J.~J.,  {Jiang} L.~Y.,
  {Wang} X.~L.,  {Han} X.~L.,   {Terheide} R.~K.,  2018, \mn@doi [\pasp]
  {10.1088/1538-3873/aaa6af}, \href
  {https://ui.adsabs.harvard.edu/abs/2018PASP..130e4202Z} {130, 054202}

\bibitem[\protect\citeauthoryear{{Zhang} et~al.,}{{Zhang}
  et~al.}{2021}]{Zhang2021NatAst}
{Zhang} B.~B.,  et~al., 2021, \mn@doi [Nature Astronomy]
  {10.1038/s41550-021-01395-z}, \href
  {https://ui.adsabs.harvard.edu/abs/2021NatAs...5..911Z} {5, 911}

\bibitem[\protect\citeauthoryear{{de Ugarte Postigo} et~al.,}{{de Ugarte
  Postigo} et~al.}{2012}]{2012A&A...548A..11D}
{de Ugarte Postigo} A.,  et~al., 2012, \mn@doi [\aap]
  {10.1051/0004-6361/201219894}, \href
  {https://ui.adsabs.harvard.edu/abs/2012A&A...548A..11D} {548, A11}

\bibitem[\protect\citeauthoryear{{de Ugarte Postigo} et~al.,}{{de Ugarte
  Postigo} et~al.}{2014a}]{deUgartePostigo2014AA}
{de Ugarte Postigo} A.,  et~al., 2014a, \mn@doi [\aap]
  {10.1051/0004-6361/201322985}, \href
  {https://ui.adsabs.harvard.edu/abs/2014A&A...563A..62D} {563, A62}

\bibitem[\protect\citeauthoryear{{de Ugarte Postigo}, {Blazek}, {Janout},
  {Sprimont}, {Th{\"o}ne}, {Gorosabel}  \& {S{\'a}nchez-Ram{\'\i}rez}}{{de
  Ugarte Postigo} et~al.}{2014b}]{2014SPIE.9152E..0BD}
{de Ugarte Postigo} A.,  {Blazek} M.,  {Janout} P.,  {Sprimont} P.,
  {Th{\"o}ne} C.~C.,  {Gorosabel} J.,   {S{\'a}nchez-Ram{\'\i}rez} R.,  2014b,
  in Software and Cyberinfrastructure for Astronomy III. p. 91520B,
  \mn@doi{10.1117/12.2055774}

\bibitem[\protect\citeauthoryear{{de Ugarte Postigo} et~al.,}{{de Ugarte
  Postigo} et~al.}{2018}]{2018A&A...620A.119D}
{de Ugarte Postigo} A.,  et~al., 2018, \mn@doi [\aap]
  {10.1051/0004-6361/201833094}, \href
  {https://ui.adsabs.harvard.edu/abs/2018A&A...620A.119D} {620, A119}

\bibitem[\protect\citeauthoryear{{de Ugarte Postigo}, {Kann}, {Izzo}, {Thoene},
  {Blazek}, {Agui Fernandez}  \& {Lombardi}}{{de Ugarte Postigo}
  et~al.}{2020}]{2020GCN.29132....1D}
{de Ugarte Postigo} A.,  {Kann} D.~A.,  {Izzo} L.,  {Thoene} C.~C.,  {Blazek}
  M.,  {Agui Fernandez} J.~F.,   {Lombardi} G.,  2020, GRB Coordinates Network,
  \href {https://ui.adsabs.harvard.edu/abs/2020GCN.29132....1D} {29132}

\bibitem[\protect\citeauthoryear{{van Dokkum}}{{van
  Dokkum}}{2001}]{2001PASP..113.1420V}
{van Dokkum} P.~G.,  2001, \mn@doi [\pasp] {10.1086/323894}, \href
  {https://ui.adsabs.harvard.edu/abs/2001PASP..113.1420V} {113, 1420}

\makeatother
\end{thebibliography}




\appendix

\section{Further analysis of the short/long nature of GRB~160410A}
\label{sec:short_long_appendix}

Further to the analysis of the Amati relation and the hardness ratio presented in Sect. \ref{sec:short_long}, we here detail an extended study of the short/long GRB nature of GRB\,160410A.

\cite{2009ApJ...703.1696Z} introduced the ``Type I'' and ``Type II'' classification scheme, independent of the classic $T_{90}$ division, which was originally motivated by the temporally long, peculiar GRB\,060614 \citep{Gehrels2006Nature,Zhang2006Nature}. Type II GRBs are those associated with the core-collapse of massive stars (so usually long and soft), whereas Type I GRBs are those that are not - usually short and hard GRBs, which are at least in part associated with the merger of compact objects, specifically inspiralling binary neutron stars \citep[e.g.][]{2017ApJ...848L..12A,2017ApJ...848L..13A}.

\begin{figure}
	\includegraphics[width=\columnwidth]{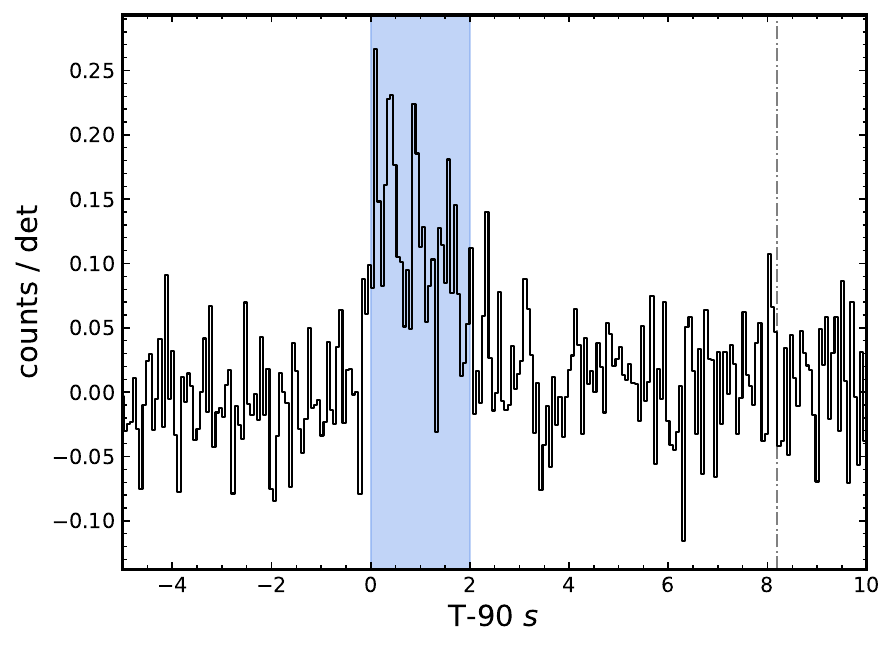}
	\caption{BAT mask-weighted lightcurve in the 15 to 350 keV range of GRB\,160410A with 64 ms binning. The blue area marks a multi-peaked structure during the first two seconds. The dash-dotted line marks the end of the $T_{90}$ duration from \protect\cite{2016GCN.19276....1S} in the same energy band.}
	\label{fig:batlc}
\end{figure}

We follow figure 8 of \cite{2009ApJ...703.1696Z} to derive a classification for GRB\,160410A in the Type I/II scheme . The GRB has (as measured by \textit{Swift}) a $T_{90}>2$ s, and also a $T_{90}/(1+z)>2$s. However, it shows a short ``Initial Pulse Complex'' followed by an ``Extended Emission'' bump, which is an often-seen feature of Type I GRBs \citep{NorrisBonnell2006ApJ}. In Fig.\,\ref{fig:batlc} we show the \textit{Swift}/BAT light curve obtained by the analysis of the {\it Swift}/BAT Gamma-Ray Burst Catalogue\footnote{http://swift.gsfc.nasa.gov/results/batgrbcat}. 

There is no information on any supernova (SN) contribution as the redshift is too high, nor are we aware of any late-time follow-up. Similarly, there is little information on the host galaxy as it is not detected even in our very deep late-time follow-up. Further decisions in the decision tree also cannot be answered with certainty, until the last step.

The bolometric isotropic energy release for this GRB is $\log{\textnormal{E}_{\gamma}/\textnormal{erg}}=53.1$, certainly not a low value, strictly leading to an ``unknown'' classification. However, we note the extremely energetic GRB\,090510 which has a similar isotropic energy release\footnote{\cite{Tsutsui2013} argue the bolometric energy release for GRB\,090510 should be measured in the 1 keV $-\;100$ MeV energy frame, resulting from the extremely high peak energy. In this frame, $E_{\textnormal{iso,bol}}$ is five times higher than in the usual frame and exceeds that of GRB\,160410A.} is also categorised as a short GRB \citep{Ackermann2010ApJ,Kann2011ApJ}. Therefore, using this scheme, we classify GRB\,160410A as a Type I candidate.

\cite{Lu2010ApJ} devise a new parameter $\epsilon(\kappa)\equiv E_{\gamma,\textnormal{iso},52}/E^\kappa_{p,z,2}$ with $E_{\gamma,\textnormal{iso},52}$ being the bolometric isotropic energy release in units of $10^{52}$ erg, and $E^\kappa_{p,z,2}$ being the intrinsic peak energy in units of $10^2$ keV. They find $\kappa=5/3$ in their study. A study of GRBs with known parameters reveals four regions separated by low/high $T_{90,z}$ and low/high $\kappa$, the delineations being $\epsilon=0.03$ ($\log\epsilon=-1.52$) and $T_{90,z}=5$s ($\log T_{90,z}=0.26$). For the parameters given by Konus-\textit{Wind} \citep{Tsvetkova2017ApJ} we derive $E_{p,z}=3853^{+1429}_{-973}$ keV and $E_{\gamma,\textnormal{iso},52}=13.21^{+0.50}_{-0.68}$. It is $T_{90,z}=3.02\pm0.59$ s for \textit{Swift} (with extended emission) and $T_{90,z}=0.58\pm0.045$ s for Konus-\textit{Wind} (only the initial pulse complex is detected). Using these values, we derive $\log\epsilon=-1.71$, which places it into the sector of Type I GRBs, though in the bottom left quadrant of Fig. 1 in \cite{Lu2010ApJ} and, within that sector, in the top right corner, close to the bulk of the Type I GRBs with EE. Only for the combination of lowest peak energy and highest isotropic energy release would it fall slightly into the sector of intrinsically short Type II GRBs.

\cite{Lu2014MNRAS} discuss the ``amplitude parameter'' $f$ which they define as the ratio between the 1 s peak count rate and the background count rate over the same time span. They simulate what $f$ value will be derived for long GRBs when adding noise to the point that the $T_{90}<2$ s, a value they designate as $f_{\textnormal{eff}}$. They find $f_{\textnormal{eff}}$ values for such noised long GRBs have a mean value $\overline{f}_{\textnormal{eff}}=1.24$, whereas $f\gtrsim2$ for Type I GRBs. For GRB\,160410A, $f=2.13$ is derived, with $f_{\textnormal{eff}}$ being essentially the same value as the extended emission is very faint (H.-J. L\"u, priv. comm.). This indicates GRB\,160410A is a true Type I GRB, but the evidence is marginal.

\cite{LZY2020} create a method to use multiwavelength data to determine the probability whether a GRB is of Type I or Type II. This uses both prompt emission data as well as environment/host galaxy data (metallicity, offset, stellar mass). Their data collection is based on \cite{Li2016ApJS} which does not include GRB\,160410A. We used their webpage\footnote{\url{http://www.physics.unlv.edu/~liye/GRB/grb_cls.html}} to input all known parameters of GRB\,160410A (the mean values as determined by \citealt{LZY2020} are used where we do not have information, such as on the host offset). Using all prompt-emission parameters and the upper limit on the host mass (see Sect.\,\ref{sec:nohost}), we derive a 97\% probability that this is a Type I GRB. However, if the extremely low metallicity is added, the probability switches completely, being vanishingly small for a Type I GRB. Considering all the other results in this section, this mainly points to the host environment of GRB\,160410A being extreme within the known parameter range of Type I GRB host galaxies (likely coupled to the high redshift), in a way that is unaccounted for in the classifier.

From an analysis of large samples of long and short GRBs, \cite{ShahmoradiNemiroff2015MNRAS} deduce the statistically most significant indicator of classification is $E_p/T_{90} [\textnormal{keV,s}^{-1}]$ (observer-frame values), with 99\% of long GRBs having values $\lesssim50$ and 95\% of short GRBs having values $\gtrsim50$. For GRB\,160410A, we find $E_p/T_{90}=173$ for the \textit{Swift}-BAT $T_{90}$ value, and $E_p/T_{90}=893$ for the Konus-\textit{Wind} duration, favoring that this is a short GRB.

\cite{Jespersen_2020} present a method to classify GRBs based on prompt emission characteristics alone. They find that GRB\,160410A is a long GRB. In their result table, they list the $T_{90}$ duration to be 96 s, significantly in excess of the value of 8.2 s \citep{2016GCN.19276....1S}. (We note that the \textit{Swift} automatic BAT analysis page for this GRB\footnote{\url{https://gcn.gsfc.nasa.gov/notices_s/682269/BA/}} even gives a value of $\sim320$ s.) \textit{Swift} entered the South Atlantic anomaly after several hundred seconds, which may influence the analysis and the significance of the Extended Emission. This, then, may also influence the classification of \cite{Jespersen_2020} (C. K. Jespersen, priv. comm.).

\section{Additional Type I GRBs and Analysis}
\label{SGRBS}
Further to several Type I GRBs taken from the sample of \cite{Kann2011ApJ}, we include the following events for our comparison with GRB\,160410A light curve:

\subsection{GRB~130603B}
Data are taken from \cite{deUgartePostigo2014AA,Pandey2019MNRAS,Cucchiara2013ApJ,Berger2013ApJ}. This bright Type I GRB is famous for showing the first clear evidence of a kilonova signature \citep{Tanvir2013Nature,Berger2013ApJ}, having the first high S/N afterglow spectrum of a short GRB \citep{deUgartePostigo2014AA} and also showing clear evidence for a jet break \citep{deUgartePostigo2014AA,Fong2014ApJ}. We fit the joint light curve with a smoothly broken power-law, finding $\alpha_1=-0.24\pm0.20$, $\alpha_2=2.55\pm0.15$, $t_b=0.281\pm0.028$ days. $n=1$ has been fixed, and with the exception of UVOT $uvm2$ and $u$ data, the host galaxy has been subtracted by the authors of the data sources. We fit the SED ($uvm2,\ u,\ g^\prime,\ V,\ r^\prime,\ i^\prime,\ z^\prime,\ J,\ K$) with an intrinsic spectral slope of $\beta=0.65$, following \cite{deUgartePostigo2014AA}, and find a large extinction $A_V=0.84\pm0.11$ mag, in full agreement with \cite{deUgartePostigo2014AA}. Note that this is lower than the result found by \cite{Japelj2015A&A}, who find $A_V=1.19^{+0.23}_{-0.12}$ mag, and SMC dust, albeit for a bluer spectral slope $\beta=0.42^{+0.12}_{-0.22}$.

\subsection{GRB~150424A}
Data are taken from \cite{2017A&A...607A..84K,2018ApJ...857..128J} as well as GCNs \citep{Malesani2015GCN17756,Butler2015GCN17762}. The redshift for this GRB is likely unknown, \cite{2019ApJ...887..206K} report spectroscopy of a nearby galaxy yields $z=0.2981$, however, deep HST imaging reveals a faint extended red object under the afterglow, likely the host galaxy \citep{2017A&A...607A..84K,2018ApJ...857..128J}. \cite{2017A&A...607A..84K} estimate $z\approx1$ from the afterglow SED, a value we adopt here.

We initially fit the afterglow with a broken power-law, and find $\alpha_1=-0.01\pm0.03$, $\alpha_2=1.60\pm0.03$, $t_b=0.396\pm0.017$ days, $n=10$ has been fixed, and no host is included (the late HST data from \citealt{2018ApJ...857..128J} are host-subtracted, and the host is very faint compared to the early data). However, this fit is statistically bad ($\chi^2/\textnormal{d.o.f.}=3.38$), and we find that especially the late-time HST data decay steeper than the extrapolation of the GROND data, which already indicates a steep decay, as initially reported by \cite{Kann2015GCN17757}. Using data only after the break ($t>0.5$ days), and setting $J_G=F125W$, we find a best fit with another broken power-law $\alpha_{2,1}=1.48\pm0.04$, $\alpha_{2,2}=2.54\pm0.21$, $t_b=4.139\pm0.739$ days, $n=10$ fixed, no host. This is fully in agreement with the results of \cite{2018ApJ...857..128J}. For this fit, the SED ($g^\prime_G,\  F606W,\ r^\prime_G,\ i^\prime_G,\ z^\prime_G,\ J_G,\ F160W$) is well-fit by a straight power-law with $\beta=0.60\pm0.36$, in accordance with the value \cite{2018ApJ...857..128J} assumed. We note that the SED derived from our initial fit shows more scatter and a shallower slope, $\beta=0.31\pm0.07$. Neither show evidence for extinction. As we assume $z=1$, $dRc=0$. Finally, we create a further SED by carefully aligning early UVOT data to the $white$ ``backbone'', and connecting that to late data by the early Keck observation \citep{2017A&A...607A..84K}. This SED yields $\beta=0.50\pm0.09$, in good agreement with our late-time SED. This likely indicates some intrinsic afterglow variability during the early plateau phase which is not captured by our broken power-law fit.

\subsection{GRB~160821B}

Data have been taken from \cite{Lamb2019ApJ,Troja2019MNRAS,Kasliwal2017ApJ,2018ApJ...857..128J} and the GCN Circulars \citep{Breeveld2016GCN19839}. This short GRB at $z=0.16$ \citep{Lamb2019ApJ} is known for its secure detection of a kilonova \citep{Lamb2019ApJ,Troja2019MNRAS}. It is also the only short GRB so far with an (albeit tentative) detection of VHE emission \citep{MAGIC2021ApJ}. As much of the light curve is dominated by kilonova light, we use the X-ray observations of the afterglow (G. P. Lamb, priv. comm.) as a stand-in, similar to the analysis of \cite{Lamb2019ApJ}. We convert the flux densities to pseudo-magnitudes and shift the earliest (post-extended emission) point to the contemporaneous early optical observations, which are unlikely to be influenced by kilonova emission ($t<0.1$ days). We find that at later times ($t\approx5-10$ days) the optical emission in the bluest available bands ($g^\prime$ and $F606W$) is in agreement with the shifted X-ray emission, indicating the kilonova emission has become very red and faded under the afterglow level in these bands. Spectral information is sparse, but a $r^\prime i^\prime z^\prime $ SED from early-time data is fit well by a simple power-law with slope $\beta=0.62\pm0.13$, therefore we assume no dust, in agreement with the localisation offset from its host galaxy.

\subsection{GRB~180418A}

Data are taken from \cite{Becerra2019ApJ,RoucoEscorial2021ApJ} as well as GCN Circulars \citep{Guidorzi2018GCN22648,Choi2018GCN22668,Horiuchi2018GCN22670,Misra2018GCN22663,Malesani2018GCN22660,Schady2018GCN22662,Schady2018GCN22666,Troja2018GCN22664}. This is an event with a very bright early afterglow \citep{Becerra2019ApJ} whose classification is unclear, however, the arguments presented in \cite{RoucoEscorial2021ApJ} indicate it is likely a short GRB, therefore we include it in this sample. The redshift is also unknown, but the host galaxy underlying the afterglow is very faint, and \cite{RoucoEscorial2021ApJ} estimate $z\approx1.0-1.5$; similar to GRB\,150424A, we adopt $z=1$ here. Using data starting 0.00417 days after the trigger (the early emission is dominated by what is likely a reverse-shock flash, \citealt{Becerra2019ApJ}), and host-subtracting the data from sources other than \cite{RoucoEscorial2021ApJ}, we find that using a single-power law decay yields a decay slope of $\alpha=0.926\pm0.003$, in general agreement with the decay slopes found in \cite{Becerra2019ApJ,RoucoEscorial2021ApJ}. Scatter combined with small error bars leads to a statistically bad fit ($\chi^2/\textnormal{d.o.f.}=5.41$). We find this fit can be improved significantly ($\chi^2/\textnormal{d.o.f.}=3.32$) by a broken power-law with $\alpha_1=0.837\pm0.009$, $\alpha_2=1.121\pm0.014$, $t_b=0.033\pm0.004$ days, $n=10$ fixed, and no host. The $\Delta\alpha$ is likely too small, and the post-break decay too shallow, for this to be a jet break. The broad SED ($uvw2,\,uvm2,\,uvw1,\,u,\,b,\,g^\prime,\,v,\,r^\prime,\,i^\prime,\,z^\prime$) shows some scatter, and is blue ($\beta=0.43\pm0.17$) with no evidence for dust. The UVOT lenticular filters are somewhat depressed compared to the rest of the data, especially $uvw2$, which is what one would expect for $z\approx1$.

\subsection{GRB~181123B}

GRB\,181123B was a high-redshift short GRB at $z=1.754$ with a bright host galaxy and a faint afterglow detection \citep{Paterson2020ApJ}. It shows extended emission \citep{2021ApJ...911L..28D} and was rapidly observed in the radio bands \citep{Rowlinson20201MNRAS,Anderson2021MNRAS}. As there is only a single $i^\prime$ detection, we have no information on colour or potential line-of-sight extinction. However, as the redshift is nearly identical to that of GRB\,160410A, we assume the same dRc to plot it in the $z = 1$ frame. Generally, the afterglow is significantly fainter and the host significantly brighter than in the case of GRB\,160410A.

\subsection{Observed GRB Afterglows}

\begin{figure}
	\includegraphics[width=\columnwidth]{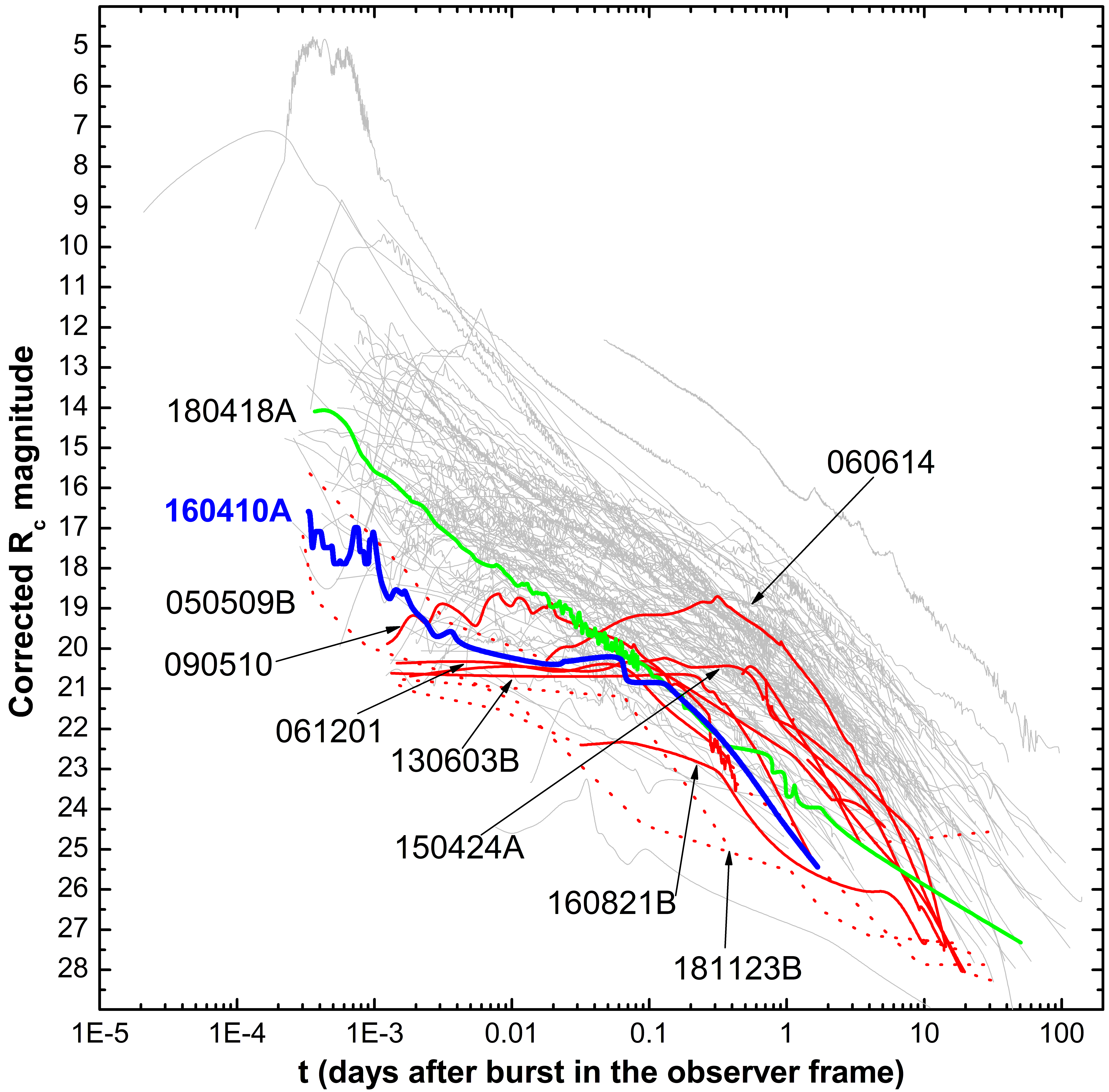}
	\caption{The afterglow of GRB\,160410A (thick blue line) in the context of a large sample of GRB afterglows. Light curves are corrected for Galactic extinction but otherwise as observed. Thin grey lines are afterglows of Type II GRBs. Thicker red lines are a selection of afterglows of other Type I GRB. We have highlighted several. The optical afterglow of GRB\,050509B was undetected, the ``light curve'' consists of deep upper limits only. GRB\,060614 is a peculiar temporally long-lasting likely Type I GRB, the afterglow shows a very late peak. The other highlighted afterglows all show early plateau phases. The afterglow of GRB\,160410A, seen to be one of the brightest ever detected (among afterglows of type I GRBs) at very early times, also evinces a plateau phase, but after an early steep decay.
}
\label{fig:KPObs}
\end{figure}

Observationally, the afterglow of GRB\,180418A, where the classification is unclear (but with more evidence pointing toward it being a Type I GRB) is the brightest at early times, followed by that of GRB\,160410A. The afterglow of GRB\,180418A lacks a plateau phase and at 0.1 days, the two afterglows are of the same magnitude; however, later on, GRB\,180418A decays less rapidly. Several other afterglows of Type I GRBs, such as the late-rising GRB 060614 and the long-plateau GRB\,150424A, are also brighter at late times.
\section{Photometric data on GRB~160410A}
\label{sec:photometry_table}
Our photometry for GRB\,160410A is given in Tab.\,\ref{tab:OAPhot} as used for the light-curve and subsequent SED fitting.

\newpage\clearpage\onecolumn
\begin{longtable}{llcll}
\caption{\label{tab:OAPhot} Photometry of the afterglow of GRB 160410A. Data are given in AB magnitudes and are not corrected for Galactic foreground extinction. Note that for UVOT data, the given exposure time is the total time coverage of each observation, for late-time data these values exceed the actual exposure time.}\\ \hline\hline
%

Time after burst & Magnitude & Bin Width/ & Filter & Telescope/Instrument \\
(days) & (AB) & Exposure Time & & \\
\hline
\endfirsthead
\caption{continued.}\\
\hline\hline 
Time after burst & Magnitude & Bin Width/ & Filter & Telescope/Instrument \\
(days) & (AB) & Exposure Time & & \\
\hline  
\endhead
\hline
\endfoot
\vspace{1mm}
0.062233	& $	21.884	^{+	0.670	}_{-	0.411	}$ &	200	&	$uvw1$	&	\textit{Swift}/UVOT	\\ \vspace{1mm}
0.131352	& $	22.564	^{+	0.479	}_{-	0.331	}$ &	900	&	$uvw1$	&	\textit{Swift}/UVOT	\\ \vspace{1mm}
0.641592	& $ >	23.195					$ &	3001	&	$uvw1$	&	\textit{Swift}/UVOT	\\
\hline \vspace{1mm}													
0.003639	& $	20.349	^{+	1.399	}_{-	0.592	}$ &	60	&	$u$	&	\textit{Swift}/UVOT	\\ \vspace{1mm}
0.003870	& $	19.848	^{+	0.713	}_{-	0.427	}$ &	20	&	$u$	&	\textit{Swift}/UVOT	\\ \vspace{1mm}
0.004102	& $	20.280	^{+	1.200	}_{-	0.556	}$ &	20	&	$u$	&	\textit{Swift}/UVOT	\\ \vspace{1mm}
0.004333	& $ >	19.357					$ &	20	&	$u$	&	\textit{Swift}/UVOT	\\ \vspace{1mm}
0.064602	& $	21.289	^{+	0.468	}_{-	0.326	}$ &	200	&	$u$	&	\textit{Swift}/UVOT	\\ \vspace{1mm}
0.140726	& $	21.893	^{+	0.394	}_{-	0.288	}$ &	705	&	$u$	&	\textit{Swift}/UVOT	\\ \vspace{1mm}
0.532774	& $ >	22.269					$ &	907	&	$u$	&	\textit{Swift}/UVOT	\\ \vspace{1mm}
0.716085	& $ >	21.919					$ &	454	&	$u$	&	\textit{Swift}/UVOT	\\
\hline \vspace{1mm}													
0.050357	& $ >	19.841					$ &	200	&	$b$	&	\textit{Swift}/UVOT	\\ \vspace{1mm}
0.066983	& $	20.842	^{+	0.680	}_{-	0.415	}$ &	200	&	$b$	&	\textit{Swift}/UVOT	\\ \vspace{1mm}
0.540848	& $	22.142	^{+	2.779	}_{-	0.710	}$ &	477	&	$b$	&	\textit{Swift}/UVOT	\\
\hline 
0.023807	& $	20.610	\pm	0.057			$ &	198	&	$g^\prime_{GROND}$	&	2.2m MPG/GROND	\\
0.846850	& $	24.440	\pm	0.207			$ &	9715	&	$g^\prime_{GROND}$	&	2.2m MPG/GROND	\\
\hline \vspace{1mm}													
0.057488	& $ >	19.510					$ &	200	&	$v$	&	\textit{Swift}/UVOT	\\ \vspace{1mm}
0.074104	& $	20.371	^{+	1.125	}_{-	0.541	}$ &	200	&	$v$	&	\textit{Swift}/UVOT	\\
\hline \vspace{1mm}													
0.001169	& $	19.823	^{+	0.294	}_{-	0.231	}$ &	20	&	$white$	&	\textit{Swift}/UVOT	\\ \vspace{1mm}
0.001400	& $	19.676	^{+	0.257	}_{-	0.207	}$ &	20	&	$white$	&	\textit{Swift}/UVOT	\\ \vspace{1mm}
0.001632	& $	20.034	^{+	0.351	}_{-	0.265	}$ &	20	&	$white$	&	\textit{Swift}/UVOT	\\ \vspace{1mm}
0.001864	& $	20.362	^{+	0.514	}_{-	0.348	}$ &	20	&	$white$	&	\textit{Swift}/UVOT	\\ \vspace{1mm}
0.002095	& $ >	20.081					$ &	20	&	$white$	&	\textit{Swift}/UVOT	\\ \vspace{1mm}
0.002440	& $	20.572	^{+	0.391	}_{-	0.287	}$ &	40	&	$white$	&	\textit{Swift}/UVOT	\\ \vspace{1mm}
0.002731	& $	21.130	^{+	3.621	}_{-	0.733	}$ &	10	&	$white$	&	\textit{Swift}/UVOT	\\ \vspace{1mm}
0.052726	& $	21.122	^{+	0.330	}_{-	0.253	}$ &	200	&	$white$	&	\textit{Swift}/UVOT	\\ \vspace{1mm}
0.069352	& $	22.172	^{+	0.508	}_{-	0.345	}$ &	200	&	$white$	&	\textit{Swift}/UVOT	\\
\hline 
0.0003298	& $	16.8	^{+	0.5	}_{-	0.5	} $ &	1	&	$r^\prime$	&	0.25m TAROT	\\ \vspace{1mm}
0.0003414	& $	16.8	^{+	0.5	}_{-	0.5	} $ &	1	&	$r^\prime$	&	0.25m TAROT	\\ \vspace{1mm}
0.0003530	& $	18.0	^{+	0.5	}_{-	0.5	} $ &	1	&	$r^\prime$	&	0.25m TAROT	\\ \vspace{1mm}
0.0003645	& $	17.3	^{+	0.7	}_{-	0.5	} $ &	1	&	$r^\prime$	&	0.25m TAROT	\\ \vspace{1mm}
0.0003761	& $	17.3	^{+	0.7	}_{-	0.5	} $ &	1	&	$r^\prime$	&	0.25m TAROT	\\ \vspace{1mm}
0.0003877	& $	17.3	^{+	0.7	}_{-	0.5	} $ &	1	&	$r^\prime$	&	0.25m TAROT	\\ \vspace{1mm}
0.0003993	& $	17.3	^{+	0.7	}_{-	0.5	} $ &	1	&	$r^\prime$	&	0.25m TAROT	\\ \vspace{1mm}
0.0004108	& $	17.3	^{+	0.7	}_{-	0.5	} $ &	1	&	$r^\prime$	&	0.25m TAROT	\\ \vspace{1mm}
0.0004224	& $	17.7	^{+	0.7	}_{-	0.5	} $ &	1	&	$r^\prime$	&	0.25m TAROT	\\ \vspace{1mm}
0.0004340	& $	17.7	^{+	0.7	}_{-	0.4	} $ &	1	&	$r^\prime$	&	0.25m TAROT	\\ \vspace{1mm}
0.0004456	& $	17.7	^{+	0.7	}_{-	0.4	} $ &	1	&	$r^\prime$	&	0.25m TAROT	\\ \vspace{1mm}
0.0004571	& $	17.7	^{+	0.7	}_{-	0.4	} $ &	1	&	$r^\prime$	&	0.25m TAROT	\\ \vspace{1mm}
0.0004687	& $	17.7	^{+	0.7	}_{-	0.4	} $ &	1	&	$r^\prime$	&	0.25m TAROT	\\ \vspace{1mm}
0.0004803	& $	17.7	^{+	0.7	}_{-	0.4	} $ &	1	&	$r^\prime$	&	0.25m TAROT	\\ \vspace{1mm}
0.0004919	& $	17.6	^{+	0.7	}_{-	0.4	} $ &	1	&	$r^\prime$	&	0.25m TAROT	\\ \vspace{1mm}
0.0005034	& $	18.1	^{+	0.2	}_{-	0.6	} $ &	1	&	$r^\prime$	&	0.25m TAROT	\\ \vspace{1mm}
0.0005150	& $	18.1	^{+	0.2	}_{-	0.6	} $ &	1	&	$r^\prime$	&	0.25m TAROT	\\ \vspace{1mm}
0.0005266	& $	18.1	^{+	0.2	}_{-	0.6	} $ &	1	&	$r^\prime$	&	0.25m TAROT	\\ \vspace{1mm}
0.0005382	& $	18.1	^{+	0.2	}_{-	0.6	} $ &	1	&	$r^\prime$	&	0.25m TAROT	\\ \vspace{1mm}
0.0005497	& $	18.1	^{+	0.2	}_{-	0.6	} $ &	1	&	$r^\prime$	&	0.25m TAROT	\\ \vspace{1mm}
0.0005613	& $	18.0	^{+	0.2	}_{-	0.6	} $ &	1	&	$r^\prime$	&	0.25m TAROT	\\ \vspace{1mm}
0.0005729	& $	18.1	^{+	0.1	}_{-	0.7	} $ &	1	&	$r^\prime$	&	0.25m TAROT	\\ \vspace{1mm}
0.0005845	& $	18.1	^{+	0.1	}_{-	0.7	} $ &	1	&	$r^\prime$	&	0.25m TAROT	\\ \vspace{1mm}
0.0005960	& $	18.1	^{+	0.1	}_{-	0.7	} $ &	1	&	$r^\prime$	&	0.25m TAROT	\\ \vspace{1mm}
0.0006076	& $	18.1	^{+	0.1	}_{-	0.7	} $ &	1	&	$r^\prime$	&	0.25m TAROT	\\ \vspace{1mm}
0.0006192	& $	18.1	^{+	0.1	}_{-	0.7	} $ &	1	&	$r^\prime$	&	0.25m TAROT	\\ \vspace{1mm}
0.0006308	& $	18.0	^{+	0.1	}_{-	0.7	} $ &	1	&	$r^\prime$	&	0.25m TAROT	\\ \vspace{1mm}
0.0006423	& $ >	18.1					$ &	1	&	$r^\prime$	&	0.25m TAROT	\\ \vspace{1mm}
0.0006539	& $ >	18.1					$ &	1	&	$r^\prime$	&	0.25m TAROT	\\ \vspace{1mm}
0.0006655	& $ >	18.1					$ &	1	&	$r^\prime$	&	0.25m TAROT	\\ \vspace{1mm}
0.0006771	& $ >	18.1					$ &	1	&	$r^\prime$	&	0.25m TAROT	\\ \vspace{1mm}
0.0006886	& $ >	18.1					$ &	1	&	$r^\prime$	&	0.25m TAROT	\\ \vspace{1mm}
0.0007002	& $ >	18.1					$ &	1	&	$r^\prime$	&	0.25m TAROT	\\ \vspace{1mm}
0.0007118	& $	17.2	^{+	0.5	}_{-	0.4	} $ &	1	&	$r^\prime$	&	0.25m TAROT	\\ \vspace{1mm}
0.0007234	& $	17.2	^{+	0.5	}_{-	0.4	} $ &	1	&	$r^\prime$	&	0.25m TAROT	\\ \vspace{1mm}
0.0007349	& $	17.2	^{+	0.5	}_{-	0.4	} $ &	1	&	$r^\prime$	&	0.25m TAROT	\\ \vspace{1mm}
0.0007465	& $	17.2	^{+	0.5	}_{-	0.4	} $ &	1	&	$r^\prime$	&	0.25m TAROT	\\ \vspace{1mm}
0.0007581	& $	17.2	^{+	0.5	}_{-	0.4	} $ &	1	&	$r^\prime$	&	0.25m TAROT	\\ \vspace{1mm}
0.0007697	& $	17.4	^{+	0.6	}_{-	0.3	} $ &	1	&	$r^\prime$	&	0.25m TAROT	\\ \vspace{1mm}
0.0007812	& $	17.8	^{+	0.6	}_{-	0.5	} $ &	1	&	$r^\prime$	&	0.25m TAROT	\\ \vspace{1mm}
0.0007928	& $	17.8	^{+	0.6	}_{-	0.5	} $ &	1	&	$r^\prime$	&	0.25m TAROT	\\ \vspace{1mm}
0.0008044	& $	17.8	^{+	0.6	}_{-	0.5	} $ &	1	&	$r^\prime$	&	0.25m TAROT	\\ \vspace{1mm}
0.0008160	& $	17.8	^{+	0.6	}_{-	0.5	} $ &	1	&	$r^\prime$	&	0.25m TAROT	\\ \vspace{1mm}
0.0008275	& $	17.8	^{+	0.6	}_{-	0.5	} $ &	1	&	$r^\prime$	&	0.25m TAROT	\\ \vspace{1mm}
0.0008391	& $	17.8	^{+	0.6	}_{-	0.5	} $ &	1	&	$r^\prime$	&	0.25m TAROT	\\ \vspace{1mm}
0.0008507	& $	18.1	^{+	0.3	}_{-	0.6	} $ &	1	&	$r^\prime$	&	0.25m TAROT	\\ \vspace{1mm}
0.0008622	& $	18.1	^{+	0.3	}_{-	0.6	} $ &	1	&	$r^\prime$	&	0.25m TAROT	\\ \vspace{1mm}
0.0008738	& $	18.1	^{+	0.3	}_{-	0.6	} $ &	1	&	$r^\prime$	&	0.25m TAROT	\\ \vspace{1mm}
0.0008854	& $	18.1	^{+	0.3	}_{-	0.6	} $ &	1	&	$r^\prime$	&	0.25m TAROT	\\ \vspace{1mm}
0.0008970	& $	18.1	^{+	0.3	}_{-	0.6	} $ &	1	&	$r^\prime$	&	0.25m TAROT	\\ \vspace{1mm}
0.0009085	& $	18.1	^{+	0.3	}_{-	0.6	} $ &	1	&	$r^\prime$	&	0.25m TAROT	\\ \vspace{1mm}
0.0009201	& $	17.5	^{+	0.7	}_{-	0.4	} $ &	1	&	$r^\prime$	&	0.25m TAROT	\\ \vspace{1mm}
0.0009317	& $	17.5	^{+	0.7	}_{-	0.4	} $ &	1	&	$r^\prime$	&	0.25m TAROT	\\ \vspace{1mm}
0.0009433	& $	17.5	^{+	0.7	}_{-	0.4	} $ &	1	&	$r^\prime$	&	0.25m TAROT	\\ \vspace{1mm}
0.0009548	& $	17.5	^{+	0.7	}_{-	0.4	} $ &	1	&	$r^\prime$	&	0.25m TAROT	\\ \vspace{1mm}
0.0009664	& $	17.5	^{+	0.7	}_{-	0.4	} $ &	1	&	$r^\prime$	&	0.25m TAROT	\\ \vspace{1mm}
0.0009780	& $	17.1	^{+	0.6	}_{-	0.4	} $ &	1	&	$r^\prime$	&	0.25m TAROT	\\ \vspace{1mm}
0.0009896	& $ >	17.9					$ &	1	&	$r^\prime$	&	0.25m TAROT	\\ \vspace{1mm}
0.0010011	& $ >	17.9					$ &	1	&	$r^\prime$	&	0.25m TAROT	\\ \vspace{1mm}
0.0010127	& $ >	17.9					$ &	1	&	$r^\prime$	&	0.25m TAROT	\\ \vspace{1mm}
0.005347	& $	20.249	\pm	0.037			$ &	5	&	$r^\prime$	&	8.2m VLT/X-shooter	\\
0.713489	& $	23.944	\pm	0.042			$ &	2700	&	$r^\prime$	&	2.5m NOT/ALFOSC	\\
1.677059	& $	25.649	\pm	0.291			$ &	3600	&	$r^\prime$	&	2.5m NOT/ALFOSC	\\
44.683899	& $ >	27.450					$ &	1800	&	$r^\prime$	&	10.4m GTC/OSIRIS	\\
\hline 
0.023807	& $	20.533	\pm	0.065			$ &	198	&	$r^\prime_{GROND}$	&	2.2m MPG/GROND	\\
0.849788	& $	24.573	\pm	0.312			$ &	10440	&	$r^\prime_{GROND}$	&	2.2m MPG/GROND	\\
\hline 
0.023807	& $	20.726	\pm	0.089			$ &	198	&	$i^\prime_{GROND}$	&	2.2m MPG/GROND	\\
0.849788	& $	23.702	\pm	0.294			$ &	10875	&	$i^\prime_{GROND}$	&	2.2m MPG/GROND	\\
\hline 
0.023807	& $	20.194	\pm	0.108			$ &	198	&	$z^\prime_{GROND}$	&	2.2m MPG/GROND	\\
0.853656	& $ >	23.395					$ &	10440	&	$z^\prime_{GROND}$	&	2.2m MPG/GROND	\\
\hline 
0.023906	& $ >	20.383					$ &	180	&	$J^\prime_{GROND}$	&	2.2m MPG/GROND	\\
0.849844	& $ >	21.510					$ &	9600	&	$J^\prime_{GROND}$	&	2.2m MPG/GROND	\\
\hline 
0.023906	& $ >	19.905					$ &	180	&	$H^\prime_{GROND}$	&	2.2m MPG/GROND	\\
0.865926	& $ >	20.976					$ &	9720	&	$H^\prime_{GROND}$	&	2.2m MPG/GROND	\\
\hline 
0.023906	& $ >	19.116					$ &	180	&	$K^\prime_{GROND}$	&	2.2m MPG/GROND	\\
0.849844	& $ >	19.066					$ &	9360	&	$K^\prime_{GROND}$	&	2.2m MPG/GROND	\\
\hline
499.651238 & $ > 24.740 $ & 3600 & $3.6\,\mu$m & \textit{Spitzer}/IRAC \\

\hline
\hline
\end{longtable}


\section{SED fit for the GRB~201221D host}
\label{sec:sed_fit}
\begin{figure}
	\includegraphics[width=\columnwidth]{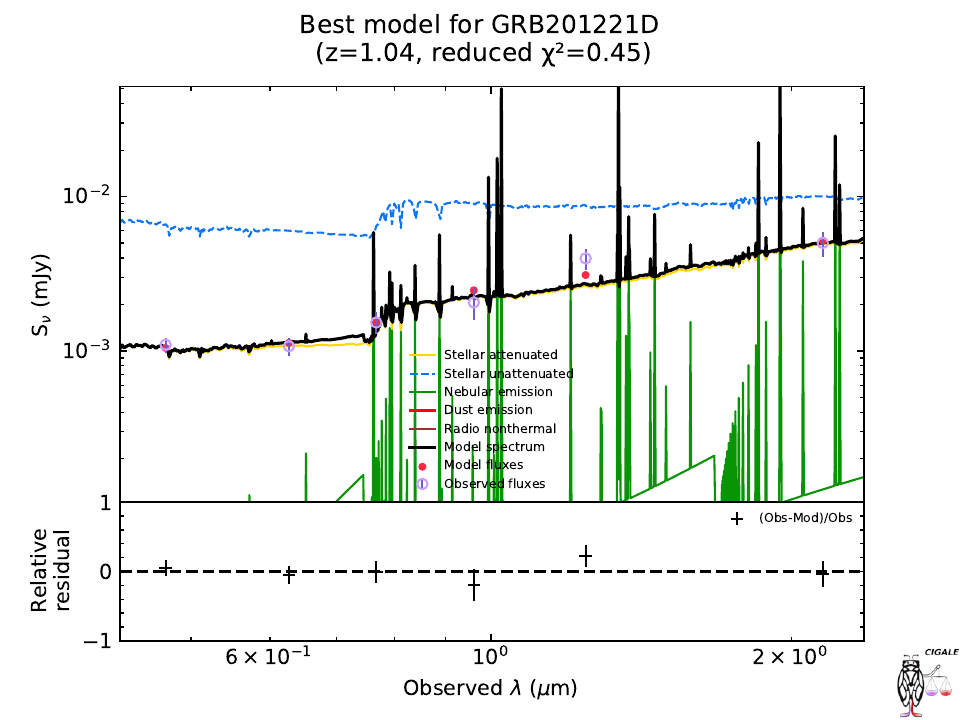}
	\caption{Best-fit SED modelling of the host of GRB\,201221D. The upper panel shows the flux density distribution for the performed SED fitting with the photometry presented in Tab.\ref{tab:201221D_host} and the lower panel the corresponding residuals.}
\label{fig:sed_fit}
\end{figure}
\bsp	
\label{lastpage}
\end{document}